%% file: main.tex
\documentclass[apjl]{emulateapj}

\usepackage{comment}
\usepackage{ifthen}

\input{newcommand_gen.tex}

\newcommand{\loopand}{\ifnum\value{planetcounter}=4 and \else\fi}
\newcommand{\loopcomma}{\ifnum\value{planetcounter}<4 ,\else. \fi}
\newcommand{\loopcommanoperiod}{\ifnum\value{planetcounter}<4 ,\else 
\space\fi}
\newcommand{\loopcommanospace}{\ifnum\value{planetcounter}<4 ,\else \fi}

\input{newcommand_spec_multi.tex}

\newcounter{planetcounter}

\newboolean{emulateapj}
\setboolean{emulateapj}{true}

\newboolean{rvtablelong}
\setboolean{rvtablelong}{true}

\newboolean{astroph}
\setboolean{astroph}{true}

\shortauthors{Bakos et al.}
\shorttitle{
\setcounter{planetcounter}{1}
\loopand\hatcur{20}\lowercase{b}\loopcommanospace
\setcounter{planetcounter}{2}
\loopand\hatcur{21}\lowercase{b}\loopcommanospace
\setcounter{planetcounter}{3}
\loopand\hatcur{22}\lowercase{b}\loopcommanospace
\setcounter{planetcounter}{4}
\loopand\hatcur{23}\lowercase{b}\loopcommanospace
}
\ifthenelse{\boolean{emulateapj}}{
    \newcommand{\titledag}{$\dagger$}
}{
    \newcommand{\titledag}{\dagger}
}

\begin{document}

\title{
\hatcur{20}\lowercase{b}--\hatcur{23}\lowercase{b}: 
Four massive transiting extrasolar planets\altaffilmark{\titledag}
}

\author{ 
    G.~\'A.~Bakos\altaffilmark{1,2},
    J.~Hartman\altaffilmark{1},
    G.~Torres\altaffilmark{1},
    D.~W.~Latham\altaffilmark{1},
    G\'eza~Kov\'acs\altaffilmark{3},
    R.~W.~Noyes\altaffilmark{1},
    D.~A.~Fischer\altaffilmark{4,5},
    J.~A.~Johnson\altaffilmark{6},
    G.~W.~Marcy\altaffilmark{7},
    A.~W.~Howard\altaffilmark{7},
    D.~Kipping\altaffilmark{1,8},
    G.~A.~Esquerdo\altaffilmark{1},
    A.~Shporer\altaffilmark{9},
    B.~B\'eky\altaffilmark{1},
    L.~A.~Buchhave\altaffilmark{10}
    G.~Perumpilly\altaffilmark{1},
    M.~Everett\altaffilmark{1},
    D.~D.~Sasselov\altaffilmark{1},
    R.~P.~Stefanik\altaffilmark{1},
    J.~L\'az\'ar\altaffilmark{11},
    I.~Papp\altaffilmark{11},
    P.~S\'ari\altaffilmark{11}
}
\altaffiltext{1}{Harvard-Smithsonian Center for Astrophysics,
    Cambridge, MA; email: gbakos@cfa.harvard.edu}

\altaffiltext{2}{NSF Fellow}

\altaffiltext{3}{Konkoly Observatory, Budapest, Hungary}

\altaffiltext{4}{Astronomy Department, Yale University,
	New Haven, CT}

\altaffiltext{5}{Department of Physics and Astronomy, San Francisco
    State University, San Francisco, CA}

\altaffiltext{6}{California Institute of Technology, Department of
    Astrophysics, MC 249-17, Pasadena, CA}

\altaffiltext{7}{Department of Astronomy, University of California,
    Berkeley, CA}

\altaffiltext{8}{University College London, Dept.~of Physics, Gower
	St., London, WC1E 6BT}

\altaffiltext{9}{
	LCOGT, 6740 Cortona Drive, Santa Barbara, CA, \& Department of Physics,
	Broida Hall, UC Santa Barbara, CA}

\altaffiltext{10}{Niels Bohr Institute, Copenhagen University, DK-2100
	Copenhagen, Denmark}

\altaffiltext{11}{Hungarian Astronomical Association, Budapest, 
    Hungary}

\altaffiltext{$\dagger$}{
    Based in part on observations obtained at the W.~M.~Keck
    Observatory, which is operated by the University of California and
    the California Institute of Technology. Keck time has been
    granted by NOAO and NASA.
}


\begin{abstract}

\setcounter{footnote}{11}
\setcounter{planetcounter}{1}
We report the discovery of four relatively massive (2--7\,\mjup)
transiting extrasolar planets.
{\bf \hatcurb{20}} orbits the moderately bright
V=\hatcurCCtassmv{20}\ \hatcurISOspec{20}\ dwarf
star \hatcurCCgsc{20} on a circular orbit, with a period
$P=\hatcurLCP{20}$\,d, transit epoch $T_c =
\hatcurLCT{20}$ ($\mathrm{BJD_{UTC}}$),
 and transit duration
\hatcurLCdur{20}\,d. The host star has a mass of
\hatcurISOm{20}\,\msun, radius of
\hatcurISOr{20}\,\rsun, effective temperature
\hatcurSMEteff{20}\,K, and metallicity $\feh =
\hatcurSMEzfeh{20}$. The planetary companion has a mass of
\hatcurPPmlong{20}\,\mjup, and radius of
\hatcurPPrlong{20}\,\rjup\ yielding a mean density of
\hatcurPPrho{20}\,\gcmc, which is the second highest value among all known
exoplanets. 
\setcounter{planetcounter}{2}
{\bf \hatcurb{21}} orbits the
V=\hatcurCCtassmv{21}\ \hatcurISOspec{21}\ dwarf
star \hatcurCCgsc{21} on an eccentric ($e=\hatcurRVeccen{21})$ orbit, 
with a period $P=\hatcurLCP{21}$\,d, transit epoch $T_c =
\hatcurLCT{21}$, and transit duration
\hatcurLCdur{21}\,d. The host star has a mass of
\hatcurISOm{21}\,\msun, radius of
\hatcurISOr{21}\,\rsun, effective temperature
\hatcurSMEteff{21}\,K, and metallicity $\feh =
\hatcurSMEzfeh{21}$. The planetary companion has a mass of
\hatcurPPmlong{21}\,\mjup, and radius of
\hatcurPPrlong{21}\,\rjup\ yielding a mean density of
\hatcurPPrho{21}\,\gcmc. 
\hatcurb{21} is a border-line object between the pM and pL class planets, 
and the transits occur near apastron.
{\bf \hatcurb{22}} orbits the bright
V=\hatcurCCtassmv{22}\ \hatcurISOspec{22}\ dwarf
star \hatcurCChd{22} on a circular orbit, with a period
$P=\hatcurLCP{22}$\,d, transit epoch $T_c =
\hatcurLCT{22}$, and transit duration
\hatcurLCdur{22}\,d. The host star has a mass of
\hatcurISOm{22}\,\msun, radius of
\hatcurISOr{22}\,\rsun, effective temperature
\hatcurSMEteff{22}\,K, and metallicity $\feh =
\hatcurSMEzfeh{22}$. The planet has a mass of
\hatcurPPmlong{22}\,\mjup, and compact radius of
\hatcurPPrlong{22}\,\rjup\ yielding a mean density of
\hatcurPPrho{22}\,\gcmc. 
The host star also harbors an M-dwarf companion at a wide separation. 
Finally, {\bf \hatcurb{23}} orbits the
V=\hatcurCCtassmv{23}\ \hatcurISOspec{23}\ dwarf
star \hatcurCCgsc{23} on a close to circular orbit, with a period
$P=\hatcurLCP{23}$\,d, transit epoch $T_c =
\hatcurLCT{23}$, and transit duration
\hatcurLCdur{23}\,d. The host star has a mass of
\hatcurISOm{23}\,\msun, radius of
\hatcurISOr{23}\,\rsun, effective temperature
\hatcurSMEteff{23}\,K, and metallicity $\feh =
\hatcurSMEzfeh{23}$. The planetary companion has a mass of
\hatcurPPmlong{23}\,\mjup, and radius of
\hatcurPPrlong{23}\,\rjup\ yielding a mean density of
\hatcurPPrho{23}\,\gcmc.
\hatcurb{23} is an inflated and massive hot Jupiter on a very short
period orbit, and has one of the shortest characteristic in-fall times
(\hatcurPPtinfall{23}\,Myr) before it gets engulfed by the star.
\setcounter{footnote}{0}
\end{abstract}

\keywords{
    planetary systems ---
    stars: individual (
\setcounter{planetcounter}{1}
\hatcur{20},
\hatcurCCgsc{20}\loopcommanoperiod
\setcounter{planetcounter}{2}
\hatcur{21},
\hatcurCCgsc{21}\loopcommanoperiod
\setcounter{planetcounter}{3}
\hatcur{22},
\hatcurCChd{22}\loopcommanoperiod
\setcounter{planetcounter}{4}
\hatcur{23},
\hatcurCCgsc{23}\loopcommanoperiod
) 
    techniques: spectroscopic, photometric
}


\section{Introduction}
\label{sec:introduction}

The majority of the $\sim90$ known transiting extrasolar planets (TEPs) have
been found to lie in the 0.5\,\mjup\ to 2.0\,\mjup\ mass range.  The
apparent drop in their mass distribution at $\sim$2\,\mjup\ has been
noted by, e.g., \citet{southworth:2009}, and by \citet{torres:2010}. 
In the currently known sample, 75\% of the TEPs have planetary mass
$\mpl<2.0\,\mjup$, and there appears to be a minor peak in their
occurrence rate at $\mpl\approx2\,\mjup$, which then sharply falls off
towards higher masses.  Are there any biases present against
discovering massive planets?  Such planets tend to be less inflated,
and theory dictates that their radii shrink as their mass increases
towards the brown dwarf regime.  According to \citet{baraffe:2010},
this reversal of the $\mpl-\rpl$ relation happens around
$\mpl\approx2-3\,\mjup$, and falls off as $\rpl\propto\mpl^{-1/8}$
\citep[see e.g.][]{fortney:2009}.  The smaller radii for massive
planets yield a minor bias against discovering them via the transit
method, since they produce shallower transits.  Very massive planets
can induce stellar variability of their host stars
\citep{shkolnik:2009}, somewhat decreasing the efficiency of detecting
their shallow transits via simple algorithms that expect constant
out-of-transit \lcs.  Also, the host stars of massive planets are
typically more rapid rotators: the average $\vsini$ for host stars with
planets $\mpl<2\mjup$ is 3.9\,\ms\ (with $3.8\,\ms$ standard deviation
around the mean), whereas the same values for the massive planet hosts
stars, are $10.9$\,\ms (with $12.7\,\ms$ standard deviation around the
mean)\footnote{This includes those 4 planets announced in this paper.}. 
The five fastest rotators all harbor planets more massive than
2\,\mjup.  This presents a bias against discovering them either via
radial velocity (RV) searches, which are more efficient around quiet
non-rotating dwarfs, or via transit searches, where the targets may be
discarded during the confirmation phase.  Along the same lines, the
large RV amplitude of the host star, as caused by the planetary
companion, may even lead to erroneous rejection during the
reconnaissance phase of candidate confirmation, since such systems
resemble eclipsing binaries.  Finally, there is a tendency that massive
planets are more likely to be eccentric\footnote{See
e.g.~exoplanets.org for statistics.} \citep{southworth:2009}, meaning
that they require more RV observations for proper mapping of their
orbits, and thus leading to a slower announcement rate.  On the other
hand, a strong bias {\em for} detecting such planets--compensating for
most of the effects above--is the fact that the large RV amplitudes of
the host stars are easier to detect, since they do not require internal
precisions at the \ms\ level (see HAT-P-2, where valuable data was
contributed to the RV fit by modest precision instruments yielding
$\sim1\,\kms$ precision; \citealt{bakos:2007}).  Altogether, while
there are minor biases for and against detecting massive transiting
planets, their overall effect appears to be negligible, and the drop in
frequency at $\gtrsim 2\,\mjup$ seems to be real.

Massive planets are important for many reasons. They provide very
strong constraints on formation and migration theories, which need to
explain the observed distribution of planetary system parameters in a
wide range \citep{baraffe:2008,baraffe:2010}, from 0.01\,\mjup\
\citep[Corot-7b; ][]{queloz:2009} to 26.4\,\mjup\
\citep[Corot-3b; ][]{deleuil:2008}.  Heavy mass objects necessitate the
inclusion of other physical mechanisms for the formation and migration,
such as planet-planet scattering \citep{chatterjee:2008,ford:2008}, and
the Kozai-mechanism \citep{fabrycky:2007}.
They are border-line objects between planets and brown-dwarfs, and help
us understand how these populations differ and overlap \citep[see ][for
a review]{leconte:2009}.  For example, a traditional definition of
planets is that they have no Deuterium burning, where the Deuterium
burning limit is thought to be around 13\,\mjup.  However, there
are large uncertainties on this limit due to the numerous model
parameters and solutions, and the fact that Deuterium may be able to
burn in the H/He layers above the core \citep{baraffe:2008}.  Another
possible definition of planets is based on their formation scenario,
i.e.~they are formed by accretion in a protoplanetary disk around their
young host star, as opposed to the gravitational collapse of a
molecular cloud (brown dwarfs).

Perhaps related to the formation and migration mechanisms, a number of
interesting correlations involving massive planets have been pointed
out.  \citet{udry:2002} noted that short period massive planets are
predominantly found in binary stellar systems.  \citet{southworth:2009}
noted that only 8.6\% of the low mass planets show significantly
eccentric orbits, whereas 77\% of the massive planets have
eccentric orbits (although low-mass systems have lower S/N RV curves,
rendering the detection of eccentric orbits more difficult). 
Curiously, there appears to be a lack of correlation between planetary
mass and host star metallicity, while one would naively think that the
formation of high mass planets (via core accretion) would require
higher metal content.  Until this work, there was a hint of a correlation
between planetary and stellar mass \citep[e.g.][]{deleuil:2008}, in the
sense that the most massive planets orbited $\mstar\gtrsim1.2\,\msun$
stars, and there was a (biased) tendency that lower mass planets orbit
less massive stars.


All of these observations suffer from small-number statistics and heavy
biases. One way of improving our knowledge is to expand the 
sample of well-characterized planets. In this work we report on 4 new
massive transiting planets around bright stars, namely
\setcounter{planetcounter}{1}
\loopand\hatcurCCgsc{20}\loopcomma
\setcounter{planetcounter}{2}
\loopand\hatcurCCgsc{21}\loopcomma
\setcounter{planetcounter}{3}
\loopand\hatcurCChd{22}\loopcomma
\setcounter{planetcounter}{4}
\loopand\hatcurCCgsc{23}\loopcomma 
This extends the currently known sample of bright ($V<13.5$) and
massive ($\mpl>2\,\mjup$) transiting planets by 30\% (from 13 to 17). 
These discoveries were made by the Hungarian-made Automated Telescope
Network \citep[HATNet;][]{bakos:2004} survey.  HATNet has been one of
the main contributors to the discovery of TEPs, among others such as
the ground-based SuperWASP \citep{pollacco:2006}, TrES
\citep{alonso:2004} and XO projects \citep{pmcc:2005}, and space-borne
searches such as CoRoT \citep{baglin:2006} and Kepler
\citep{borucki:2010}.  In operation since 2003, HATNet has now covered
approximately 14\% of the sky, searching for TEPs around bright stars
($8\lesssim I \lesssim 14$).  We operate six wide-field instruments:
four at the Fred Lawrence Whipple Observatory (FLWO) in Arizona, and
two on the roof of the hangar servicing the Smithsonian Astrophysical
Observatory's Submillimeter Array, in Hawaii.

The layout of the paper is as follows. In \refsecl{obs} we report the
detections of the photometric signals and the follow-up spectroscopic
and photometric observations for each of the planets.  In
\refsecl{analysis} we describe the analysis of the data, beginning
with the determination of the stellar parameters, continuing with a
discussion of the methods used to rule out nonplanetary, false
positive scenarios which could mimic the photometric and spectroscopic
observations, and finishing with a description of our global modeling
of the photometry and radial velocities.  Our findings are discussed
in \refsecl{discussion}.

\section{Observations}
\label{sec:obs}

\subsection{Photometric detection}
\label{sec:detection}

\begin{figure}[!ht]
\plotone{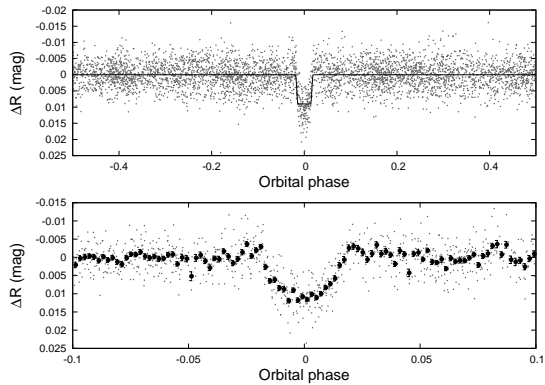}
\caption[]{
    Unbinned \lc{} of \hatcur{22} including all
    4200 instrumental \band{R} 5.5 minute cadence
    measurements obtained with the HAT-5 telescope of HATNet (see
    the text for details), and folded with the period $P =
    \hatcurLCPprec{22}$\,days resulting from the global
    fit described in \refsecl{analysis}. The solid line shows the
    ``P1P3'' transit model fit to the light curve
    (\refsecl{globmod}). The bold points
    in the lower panel show the light curve binned in phase with a
    bin-size of 0.002.
\label{fig:hatnet22}}
\end{figure}

\reftabl{photobs} summarizes the HATNet discovery observations of each
new planetary system.  The calibration of the HATNet frames was carried
out using standard procedures correcting for the CCD bias, dark-current
and flatfield structure.  The calibrated images were then subjected to
star detection and astrometry, as described in \cite{pal:2006}. 
Aperture photometry was performed on each image at the stellar
centroids derived from the Two Micron All Sky Survey
\citep[2MASS;][]{skrutskie:2006} catalog and the individual astrometric
solutions.  For certain datasets (HAT-P-20, HAT-P-22, HAT-P-23) we also
carried out an image subtraction \citep{alard:2000} based photometric
reduction using discrete kernels \citep{bramich:2008}, as described in
\citet{pal:2009b}.  The resulting \lcs\ were decorrelated (cleaned of
trends) using the External Parameter Decorrelation \citep[EPD;
  see][]{bakos:2009} technique in ``constant'' mode and the Trend
Filtering Algorithm \citep[TFA; see][]{kovacs:2005}.  The \lcs{} were
searched for periodic box-shaped signals using the Box Least-Squares
\citep[BLS; see][]{kovacs:2002} method. We detected significant signals
in the \lcs\ of the stars as summarized below:

\begin{itemize}
\item {\em \hatcur{20}} -- \hatcurCCgsc{20} (also known as
  \hatcurCCtwomass{20}; $\alpha = \hatcurCCra{20}$, $\delta =
  \hatcurCCdec{20}$; J2000; V=\hatcurCCtassmv{20}).  A signal was
  detected for this star with an apparent depth of
  $\sim$\hatcurLCdip{20}\,mmag, and a period of
  $P=$\hatcurLCPshort{20}\,days (see \reffigl{hatnet20}).
  Note that the depth was attenuated by the presence of the fainter
  neighbor star that is not resolved on the coarse resolution (14\pxs)
  HATNet pixels.  Also, the depth by fitting a trapese instead of the
  correct \citet{mandel:2002} model, is somewhat shallower than the
  maximum depth in the \citet{mandel:2002} model fit (which was
  19.6\,mmag; see later in \refsec{globmod}).  The drop in brightness
  had a first-to-last-contact duration, relative to the total period,
  of $q = $\hatcurLCq{20}, corresponding to a total duration of $Pq =
  $\hatcurLCdurhr{20}~hr.  \hatcur{20} has a red companion (2MASS
  07273963+2420171, $J-K = 0.92$) at 6.86\arcsec\ separation that is
  fainter than \hatcur{20} by $\Delta R = 1.36$\,mag.
\item {\em \hatcur{21}} -- \hatcurCCgsc{21} (also known as
  \hatcurCCtwomass{21}; $\alpha = \hatcurCCra{21}$, $\delta =
  \hatcurCCdec{21}$; J2000; V=\hatcurCCtassmv{21}).  A signal was
  detected for this star with an apparent depth of
  $\sim$\hatcurLCdip{21}\,mmag, and a period of
  $P=$\hatcurLCPshort{21}\,days (see \reffigl{hatnet21}).  The drop in
  brightness had a first-to-last-contact duration, relative to the
  total period, of $q = $\hatcurLCq{21}, corresponding to a total
  duration of $Pq = $\hatcurLCdurhr{21}~hr.
\item {\em \hatcur{22}} -- \hatcurCChd{22} (also known as
  \hatcurCCgsc{22} and \hatcurCCtwomass{22}; $\alpha =
  \hatcurCCra{22}$, $\delta = \hatcurCCdec{22}$; J2000;
  V=\hatcurCCtassmv{22}).  A signal was detected for this star with an
  apparent depth of $\sim$\hatcurLCdip{22}\,mmag, and a period of
  $P=$\hatcurLCPshort{22}\,days (see \reffigl{hatnet22}).  The drop in
  brightness had a first-to-last-contact duration, relative to the
  total period, of $q = $\hatcurLCq{22}, corresponding to a total
  duration of $Pq = $\hatcurLCdurhr{22}~hr.  \hatcur{22} has a close
  red companion star (2MASS 10224397+5007504, $J-K = 0.86$) at
  9.1\arcsec\ separation and $\Delta i = 2.58$ magnitude fainter. 
\item {\em \hatcur{23}} -- \hatcurCCgsc{23} (also known as
  \hatcurCCtwomass{23}; $\alpha = \hatcurCCra{23}$, $\delta =
  \hatcurCCdec{23}$; J2000; V=\hatcurCCtassmv{23}).  A signal was
  detected for this star with an apparent depth of $\sim 11.5$\,mmag,
  and a period of $P=$\hatcurLCPshort{23}\,days (see
  \reffigl{hatnet23}).  Similarly to \hatcur{20}, the depth was
  attenuated by close-by faint neighbors.  The drop in brightness had a
  first-to-last-contact duration, relative to the total period, of $q =
  $\hatcurLCq{23}, corresponding to a total duration of $Pq =
  $\hatcurLCdurhr{23}~hr.
\end{itemize}
%
%
\begin{figure}[!ht]
\plotone{\hatcurhtr{20}-hatnet.eps}
\caption[]{
    Unbinned \lc{} of \hatcur{20} including all 2600 instrumental
    \band{R} 5.5 minute cadence measurements obtained with the HAT-7
    and HAT-8 telescopes of HATNet (see \reftabl{photobs} for
    details), and folded with the period $P =
    \hatcurLCPprec{20}$\,days resulting from the global fit described
    in \refsecl{analysis}.  The solid line shows the ``P1P3'' transit
    model fit to the light curve (\refsecl{globmod}). The bold points
    in the lower panel show the light curve binned in phase with a
    bin-size of 0.002.
\label{fig:hatnet20}}
\end{figure}
%
\begin{figure}[!ht]
\plotone{\hatcurhtr{21}-hatnet.eps}
\caption[]{
    Unbinned \lc{} of \hatcur{21} including all 28,000 instrumental
    \band{I} and \band{R} 5.5 minute cadence measurements obtained
    with the HAT-5, HAT-6, HAT-8 and HAT-9 telescopes of HATNet (see
    \reftabl{photobs} for details), and folded with the period $P =
    \hatcurLCPprec{21}$\,days resulting from the global fit described
    in \refsecl{analysis}. The solid line shows the ``P1P3'' transit
    model fit to the light curve (\refsecl{globmod}). The bold points
    in the lower panel show the light curve binned in phase with a
    bin-size of 0.002.
\label{fig:hatnet21}}
\end{figure}
%
%
\begin{figure}[!ht]
\plotone{\hatcurhtr{23}-hatnet.eps}
\caption[]{
    Unbinned \lc{} of \hatcur{23} including all 3500 instrumental
    \band{R} 5.5 minute cadence measurements obtained with the HAT-6
    and HAT-9 telescopes of HATNet (see \reftabl{photobs} for
    details), and folded with the period $P =
    \hatcurLCPprec{23}$\,days resulting from the global fit described
    in \refsecl{analysis}. The solid line shows the ``P1P3'' transit
    model fit to the light curve (\refsecl{globmod}). The bold points
    in the lower panel show the light curve binned in phase with a
    bin-size of 0.002.
\label{fig:hatnet23}}
\end{figure}

\ifthenelse{\boolean{emulateapj}}{
    \begin{deluxetable*}{llrrr}
}{
    \begin{deluxetable}{llrrr}
}
\tablewidth{0pc}
\tabletypesize{\scriptsize}
\tablecaption{
    Summary of photometric observations
    \label{tab:photobs}
}
\tablehead{
    \colhead{~~~~~~~~Instrument/Field~~~~~~~~}  &
    \colhead{Date(s)} &
    \colhead{Number of Images} &
    \colhead{Cadence (s)} &
    \colhead{Filter}
}
\startdata
\sidehead{\textbf{\hatcur{20}}}
~~~~HAT-7/G267 & 2007 Dec--2008 May &  802 & 330 & $R$ \\
~~~~HAT-8/G267 & 2007 Oct--2008 May & 1850 & 330 & $R$ \\
~~~~KeplerCam  & 2009 Mar 11        &  268 &  43 & Sloan $i$ \\
~~~~KeplerCam  & 2009 Oct 21        &  343 &  32 & Sloan $i$ \\
\sidehead{\textbf{\hatcur{21}}}
~~~~HAT-6/G183 & 2006 Dec--2007 May & 4528 & 330 & $I$ \\
~~~~HAT-9/G183 & 2006 Nov--2007 Jun & 4586 & 330 & $I$ \\
~~~~HAT-5/G184 & 2006 Dec--2007 Jun & 4040 & 330 & $I$ \\
~~~~HAT-8/G184 & 2006 Dec--2007 Jun & 5606 & 330 & $I$ \\
~~~~HAT-6/G141 & 2008 Jan--2008 Jun & 5142 & 330 & $R$ \\
~~~~HAT-9/G141 & 2008 Jan--2008 Jun & 3964 & 330 & $R$ \\
~~~~KeplerCam  & 2009 Apr 20        &  243 &  53 & Sloan $i$ \\
~~~~KeplerCam  & 2010 Feb 15        &  412 &  43 & Sloan $i$ \\
~~~~FTN/LCOGT\tablenotemark{a}  & 2010 Feb 19        &  511 &  31 & Sloan $i$ \\
\sidehead{\textbf{\hatcur{22}}}
~~~~HAT-5/G139 & 2007 Dec--2008 May & 4288 & 330 & $R$ \\
~~~~KeplerCam  & 2009 Feb 28        &  532 &  28 & Sloan $z$ \\
~~~~KeplerCam  & 2009 Apr 30        &  353 &  33 & Sloan $g$ \\
\sidehead{\textbf{\hatcur{23}}}
~~~~HAT-6/G341 & 2007 Sep--2007 Dec & 1178 & 330 & $R$ \\
~~~~HAT-9/G341 & 2007 Sep--2007 Nov & 2351 & 330 & $R$ \\
~~~~KeplerCam  & 2008 Jun 14        &  147 &  73 & Sloan $i$ \\
~~~~KeplerCam  & 2008 Sep 08        &  246 &  73 & Sloan $i$ \\
~~~~KeplerCam  & 2008 Sep 13        &  265 &  73 & Sloan $i$ \\
~~~~KeplerCam  & 2008 Nov 03        &  117 &  89 & Sloan $i$ \\
~~~~KeplerCam  & 2009 Apr 19        &   46 & 150 & Sloan $g$ \\
~~~~KeplerCam  & 2009 Jul 13        &  150 &  73 & Sloan $i$
\enddata
\tablenotetext{a}{
	Observations were performed without guiding due to a technical
	problem with the guiding system, and resulted in decreased data
	quality.
}
\ifthenelse{\boolean{emulateapj}}{
    \end{deluxetable*}
}{
    \end{deluxetable}
}

\subsection{Reconnaissance Spectroscopy}
\label{sec:recspec}

As is routine in the HATNet project, all candidates are subjected to
careful scrutiny before investing valuable time on large telescopes. 
This includes spectroscopic observations at relatively modest
facilities to establish whether the transit-like feature in the light
curve of a candidate might be due to astrophysical phenomena other than
a planet transiting a star.  Many of these false positives are
associated with large radial-velocity variations in the star (tens of
\kms) that are easily recognized.  The reconnaissance spectroscopic
observations and results for each system are summarized in
\reftabl{reconspecobs}; below we provide a brief description of the
instruments used, the data reduction, and the analysis procedure.

%
One of the tools we have used for this purpose is the
Harvard-Smithsonian Center for Astrophysics (CfA) Digital Speedometer
\citep[DS;][]{latham:1992}, an echelle spectrograph mounted on the
\flwos\ telescope. This instrument delivers high-resolution spectra
($\lambda/\Delta\lambda \approx 35,\!000$) over a single order
centered on the \ion{Mg}{1}\,b triplet ($\sim$5187\,\AA), with
typically low signal-to-noise (S/N) ratios that are nevertheless
sufficient to derive radial velocities (RVs) with moderate precisions
of 0.5--1.0\,\kms\ for slowly rotating stars.  The same spectra can be
used to estimate the effective temperature, surface gravity, and
projected rotational velocity of the host star, as described by
\cite{torres:2002}.  With this facility we are able to reject many
types of false positives, such as F dwarfs orbited by M dwarfs,
grazing eclipsing binaries, or triple or quadruple star
systems. Additional tests are performed with other spectroscopic
material described in the next section.

%
Another of the tools we have used for this purpose is the FIbre-fed
\'Echelle Spectrograph (FIES) at the 2.5\,m Nordic Optical Telescope
(NOT) at La Palma, Spain \citep{djupvik:2010}. We used the
medium-resolution fiber which produces spectra at a resolution of
$\lambda/\Delta\lambda \approx 46,\!000$ and a wavelength coverage of
$\sim$\,3600-7400\,\AA\ to observe \hatcur{20}. The spectrum was
extracted and analyzed to measure the radial velocity, effective
temperature, surface gravity, and projected rotation velocity of the
host star, following the procedures described by \cite{buchhave:2010}.

Based on the observations summarized in \reftabl{reconspecobs} we find
that \hatcur{21}, \hatcur{22} and \hatcur{23} have rms residuals
consistent with no detectable RV variation within the precision of the
measurements.  Curiously, \hatcur{20} showed significant RV variations,
even at the modest ($\sim0.5$\,\kms) precision of the Digital
Speedometer, and the reconnaissance RV variations (including the FIES
spectrum; see later) phased up with the photometric ephemeris.  All
spectra were single-lined, i.e., there is no evidence that any of these
targets consist of more than one star.  The gravities for all of the
stars indicate that they are dwarfs.

\ifthenelse{\boolean{emulateapj}}{
    \begin{deluxetable*}{llrrrrr}
}{
    \begin{deluxetable}{llrrrrr}
}
\tablewidth{0pc}
\tabletypesize{\scriptsize}
\tablecaption{
    Summary of reconnaissance spectroscopy observations
    \label{tab:reconspecobs}
}
\tablehead{
    \multicolumn{1}{c}{Instrument}          &
    \multicolumn{1}{c}{Date(s)}             &
    \multicolumn{1}{c}{Number of Spectra}   &
    \multicolumn{1}{c}{$\teffstar$}         &
    \multicolumn{1}{c}{$\loggstar$}         &
    \multicolumn{1}{c}{$\vsini$}            &
    \multicolumn{1}{c}{$\gamma_{\rm RV}$\tablenotemark{a}} \\
    &
    &
    &
    \multicolumn{1}{c}{(K)}                 &
    \multicolumn{1}{c}{(cgs)}               &
    \multicolumn{1}{c}{(\kms)}              &
    \multicolumn{1}{c}{(\kms)}
}
\startdata
\sidehead{\textbf{\hatcur{20}}}
~~~~DS                & 2009 Feb 11--2009 Feb 15 & 3 & \hatcurDSteff{20}  & \hatcurDSlogg{20} & \hatcurDSvsini{20} & \hatcurDSgamma{20} \\
~~~~FIES              & 2009 Oct 07              & 1 & \hatcurFIESteff{20}  & \hatcurFIESlogg{20} & \hatcurFIESvsini{20} & \hatcurFIESgamma{20} \\
\sidehead{\textbf{\hatcur{21}}}
~~~~DS                & 2009 Mar 08--2009 Apr 05 & 3 & \hatcurDSteff{21}  & \hatcurDSlogg{21} & \hatcurDSvsini{21} & \hatcurDSgamma{21} \\
\sidehead{\textbf{\hatcur{22}}}
~~~~DS                & 2009 Feb 11--2009 Feb 16 & 4 & \hatcurDSteff{22}  & \hatcurDSlogg{22} & \hatcurDSvsini{22} & \hatcurDSgamma{22} \\
\sidehead{\textbf{\hatcur{23}}}
~~~~DS                & 2008 May 19--2008 Sep 14 & 5 & \hatcurDSteff{23}  & \hatcurDSlogg{23} & \hatcurDSvsini{23} & \hatcurDSgamma{23}
\enddata 
\tablenotetext{a}{
    The mean heliocentric RV of the target (in the IAU system). The
    error gives the rms of the individual velocity measures for the
    target with the given instrument.
}
\ifthenelse{\boolean{emulateapj}}{
    \end{deluxetable*}
}{
    \end{deluxetable}
}

\subsection{High resolution, high S/N spectroscopy}
\label{sec:hispec}

We proceeded with the follow-up of each candidate by obtaining
high-resolution, high-S/N spectra to characterize the RV variations,
and to refine the determination of the stellar parameters.  These
observations are summarized in \reftabl{highsnspecobs}.  The RV
measurements and uncertainties are given in \reftabl{rvs20},
\reftabl{rvs21}, \reftabl{rvs22} and \reftabl{rvs23} for \hatcur{20}
through \hatcur{23}, respectively.  The period-folded data, along with
our best fits described below in \refsecl{analysis}, are displayed in
\reffigl{rvbis20} through \reffigl{rvbis23} for \hatcur{20} through
\hatcur{23}.  Below we briefly describe the instruments used, the data
reduction, and the analysis procedure.

\ifthenelse{\boolean{emulateapj}}{
    \begin{deluxetable}{llrr}
}{
    \begin{deluxetable}{llrr}
}
\tablewidth{0pc}
\tabletypesize{\scriptsize}
\tablecaption{
    Summary of high-resolution/high-SN spectroscopic observations
    \label{tab:highsnspecobs}
}
\tablehead{
    \multicolumn{1}{c}{Instrument}  &
    \multicolumn{1}{c}{Date(s)}     &
    \multicolumn{1}{c}{Number of}   \\
    &
    &
    \multicolumn{1}{c}{RV obs.}
}
\startdata
\sidehead{\textbf{\hatcur{20}}}
~~~~Keck/HIRES           & 2009 Apr--2009 Dec & 10  \\
\sidehead{\textbf{\hatcur{21}}}
~~~~Keck/HIRES           & 2009 May--2010 Feb & 15  \\
\sidehead{\textbf{\hatcur{22}}}
~~~~Keck/HIRES           & 2009 Apr--2009 Dec & 12  \\
\sidehead{\textbf{\hatcur{23}}}
~~~~Keck/HIRES           & 2008 Jun--2009 Dec & 13
\enddata 
\ifthenelse{\boolean{emulateapj}}{
    \end{deluxetable}
}{
    \end{deluxetable}
}

\begin{figure}[ht]
\plotone{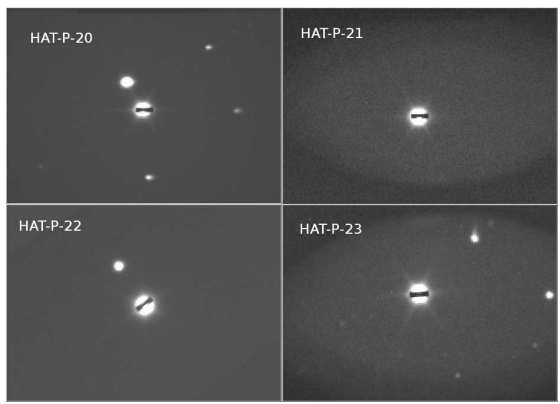}
\caption{
	Keck/HIRES guider camera snapshots of HAT-P-20 through HAT-P-23
	(labeled). North is up and East is to the left. The snapshots cover an
	area of approximately $30\times20\arcsec$. The slit is also
	visible, as positioned on the planet host stars.
}
\label{fig:kecksnap}
\end{figure}

Observations were made of all four planet host stars with the HIRES
instrument \citep{vogt:1994} on the Keck~I telescope located on Mauna
Kea, Hawaii.  The width of the spectrometer slit was $0\farcs86$,
resulting in a resolving power of $\lambda/\Delta\lambda \approx
55,\!000$, with a wavelength coverage of $\sim$3800--8000\,\AA\@.  We
typically used the B5 decker yielding a
$3.5\arcsec(H)\times0.861\arcsec(W)$ slit, and for the last few
observations on each target we used the C2 decker that enables a better
sky subtraction due to the longer slit
$14.0\arcsec(H)\times0.861\arcsec(W)$.  The slit height was oriented
with altitude (vertical), except for rare cases, when the slit would
have run through the faint companion to HAT-P-20 or HAT-P-22.  A
Keck/HIRES snapshot for each planet host star is shown in
\reffigl{kecksnap}.  Spectra were obtained through an iodine gas
absorption cell, which was used to superimpose a dense forest of
$\mathrm{I}_2$ lines on the stellar spectrum and establish an accurate
wavelength fiducial \citep[see][]{marcy:1992}.  For each target an
additional exposure was taken without the iodine cell, for use as a
template in the reductions.  Relative RVs in the solar system
barycentric frame were derived as described by \cite{butler:1996},
incorporating full modeling of the spatial and temporal variations of
the instrumental profile.

\setcounter{planetcounter}{1}
%
\begin{figure}[ht]
\ifthenelse{\boolean{emulateapj}}{
	\plotone{\hatcurhtr{20}-rv.eps}
}{
    \includegraphics[scale=0.8]{\hatcurhtr{20}-rv.eps}
}
\ifthenelse{\value{planetcounter}=1}{
\caption{
    {\em Top panel:} Keck/HIRES RV measurements for
        \hbox{\hatcur{20}{}} shown as a function of orbital
        phase, along with our best-fit eccentric model (see
        \reftabl{planetparam}).  Zero phase corresponds to the
        time of mid-transit.  The center-of-mass velocity has been
        subtracted. The rms around the best orbital fit is
        \hatcurRVfitrms{20}\,\ms.
    {\em Second panel:} Velocity $O\!-\!C$ residuals from the best
        fit. The error bars for both the top and second panel
        include a component from the
        jitter (\hatcurRVjitter{20}\,\ms) added in quadrature to
        the formal errors (see \refsecl{globmod}).
    {\em Third panel:} Bisector spans (BS), with the mean value
        subtracted. The measurement from the template spectrum is
        included (see \refsecl{blend}).
    {\em Bottom panel:} Relative chromospheric activity index $S$
        measured from the Keck spectra.
}}{
\caption{
    Keck/HIRES observations of \hatcur{20}. The panels are as in
    \reffigl{rvbis20}.  The parameters used in the
    best-fit model are given in \reftabl{planetparam}.
}}
\label{fig:rvbis20}
\end{figure}
\setcounter{planetcounter}{2}
%
\begin{figure}[ht]
\ifthenelse{\boolean{emulateapj}}{
	\plotone{\hatcurhtr{21}-rv.eps}
}{
    \includegraphics[scale=0.8]{\hatcurhtr{21}-rv.eps}
}
\ifthenelse{\value{planetcounter}=1}{
\caption{
    {\em Top panel:} Keck/HIRES RV measurements for
        \hbox{\hatcur{21}{}} shown as a function of orbital
        phase, along with our best-fit model (see
        \reftabl{planetparam}).  Zero phase corresponds to the
        time of mid-transit.  The center-of-mass velocity has been
        subtracted.
    {\em Second panel:} Velocity $O\!-\!C$ residuals from the best
        fit. The error bars include a component from astrophysical
        jitter (\hatcurRVjitter{21}\,\ms) added in quadrature to
        the formal errors (see \refsecl{globmod}).
    {\em Third panel:} Bisector spans (BS), with the mean value
        subtracted. The measurement from the template spectrum is
        included (see \refsecl{blend}).
    {\em Bottom panel:} Relative chromospheric activity index $S$
        measured from the Keck spectra.
}}{
\caption{
    Keck/HIRES observations of \hatcur{21}.  The panels are as in
	\reffigl{rvbis20}.  The parameters used in the best-fit model are given
	in \reftabl{planetparam}, the RV jitter was
	\hatcurRVjitter{21}\,\ms, and the fit rms was \hatcurRVfitrms{21}\,\ms.
	Observations shown twice are represented with open symbols.
}}
\label{fig:rvbis21}
\end{figure}
\setcounter{planetcounter}{3}
%
\begin{figure}[ht]
\ifthenelse{\boolean{emulateapj}}{
	\plotone{\hatcurhtr{22}-rv.eps}
}{
    \includegraphics[scale=0.8]{\hatcurhtr{22}-rv.eps}
}
\ifthenelse{\value{planetcounter}=1}{
\caption{
    {\em Top panel:} Keck/HIRES RV measurements for
        \hbox{\hatcur{22}{}} shown as a function of orbital
        phase, along with our best-fit model (see
        \reftabl{planetparam}).  Zero phase corresponds to the
        time of mid-transit.  The center-of-mass velocity has been
        subtracted.
    {\em Second panel:} Velocity $O\!-\!C$ residuals from the best
        fit. The error bars include a component from astrophysical
        jitter (\hatcurRVjitter{22}\,\ms) added in quadrature to
        the formal errors (see \refsecl{globmod}).
    {\em Third panel:} Bisector spans (BS), with the mean value
        subtracted. The measurement from the template spectrum is
        included (see \refsecl{blend}).
    {\em Bottom panel:} Relative chromospheric activity index $S$
        measured from the Keck spectra.
    Note the different vertical scales of the panels. Observations
    shown twice are represented with open symbols.
}}{
\caption{
    Keck/HIRES observations of \hatcur{22}. The panels are as in
    \reffigl{rvbis20}.  The parameters used in the
    best-fit model are given in \reftabl{planetparam}, the RV jitter 
	was \hatcurRVjitter{22}\,\ms, and the fit rms was 
	\hatcurRVfitrms{22}\,\ms.
	Observations
    shown twice are represented with open symbols.
}}
\label{fig:rvbis22}
\end{figure}
\setcounter{planetcounter}{4}
%
\begin{figure}[ht]
\ifthenelse{\boolean{emulateapj}}{
	\plotone{\hatcurhtr{23}-rv.eps}
}{
    \includegraphics[scale=0.8]{\hatcurhtr{23}-rv.eps}
}
\ifthenelse{\value{planetcounter}=1}{
\caption{
    {\em Top panel:} Keck/HIRES RV measurements for
        \hbox{\hatcur{23}{}} shown as a function of orbital
        phase, along with our best-fit model (see
        \reftabl{planetparam}).  Zero phase corresponds to the
        time of mid-transit.  The center-of-mass velocity has been
        subtracted.
    {\em Second panel:} Velocity $O\!-\!C$ residuals from the best
        fit. The error bars include a component from astrophysical
        jitter (\hatcurRVjitter{23}\,\ms) added in quadrature to
        the formal errors (see \refsecl{globmod}).
    {\em Third panel:} Bisector spans (BS), with the mean value
        subtracted. The measurement from the template spectrum is
        included (see \refsecl{blend}).
    {\em Bottom panel:} Relative chromospheric activity index $S$
        measured from the Keck spectra.
    Note the different vertical scales of the panels. Observations
    shown twice are represented with open symbols.
}}{
\caption{
    Keck/HIRES observations of \hatcur{23}. The panels are as in
    \reffigl{rvbis20}.  The parameters used in the
    best-fit model are given in \reftabl{planetparam}, and the RV jitter 
	was \hatcurRVjitter{23}\,\ms.
	Observations
    shown twice are represented with open symbols.
}}
\label{fig:rvbis23}
\end{figure}

In each of Figures \ref{fig:rvbis20}--\ref{fig:rvbis23} we show also
the relative $S$ index, which is a measure of the chromospheric
activity of the star derived from the flux in the cores of the
\ion{Ca}{2} H and K lines.  This index was computed following the
prescription given by \citet{vaughan:1978}, and as described in
\citet{hartman:2009}.
Note that our relative $S$ index has not been calibrated to the scale
of \citet{vaughan:1978}. We do not detect any significant variation of
the index correlated with orbital phase; such a correlation might have
indicated that the RV variations could be due to stellar activity,
casting doubt on the planetary nature of the candidate. 
There is no sign of emission in the cores of the \ion{Ca}{2} H and K
lines in any of our spectra, from which we conclude that all of the
targets have low chromospheric activity levels.

\ifthenelse{\boolean{emulateapj}}{
    \begin{deluxetable*}{lrrrrrr}
}{
    \begin{deluxetable}{lrrrrrr}
}
\tablewidth{0pc}
\tablecaption{
    Relative radial velocities, bisector spans, and activity index
    measurements of \hatcur{20}.
    \label{tab:rvs20}
}
\tablehead{
    \colhead{$\mathrm{BJD_{UTC}}$}\tablenotemark{a} &
    \colhead{RV\tablenotemark{b}} &
    \colhead{\ensuremath{\sigma_{\rm RV}}\tablenotemark{c}} &
    \colhead{BS} &
    \colhead{\ensuremath{\sigma_{\rm BS}}} &
    \colhead{S\tablenotemark{d}} &
    \colhead{\ensuremath{\sigma_{\rm S}}}\\
    \colhead{\hbox{(2,454,000$+$)}} &
    \colhead{(\ms)} &
    \colhead{(\ms)} &
    \colhead{(\ms)} &
    \colhead{(\ms)} &
    \colhead{} &
    \colhead{}
}
\startdata
\ifthenelse{\boolean{rvtablelong}}{
    \input{\hatcurhtr{20}_rvtable.tex}
	[-1.5ex]
}{
    \input{\hatcurhtr{20}_rvtable_short.tex}
	[-1.5ex]
}
\enddata
\tablenotetext{a}{
	Barycentric Julian dates throughout the paper are calculated from
    Coordinated Universal Time (UTC).
}
\tablenotetext{b}{
    The zero-point of these velocities is arbitrary. An overall offset
    $\gamma_{\rm rel}$ fitted to these velocities in \refsecl{globmod}
    has {\em not} been subtracted.
}
\tablenotetext{c}{
    Internal errors excluding the component of astrophysical jitter
    considered in \refsecl{globmod}.
}
\tablenotetext{d}{
    Relative chromospheric activity index, not calibrated to the scale
    of \citet{vaughan:1978}.
}
\ifthenelse{\boolean{rvtablelong}}{
    \tablecomments{
        Note that for the iodine-free template exposures we do not
        measure the RV but do measure the BS and S index.  Such
        template exposures can be distinguished by the missing RV
        value.
    }
}{
    \tablecomments{
        Note that for the iodine-free template exposures we do not
        measure the RV but do measure the BS and S index.  Such
        template exposures can be distinguished by the missing RV
        value.  This table is presented in its entirety in the
        electronic edition of the Astrophysical Journal.  A portion is
        shown here for guidance regarding its form and content.
    }
} 
\ifthenelse{\boolean{emulateapj}}{
    \end{deluxetable*}
}{
    \end{deluxetable}
}
%
\ifthenelse{\boolean{emulateapj}}{
    \begin{deluxetable*}{lrrrrrr}
}{
    \begin{deluxetable}{lrrrrrr}
}
\tablewidth{0pc}
\tablecaption{
    Relative radial velocities, bisector spans, and activity index
    measurements of \hatcur{21}.
    \label{tab:rvs21}
}
\tablehead{
    \colhead{$\mathrm{BJD_{UTC}}$} &
    \colhead{RV\tablenotemark{a}} &
    \colhead{\ensuremath{\sigma_{\rm RV}}\tablenotemark{b}} &
    \colhead{BS} &
    \colhead{\ensuremath{\sigma_{\rm BS}}} &
    \colhead{S\tablenotemark{c}} &
    \colhead{\ensuremath{\sigma_{\rm S}}}\\
    \colhead{\hbox{(2,454,000$+$)}} &
    \colhead{(\ms)} &
    \colhead{(\ms)} &
    \colhead{(\ms)} &
    \colhead{(\ms)} &
    \colhead{} &
    \colhead{}
}
\startdata
\ifthenelse{\boolean{rvtablelong}}{
    \input{\hatcurhtr{21}_rvtable.tex}
	[-1.5ex]
}{
    \input{\hatcurhtr{21}_rvtable_short.tex}
	[-1.5ex]
}
\enddata
\ifthenelse{\boolean{rvtablelong}}{
    \tablecomments{
		Notes for this table are identical to that of \reftabl{rvs20}.
    }
}{
    \tablecomments{
		Notes for this table are identical to that of \reftabl{rvs20}.
    }
} 
\ifthenelse{\boolean{emulateapj}}{
    \end{deluxetable*}
}{
    \end{deluxetable}
}
%
\ifthenelse{\boolean{emulateapj}}{
    \begin{deluxetable*}{lrrrrrr}
}{
    \begin{deluxetable}{lrrrrrr}
}
\tablewidth{0pc}
\tablecaption{
    Relative radial velocities, bisector spans, and activity index
    measurements of \hatcur{22}.
    \label{tab:rvs22}
}
\tablehead{
    \colhead{$\mathrm{BJD_{UTC}}$} &
    \colhead{RV\tablenotemark{a}} &
    \colhead{\ensuremath{\sigma_{\rm RV}}\tablenotemark{b}} &
    \colhead{BS} &
    \colhead{\ensuremath{\sigma_{\rm BS}}} &
    \colhead{S\tablenotemark{c}} &
    \colhead{\ensuremath{\sigma_{\rm S}}}\\
    \colhead{\hbox{(2,454,000$+$)}} &
    \colhead{(\ms)} &
    \colhead{(\ms)} &
    \colhead{(\ms)} &
    \colhead{(\ms)} &
    \colhead{} &
    \colhead{}
}
\startdata
\ifthenelse{\boolean{rvtablelong}}{
    \input{\hatcurhtr{22}_rvtable.tex}
	[-1.5ex]
}{
    \input{\hatcurhtr{22}_rvtable_short.tex}
	[-1.5ex]
}
\enddata
\ifthenelse{\boolean{rvtablelong}}{
    \tablecomments{
		Notes for this table are identical to that of \reftabl{rvs20}.
    }
}{
    \tablecomments{
		Notes for this table are identical to that of \reftabl{rvs20}.
    }
} 
\ifthenelse{\boolean{emulateapj}}{
    \end{deluxetable*}
}{
    \end{deluxetable}
}

%
\ifthenelse{\boolean{emulateapj}}{
    \begin{deluxetable*}{lrrrrrr}
}{
    \begin{deluxetable}{lrrrrrr}
}
\tablewidth{0pc}
\tablecaption{
    Relative radial velocities, bisector spans, and activity index
    measurements of \hatcur{23}.
    \label{tab:rvs23}
}
\tablehead{
    \colhead{$\mathrm{BJD_{UTC}}$} &
    \colhead{RV\tablenotemark{a}} &
    \colhead{\ensuremath{\sigma_{\rm RV}}\tablenotemark{b}} &
    \colhead{BS} &
    \colhead{\ensuremath{\sigma_{\rm BS}}} &
    \colhead{S\tablenotemark{c}} &
    \colhead{\ensuremath{\sigma_{\rm S}}}\\
    \colhead{\hbox{(2,454,000$+$)}} &
    \colhead{(\ms)} &
    \colhead{(\ms)} &
    \colhead{(\ms)} &
    \colhead{(\ms)} &
    \colhead{} &
    \colhead{}
}
\startdata
\ifthenelse{\boolean{rvtablelong}}{
    \input{\hatcurhtr{23}_rvtable.tex}
	[-1.5ex]
}{
    \input{\hatcurhtr{23}_rvtable_short.tex}
	[-1.5ex]
}
\enddata
\ifthenelse{\boolean{rvtablelong}}{
    \tablecomments{
		Notes for this table are identical to that of \reftabl{rvs20}.
    }
}{
    \tablecomments{
		Notes for this table are identical to that of \reftabl{rvs20}.
    }
} 
\ifthenelse{\boolean{emulateapj}}{
    \end{deluxetable*}
}{
    \end{deluxetable}
}

\subsection{Photometric follow-up observations}
\label{sec:phot}

\setcounter{planetcounter}{1}
%
\begin{figure}[!ht]
\ifthenelse{\boolean{emulateapj}}{
	\plotone{\hatcurhtr{20}-lc.eps}
}{
    \includegraphics[scale=0.8]{\hatcurhtr{20}-lc.eps}
}
\ifthenelse{\value{planetcounter}=1}{
\caption{
    Unbinned transit \lcs{} for \hatcur{20}, acquired with
    KeplerCam at the \flwof{} telescope.  The light curves have been
    EPD and TFA processed, as described in \refsec{globmod}.
    %
    %
    The dates of the events are indicated.  Curves after the first are
    displaced vertically for clarity.  Our best fit from the global
    modeling described in \refsecl{globmod} is shown by the solid
    lines.  Residuals from the fits are displayed at the bottom, in
    the same order as the top curves.  The error bars represent the
    photon and background shot noise, plus the readout noise. 
}}{
\caption{
    Similar to \reffigl{lc20}; here we show the follow-up
    \lcs{} for \hatcur{20}.
}}
\label{fig:lc20}
\end{figure}
\setcounter{planetcounter}{2}
%
\begin{figure}[!ht]
\ifthenelse{\boolean{emulateapj}}{
	\plotone{\hatcurhtr{21}-lc.eps}
}{
    \includegraphics[scale=0.8]{\hatcurhtr{21}-lc.eps}
}
\ifthenelse{\value{planetcounter}=1}{
\caption{
    Unbinned transit \lcs{} for \hatcur{21}, acquired with
    KeplerCam at the \flwof{} telescope.  The light curves have been
    EPD and TFA processed, as described in \refsec{globmod}.
    %
    %
    The dates of the events are indicated.  Curves after the first are
    displaced vertically for clarity.  Our best fit from the global
    modeling described in \refsecl{globmod} is shown by the solid
    lines.  Residuals from the fits are displayed at the bottom, in the
    same order as the top curves.  The error bars represent the photon
    and background shot noise, plus the readout noise.
}}{
\caption{
    Similar to \reffigl{lc20}; here we show the follow-up
    \lcs{} for \hatcur{21}.
}}
\label{fig:lc21}
\end{figure}
\setcounter{planetcounter}{3}
%
\begin{figure}[!ht]
\ifthenelse{\boolean{emulateapj}}{
	\plotone{\hatcurhtr{22}-lc.eps}
}{
    \includegraphics[scale=0.8]{\hatcurhtr{22}-lc.eps}
}
\ifthenelse{\value{planetcounter}=1}{
\caption{
    Unbinned transit \lcs{} for \hatcur{22}, acquired with
    KeplerCam at the \flwof{} telescope.  The light curves have been
    EPD and TFA processed, as described in \refsec{globmod}.
    %
    %
    The dates of the events are indicated.  Curves after the first are
    displaced vertically for clarity.  Our best fit from the global
    modeling described in \refsecl{globmod} is shown by the solid
    lines.  Residuals from the fits are displayed at the bottom, in the
    same order as the top curves.  The error bars represent the photon
    and background shot noise, plus the readout noise.
}}{
\caption{
    Similar to \reffigl{lc20}; here we show the follow-up
    \lcs{} for \hatcur{22}. 
}}
\label{fig:lc22}
\end{figure}
\setcounter{planetcounter}{4}
%
\begin{figure}[!ht]
\ifthenelse{\boolean{emulateapj}}{
	\plotone{\hatcurhtr{23}-lc.eps}
}{
    \includegraphics[scale=0.8]{\hatcurhtr{23}-lc.eps}
}
\ifthenelse{\value{planetcounter}=1}{
\caption{
    Unbinned transit \lcs{} for \hatcur{23}, acquired with
    KeplerCam at the \flwof{} telescope.  The light curves have been
    EPD and TFA processed, as described in \refsec{globmod}.
    %
    %
    The dates of the events are indicated.  Curves after the first are
    displaced vertically for clarity.  Our best fit from the global
    modeling described in \refsecl{globmod} is shown by the solid
    lines.  Residuals from the fits are displayed at the bottom, in the
    same order as the top curves.  The error bars represent the photon
    and background shot noise, plus the readout noise.
}}{
\caption{
    Similar to \reffigl{lc20}; here we show the follow-up
    \lcs{} for \hatcur{23}.
}}
\label{fig:lc23}
\end{figure}


In order to permit a more accurate modeling of the light curves, we
conducted additional photometric observations with the KeplerCam CCD
camera on the \flwof{} telescope for each star, and
with the Faulkes North Telescope (FTN) of the Las Cumbres Observatory
Global Network (LCOGT) at Hawaii for HAT-P-21 only.  The observations for
each target are summarized in \reftabl{photobs}.
%

%
The reduction of these images, including basic calibration, astrometry,
and aperture photometry, was performed as described by
\citet{bakos:2009}.  We found that the aperture photometry for
\hatcur{20} was significantly affected by the close-by neighbor star
2MASS 07273995+2420118 with $\Delta i = 1.1$ magnitude difference at
6.86\arcsec\ separation (\reffigl{kecksnap}).  Thus, we performed image
subtraction on the \flwof\ images with the same toolset used for the
HATNet reductions, but applied a discrete kernel with half-size of 5
pixels and no spatial variations.  Indeed, for this stellar
configuration, the image subtraction results proved to be superior to
the aperture photometry.  For all of the follow-up \lcs, we performed
EPD and TFA to remove trends simultaneously with the light curve
modeling (for more details, see \refsecl{analysis}, and
\citealt{bakos:2009}).  The final time series, together with our
best-fit transit \lc{} model, are shown in the top portion of Figures
\ref{fig:lc20} through \ref{fig:lc23}; the individual measurements are
reported in Tables \ref{tab:phfu20}---\ref{tab:phfu23}, for \hatcur{20}
through \hatcur{23}, respectively.

\begin{deluxetable}{lrrrr}
\tablewidth{0pc}
\tablecaption{
    High-precision differential photometry of
    \hatcur{20}\label{tab:phfu20}.
}
\tablehead{
    \colhead{$\mathrm{BJD_{UTC}}$} & 
    \colhead{Mag\tablenotemark{a}} & 
    \colhead{\ensuremath{\sigma_{\rm Mag}}} &
    \colhead{Mag(orig)\tablenotemark{b}} & 
    \colhead{Filter} \\
    \colhead{\hbox{~~~~(2,400,000$+$)~~~~}} & 
    \colhead{} & 
    \colhead{} &
    \colhead{} & 
    \colhead{}
}
\startdata
	\input{\hatcurhtr{20}_phfu_tab_short.tex}
	[-1.5ex]
\enddata
\tablenotetext{a}{
    The out-of-transit level has been subtracted. These magnitudes have
    been subjected to the EPD and TFA procedures, carried out
    simultaneously with the transit fit.
}
\tablenotetext{b}{
    Raw magnitude values without application of the EPD and TFA
    procedures.
}
\tablecomments{
    This table is available in a machine-readable form in the online
    journal.  A portion is shown here for guidance regarding its form
    and content.
}
\end{deluxetable}
%
\begin{deluxetable}{lrrrr}
\tablewidth{0pc}
\tablecaption{
    High-precision differential photometry of
    \hatcur{21}\label{tab:phfu21}.
}
\tablehead{
    \colhead{$\mathrm{BJD_{UTC}}$} & 
    \colhead{Mag\tablenotemark{a}} & 
    \colhead{\ensuremath{\sigma_{\rm Mag}}} &
    \colhead{Mag(orig)\tablenotemark{b}} & 
    \colhead{Filter} \\
    \colhead{\hbox{~~~~(2,400,000$+$)~~~~}} & 
    \colhead{} & 
    \colhead{} &
    \colhead{} & 
    \colhead{}
}
\startdata
	\input{\hatcurhtr{21}_phfu_tab_short.tex}
	[-1.5ex]
\enddata
\tablecomments{
    Notes for this table are identical to those of \reftabl{phfu20}. 
}
\end{deluxetable}
%
\begin{deluxetable}{lrrrr}
\tablewidth{0pc}
\tablecaption{
    High-precision differential photometry of
    \hatcur{22}\label{tab:phfu22}.
}
\tablehead{
    \colhead{$\mathrm{BJD_{UTC}}$} & 
    \colhead{Mag\tablenotemark{a}} & 
    \colhead{\ensuremath{\sigma_{\rm Mag}}} &
    \colhead{Mag(orig)\tablenotemark{b}} & 
    \colhead{Filter} \\
    \colhead{\hbox{~~~~(2,400,000$+$)~~~~}} & 
    \colhead{} & 
    \colhead{} &
    \colhead{} & 
    \colhead{}
}
\startdata
	\input{\hatcurhtr{22}_phfu_tab_short.tex}
	[-1.5ex]
\enddata
\tablecomments{
	Notes for this table are identical to those of \reftabl{phfu20}.
}
\end{deluxetable}
%
\begin{deluxetable}{lrrrr}
\tablewidth{0pc}
\tablecaption{
    High-precision differential photometry of
    \hatcur{23}\label{tab:phfu23}.
}
\tablehead{
    \colhead{$\mathrm{BJD_{UTC}}$} & 
    \colhead{Mag\tablenotemark{a}} & 
    \colhead{\ensuremath{\sigma_{\rm Mag}}} &
    \colhead{Mag(orig)\tablenotemark{b}} & 
    \colhead{Filter} \\
    \colhead{\hbox{~~~~(2,400,000$+$)~~~~}} & 
    \colhead{} & 
    \colhead{} &
    \colhead{} & 
    \colhead{}
}
\startdata
	\input{\hatcurhtr{23}_phfu_tab_short.tex}
	[-1.5ex]
\enddata
\tablecomments{
	Notes for this table are identical to those of \reftabl{phfu20}.
}
\end{deluxetable}

\section{Analysis}
\label{sec:analysis}

\subsection{Properties of the parent stars}
\label{sec:stelparam}

Fundamental parameters for each of the host stars, including the mass
(\mstar) and radius (\rstar), which are needed to infer the planetary
properties, depend strongly on other stellar quantities that can be
derived spectroscopically.  For this we have relied on our template
spectra obtained with the Keck/HIRES instrument, and the analysis
package known as Spectroscopy Made Easy \citep[SME;][]{valenti:1996},
along with the atomic line database of \cite{valenti:2005}.  For each
star, SME yielded the following {\em initial} values and uncertainties
(which we have conservatively increased to include our estimates of the
systematic errors):
\begin{itemize}
\item {\em \hatcur{20}} --
effective temperature $\teffstar=\hatcurSMEiteff{20}$\,K, 
stellar surface gravity $\loggstar=\hatcurSMEilogg{20}$\,(cgs),
metallicity $\feh=\hatcurSMEizfeh{20}$\,dex, and 
projected rotational velocity $\vsini=\hatcurSMEivsin{20}$\,\kms.
\item {\em \hatcur{21}} --
effective temperature $\teffstar=\hatcurSMEiteff{21}$\,K, 
stellar surface gravity $\loggstar=\hatcurSMEilogg{21}$\,(cgs),
metallicity $\feh=\hatcurSMEizfeh{21}$\,dex, and 
projected rotational velocity $\vsini=\hatcurSMEivsin{21}$\,\kms.
\item {\em \hatcur{22}} --
effective temperature $\teffstar=\hatcurSMEiteff{22}$\,K, 
stellar surface gravity $\loggstar=\hatcurSMEilogg{22}$\,(cgs),
metallicity $\feh=\hatcurSMEizfeh{22}$\,dex, and 
projected rotational velocity $\vsini=\hatcurSMEivsin{22}$\,\kms.
\item {\em \hatcur{23}} --
effective temperature $\teffstar=\hatcurSMEiteff{23}$\,K, 
stellar surface gravity $\loggstar=\hatcurSMEilogg{23}$\,(cgs),
metallicity $\feh=\hatcurSMEizfeh{23}$\,dex, and 
projected rotational velocity $\vsini=\hatcurSMEivsin{23}$\,\kms.
\end{itemize}

In principle the effective temperature and metallicity, along with the
surface gravity taken as a luminosity indicator, could be used as
constraints to infer the stellar mass and radius by comparison with
stellar evolution models.  However, the effect of \loggstar\ on the
spectral line shapes is rather subtle, and as a result it is typically
difficult to determine accurately, so that it is a rather poor
luminosity indicator in practice.  Unfortunately a trigonometric
parallax is not available for any of the host stars, since they were not
included among the targets of the {\it Hipparcos\/} mission
\citep{perryman:1997}.  For planetary transits, another possible 
constraint is
provided by the \arstar\ normalized semi-major axis, which is
closely related to \rhostar, the mean stellar density.  
The quantity
\arstar\ can be derived directly from the combination of the transit
\lcs\ \citep{sozzetti:2007} and the RV data (required for eccentric
cases, see \refsecl{globmod}).
This, in turn, allows us to
improve on the determination of the spectroscopic parameters by
supplying an indirect constraint on the weakly determined spectroscopic
value of \loggstar\ that removes degeneracies.  We take this approach
here, as described below.  The validity of our assumption, namely that
the adequate physical model describing our data is a planetary transit
(as opposed to a blend), is shown later in \refsecl{blend}.

For each system, our initial values of \teffstar, \loggstar, and \feh\
were used to determine auxiliary quantities needed in the global
modeling of the follow-up photometry and radial velocities
(specifically, the limb-darkening coefficients).  This modeling, the
details of which are described in \refsecl{globmod}, uses a Monte Carlo
approach to deliver the numerical probability distribution of \arstar\
and other fitted variables.  For further details we refer the reader to
\cite{pal:2009b}.  When combining \arstar\ (used as a proxy for
luminosity) with assumed Gaussian distributions for \teffstar\ and
\feh\ based on the SME determinations, a comparison with stellar
evolution models allows the probability distributions of other stellar
properties to be inferred, including \loggstar.  Here we use the
stellar evolution calculations from the Yonsei-Yale group
\citep[YY; ][]{yi:2001} for all planets presented in this work.  The
comparison against the model isochrones was carried out for each of
10,000 Monte Carlo trial sets for \hatcur{21}, \hatcur{22}, and
\hatcur{23}, and for 20,000 Monte Carlo trial sets for \hatcur{20} (see
\refsecl{globmod}).  Parameter combinations corresponding to unphysical
locations in the \hbox{H-R} diagram (26\% of the trials for
\hatcur{20}, and less than 1\% of the trials for the other objects)
were ignored, and replaced with another randomly drawn parameter set. 
For each system we carried out a second SME iteration in which we
adopted the value of \loggstar\ so determined and held it fixed in a
new SME analysis (coupled with a new global modeling of the RV and
\lcs), adjusting only \teffstar, \feh, and \vsini.  This gave:
\begin{itemize}
\item {\em \hatcur{20}}:
$\loggstar=\hatcurSMEiilogg{20}$,
$\teffstar=\hatcurSMEiiteff{20}$\,K, 
$\feh=\hatcurSMEiizfeh{20}$, and
$\vsini=\hatcurSMEiivsin{20}$\,\kms.
\item {\em \hatcur{21}}:
$\loggstar=\hatcurSMEiilogg{21}$,
$\teffstar=\hatcurSMEiiteff{21}$\,K, 
$\feh=\hatcurSMEiizfeh{21}$, and
$\vsini=\hatcurSMEiivsin{21}$\,\kms.
\item {\em \hatcur{22}}:
$\loggstar=\hatcurSMEiilogg{22}$,
$\teffstar=\hatcurSMEiiteff{22}$\,K, 
$\feh=\hatcurSMEiizfeh{22}$, and
$\vsini=\hatcurSMEiivsin{22}$\,\kms.
\item {\em \hatcur{23}}:
$\loggstar=\hatcurSMEiilogg{23}$,
$\teffstar=\hatcurSMEiiteff{23}$\,K, 
$\feh=\hatcurSMEiizfeh{23}$, and
$\vsini=\hatcurSMEiivsin{23}$\,\kms.
\end{itemize}
In each case the conservative uncertainties for $\teffstar$ and $\feh$
have been increased by a factor of two over their formal values, as
before.
For each system, a further iteration did not change
\loggstar\ significantly, so we adopted the values stated above as the
final atmospheric properties of the stars.  They are collected in
\reftabl{stellar}.

With the adopted spectroscopic parameters the model isochrones yield
the stellar mass and radius, and other properties.  These are listed
for each of the systems in
\reftabl{stellar}.  According
to these models
\setcounter{planetcounter}{1}
\loopand\hatcur{20} is a dwarf star with an
estimated age of
\hatcurISOage{20}\,Gyr\loopcomma
\setcounter{planetcounter}{2}
\loopand\hatcur{21} is a slightly evolved star with an
estimated age of
\hatcurISOage{21}\,Gyr\loopcomma
\setcounter{planetcounter}{3}
\loopand\hatcur{22} is a slightly evolved star with an
estimated age of
\hatcurISOage{22}\,Gyr\loopcomma
\setcounter{planetcounter}{4}
\loopand\hatcur{23} is a slightly evolved star with an
estimated age of
\hatcurISOage{23}\,Gyr\loopcomma
The inferred location of each star in a diagram of \arstar\ versus
\teffstar, analogous to the classical H--R diagram, is shown in
\reffigl{iso}.
In all cases the stellar properties and their 1$\sigma$ and 2$\sigma$
confidence ellipsoides are displayed against the backdrop of model
isochrones for a range of ages, and the appropriate stellar
metallicity.  For comparison, the locations implied by the initial SME
results are also shown (in each case with a triangle).

\setcounter{planetcounter}{1}
%
\ifthenelse{\boolean{emulateapj}}{
    \begin{figure*}[!ht]
}{
    \begin{figure}[!ht]
}
\plottwo{\hatcurhtr{20}-iso-ar.eps}{\hatcurhtr{21}-iso-ar.eps}
\plottwo{\hatcurhtr{22}-iso-ar.eps}{\hatcurhtr{23}-iso-ar.eps}
\caption{
    Upper left: Model isochrones from \cite{\hatcurisocite{20}} for
    the measured metallicity of \hatcur{20}, \feh =
    \hatcurSMEiizfehshort{20}, and ages of 3.0, 4.0, 5.0, 5.5, 6.0,
    6.5, 7.0, 7.5, 8.0, 8.5, 9.0, 10.0 and 11.0\,Gyr (left to right).
    The adopted values of $\teffstar$ and \arstar\ are shown together
    with their 1$\sigma$ and 2$\sigma$ confidence ellipsoids.  The
    initial values of \teffstar\ and \arstar\ from the first SME and
    \lc\ analyses are represented with a triangle. Upper right: Same
    as upper left, here we show the results for \hatcur{21}, with \feh
    = \hatcurSMEiizfehshort{21}, and ages of 5.0, 6.0, 7.0, 8.0, 9.0,
    10.0, 11.0, 12.0, and 13.0\,Gyr (left to right). Lower left: Same
    as upper left, here we show the results for \hatcur{22}, with \feh
    = \hatcurSMEiizfehshort{22}, and ages of 5.0, 6.0, 7.0, 8.0, 9.0,
    10., 11.0, 12., 13.0, and 14.0\,Gyr (left to right). Lower right:
    Same as upper left, here we show the results for \hatcur{23},
    with \feh = \hatcurSMEiizfehshort{23}, and ages of 0.2, 0.5, 1.0,
    1.5, 2.0, 2.5, 3.0, 3.5, 4.0, 4.5 and 5.0\,Gyr (left to right).
}
\label{fig:iso}
\ifthenelse{\boolean{emulateapj}}{
    \end{figure*}
}{
    \end{figure}
}

The stellar evolution modeling provides color indices (see
\reftabl{stellar}) that may be compared against the measured values as
a sanity check.  For each star, the best available measurements are the
near-infrared magnitudes from the 2MASS Catalogue
\citep{skrutskie:2006}, which are given in \reftabl{stellar}.  These
are converted to the photometric system of the models (ESO system)
using the transformations by \citet{carpenter:2001}.  The resulting
color indices are also shown in \reftabl{stellar} for \hatcur{20}
through \hatcur{23}, respectively.  Indeed, the colors from the stellar
evolution models and from the observations agree for all of the host
stars within 2-$\sigma$.  The distance to each object may be computed
from the absolute $K$ magnitude from the models and the 2MASS $K_s$
magnitudes, which has the advantage of being less affected by
extinction than optical magnitudes.  The results are given in
\reftabl{stellar}, where in all cases the uncertainty excludes possible
systematics in the model isochrones that are difficult to quantify.

\ifthenelse{\boolean{emulateapj}}{
    \begin{deluxetable*}{lccccl}
}{
    \begin{deluxetable}{lccccl}
}
\tablewidth{0pc}
\tabletypesize{\scriptsize}
\tablecaption{
    Stellar parameters for \hatcur{20}--\hatcur{23}
    \label{tab:stellar}
}
\tablehead{
    \colhead{~~~~~~~~Parameter~~~~~~~~} &
    \colhead{\hatcur{20}}               &
    \colhead{\hatcur{21}}               &
    \colhead{\hatcur{22}}               &
    \colhead{\hatcur{23}}               &
    \colhead{Source}
}
\startdata
\noalign{\vskip -3pt}
\sidehead{Spectroscopic properties}
~~~~$\teffstar$ (K)\dotfill         &  \hatcurSMEteff{20}   & \hatcurSMEteff{21} & \hatcurSMEteff{22} & \hatcurSMEteff{23} & SME\tablenotemark{a}\\
~~~~$\feh$\dotfill                  &  \hatcurSMEzfeh{20}   & \hatcurSMEzfeh{21} & \hatcurSMEzfeh{22} & \hatcurSMEzfeh{23} & SME                 \\
~~~~$\vsini$ (\kms)\dotfill         &  \hatcurSMEvsin{20}   & \hatcurSMEvsin{21} & \hatcurSMEvsin{22} & \hatcurSMEvsin{23} & SME                 \\
~~~~$\vmac$ (\kms)\dotfill          &  \hatcurSMEvmac{20}   & \hatcurSMEvmac{21} & \hatcurSMEvmac{22} & \hatcurSMEvmac{23} & SME                 \\
~~~~$\vmic$ (\kms)\dotfill          &  \hatcurSMEvmic{20}   & \hatcurSMEvmic{21} & \hatcurSMEvmic{22} & \hatcurSMEvmic{23} & SME                 \\
~~~~$\gamma_{\rm RV}$ (\kms)\dotfill&  \hatcurDSgamma{20}   & \hatcurDSgamma{21} & \hatcurDSgamma{22} & \hatcurDSgamma{23} & DS                  \\
\sidehead{Photometric properties}
~~~~$V$ (mag)\dotfill               &  \hatcurCCtassmv{20}  & \hatcurCCtassmv{21} & \hatcurCCtassmv{22} & \hatcurCCtassmv{23} & TASS                \\
~~~~$\vic$ (mag)\dotfill            &  \hatcurCCtassvi{20}  & \hatcurCCtassvi{21} & \hatcurCCtassvi{22} & \hatcurCCtassvi{23} & TASS                \\
~~~~$J$ (mag)\dotfill               &  \hatcurCCtwomassJmag{20} & \hatcurCCtwomassJmag{21} & \hatcurCCtwomassJmag{22} & \hatcurCCtwomassJmag{23} & 2MASS           \\
~~~~$H$ (mag)\dotfill               &  \hatcurCCtwomassHmag{20} & \hatcurCCtwomassHmag{21} & \hatcurCCtwomassHmag{22} & \hatcurCCtwomassHmag{23} & 2MASS           \\
~~~~$K_s$ (mag)\dotfill             &  \hatcurCCtwomassKmag{20} & \hatcurCCtwomassKmag{21} & \hatcurCCtwomassKmag{22} & \hatcurCCtwomassKmag{23} & 2MASS           \\
~~~~$J-K$ (mag,\hatcurjhkfilset{20})\dotfill       &  \hatcurCCesoJKmag{20}    & \hatcurCCesoJKmag{21}    & \hatcurCCesoJKmag{22}    & \hatcurCCesoJKmag{23}    & 2MASS           \\
\sidehead{Derived properties}
~~~~$\mstar$ ($\msun$)\dotfill      &  \hatcurISOmlong{20}   & \hatcurISOmlong{21} & \hatcurISOmlong{22} & \hatcurISOmlong{23} & \hatcurisoshort{20}+\hatcurlumind{20}+SME \tablenotemark{b}\\
~~~~$\rstar$ ($\rsun$)\dotfill      &  \hatcurISOrlong{20}   & \hatcurISOrlong{21} & \hatcurISOrlong{22} & \hatcurISOrlong{23} & \hatcurisoshort{20}+\hatcurlumind{20}+SME         \\
~~~~$\loggstar$ (cgs)\dotfill       &  \hatcurISOlogg{20}    & \hatcurISOlogg{21} & \hatcurISOlogg{22} & \hatcurISOlogg{23} & \hatcurisoshort{20}+\hatcurlumind{20}+SME         \\
~~~~$\lstar$ ($\lsun$)\dotfill      &  \hatcurISOlum{20}     & \hatcurISOlum{21} & \hatcurISOlum{22} & \hatcurISOlum{23} & \hatcurisoshort{20}+\hatcurlumind{20}+SME         \\
~~~~$M_V$ (mag)\dotfill             &  \hatcurISOmv{20}      & \hatcurISOmv{21} & \hatcurISOmv{22} & \hatcurISOmv{23} & \hatcurisoshort{20}+\hatcurlumind{20}+SME         \\
~~~~$M_K$ (mag,\hatcurjhkfilset{20})\dotfill &  \hatcurISOMK{20} & \hatcurISOMK{21} & \hatcurISOMK{22} & \hatcurISOMK{23} & \hatcurisoshort{20}+\hatcurlumind{20}+SME         \\
~~~~$J-K$ (mag,\hatcurjhkfilset{20})\dotfill &  \hatcurISOJK{20} & \hatcurISOJK{21} & \hatcurISOJK{22} & \hatcurISOJK{23} & \hatcurisoshort{20}+\hatcurlumind{20}+SME         \\
~~~~Age (Gyr)\dotfill               &  \hatcurISOage{20}     & \hatcurISOage{21} & \hatcurISOage{22} & \hatcurISOage{23} & \hatcurisoshort{20}+\hatcurlumind{20}+SME         \\
~~~~Distance (pc)\dotfill           &  \hatcurXdist{20}\phn  & \hatcurXdist{21} & \hatcurXdist{22} & \hatcurXdist{23} & \hatcurisoshort{20}+\hatcurlumind{20}+SME\\
[-1.5ex]
\enddata
\tablenotetext{a}{
    SME = ``Spectroscopy Made Easy'' package for the analysis of
    high-resolution spectra \citep{valenti:1996}.  These parameters
    rely primarily on SME, but have a small dependence also on the
    iterative analysis incorporating the isochrone search and global
    modeling of the data, as described in the text.
}
\tablenotetext{b}{
    \hatcurisoshort{20}+\hatcurlumind{20}+SME = Based on the \hatcurisoshort{20}
    isochrones \citep{\hatcurisocite{20}}, \hatcurlumind{20} as a luminosity
    indicator, and the SME results.
}
\ifthenelse{\boolean{emulateapj}}{
    \end{deluxetable*}
}{
    \end{deluxetable}
}

\subsection{Excluding blend scenarios}
\label{sec:blend}

Our initial spectroscopic analyses discussed in \refsecl{recspec} and
\refsecl{hispec} rule out the most obvious astrophysical false positive
scenarios.  However, more subtle phenomena such as blends (contamination by
an unresolved eclipsing binary, whether in the background or associated with
the target) can still mimic both the photometric and spectroscopic
signatures we see.  In the following section we investigate whether
such scenarios may have caused the observed photometric and spectroscopic
features.

\subsubsection{Spectral line-bisector analysis}
\label{sec:bisec}

\begin{figure*}[!ht]
\ifthenelse{\boolean{emulateapj}}{
	\plotone{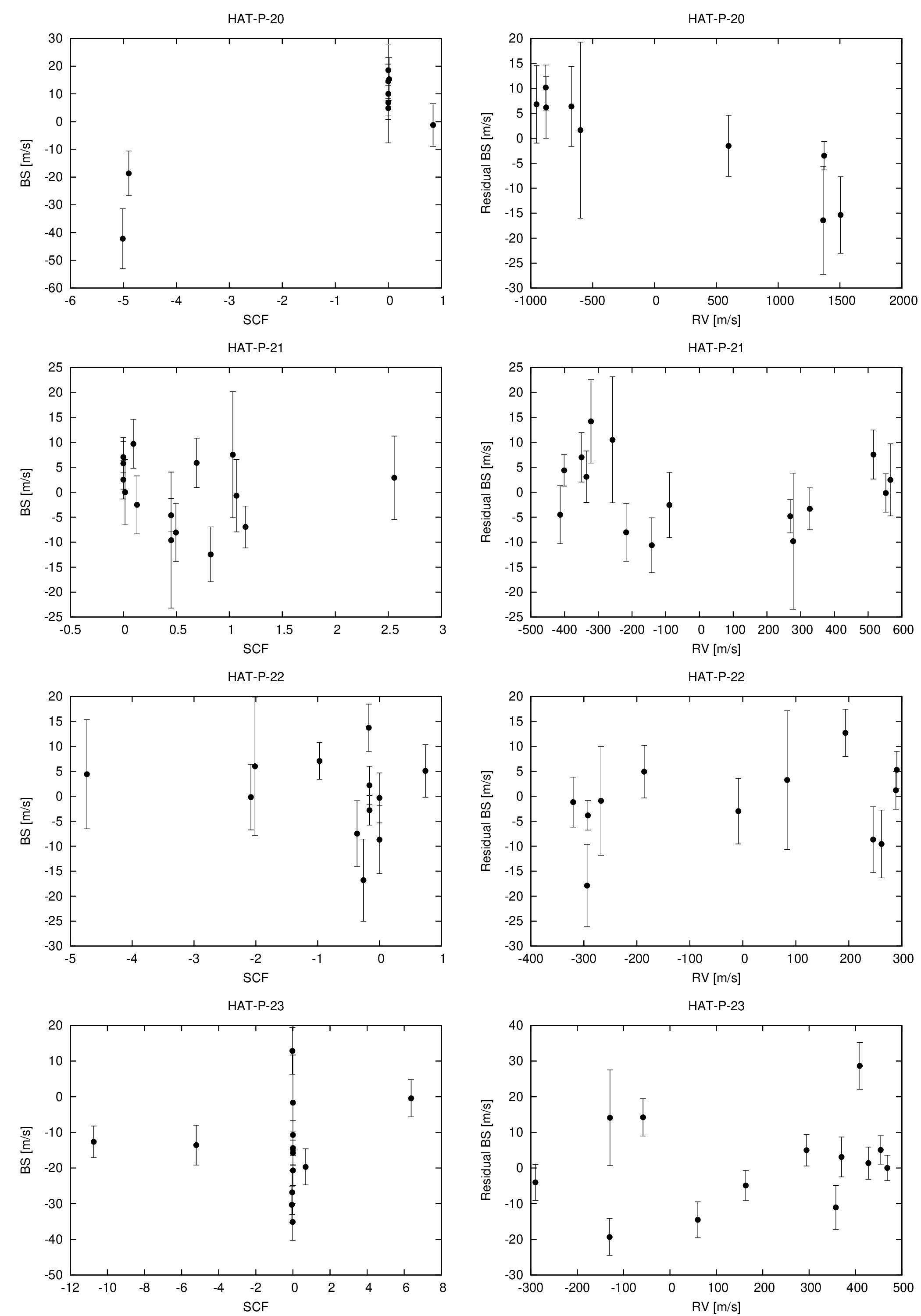}
}{
    \includegraphics[scale=0.3]{SCF_residvsRV.eps}
}
\caption[]{
	Panels on the left show the bisector spans (BS) as a function of Sky
	Contamination Factor (SCF).  Panels on the right exhibit the
	SCF-corrected BS as a function of the radial velocities. The individual
	planets are labeled. 
\label{fig:bisecscf}}
\end{figure*}

Following \cite{torres:2007}, we explored the possibility that the
measured radial velocities are not real, but are instead caused by
distortions in the spectral line profiles due to contamination from a
nearby unresolved eclipsing binary.  A bisector span (BS) analysis for
each system based on the Keck spectra was done as described in \S 5 of
\cite{bakos:2007}.  In general, none of the Keck/HIRES spectra suffer
significant sky contamination.  Nevertheless, we calculated the Sky
Contamination Factors (SCF) as described in \citet{hartman:2009}, and
corrected for the minor correlation between SCF and BS.  The results
are exhibited in \reffigl{bisecscf}, where we show the SCF--BS and
RV--$\mathrm{BS_{SCF}}$ (BS after SCF correction) plots for each
planetary system.  We also calculated the Spearman rank-order
correlation coefficients (denoted as $R_{s}$ for the RV vs.~BS
quantities and the false alarm probabilities (see \reftabl{bisec}). 
There is no correlation for HAT-P-21, HAT-P-22 and HAT-P-23, and thus
the interpretation of these systems as transiting planets is clear. 
There is an anti-correlation present for HAT-P-20, which is
strengthened when the SCF correction is applied.  A plausible
explanation for this is that the neighboring star at 6\arcsec\
separation (see \reffigl{kecksnap}) is bleeding into the slit, even
though we were careful during the observations to keep the slit
centered on the main target, and adjusted the slit orientation to be
perpendicular to the direction to the neighbor.  We simulated this
scenario, and calculated the expected BS as a function of RV due to the
neighbor, assuming that the two stars have the same systemic velocity
and the seeing is 1\arcsec.  Indeed, we get a slight anti-correlation
from this simulation, and the range of magnitude in the BS variation is
consistent with the observations.  It is also possible that some of the
anti-correlation is due to the fact that the slit was not in vertical
angle for many of the observations.  The non-vertical slit mode may
result in wavelength-dependent slit losses due to atmospheric
dispersion, and this could bring in a correlation with the sky
background, and change the shape of the spectral lines.

\ifthenelse{\boolean{emulateapj}}{
    \begin{deluxetable}{lrrrr}
}{
    \begin{deluxetable}{lrrrr}
}
\tablewidth{0pc}
\tabletypesize{\scriptsize}
\tablecaption{
    Summary of RV vs.~BS correlations.
    \label{tab:bisec}
}
\tablehead{
    \colhead{Name}  &
    \colhead{$R_{s1}$\tablenotemark{a}} &
    \colhead{$\mathrm{FAP_1}$\tablenotemark{b}} &
    \colhead{$R_{s2}$\tablenotemark{c}} &
    \colhead{$\mathrm{FAP_2}$\tablenotemark{d}}
}
\startdata
HAT-P-20  &    -0.73    &         2.46\%    &           -0.92     &           0.05\% \\
HAT-P-21  &    -0.24    &         39\%      &           -0.20     &           47\%   \\
HAT-P-22  &    0.33     &         30\%      &            0.27     &           37\%   \\
HAT-P-23  &    0.27     &         37\%      &            0.27     &           37\%   \\
[-1.5ex]
\enddata
\tablenotetext{a}{
	The Spearman correlation coefficient between the bisector
	(BS) variations and the radial velocities (RV).
}
\tablenotetext{b}{
	False alarm probability for $R_{s1}$.
}
\tablenotetext{c}{
	The Spearman correlation coefficient between BS corrected for the sky
	contamination factor (SCF) and the RVs.
}
\tablenotetext{d}{
	False alarm probability for $R_{s2}$.
}
\ifthenelse{\boolean{emulateapj}}{
    \end{deluxetable}
}{
    \end{deluxetable}
}

\subsection{Global modeling of the data}
\label{sec:globmod}

This section describes the procedure we followed for each system to
model the HATNet photometry, the follow-up photometry, and the radial
velocities simultaneously.  Our model for the follow-up \lcs\ used
analytic formulae based on \citet{mandel:2002} for the eclipse of a
star by a planet, with limb darkening being prescribed by a quadratic
law.  The limb darkening coefficients for the Sloan \band{g}, Sloan
\band{i}, and Sloan \band{z} were interpolated from the tables by
\citet{claret:2004} for the spectroscopic parameters of each star as
determined from the SME analysis (\refsecl{stelparam}).  The transit
shape was parametrized by the normalized planetary radius $p\equiv
\rpl/\rstar$, the square of the impact parameter $b^2$, and the
reciprocal of the half duration of the transit $\zrstar$.  We chose
these parameters because of their simple geometric meanings and the
fact that these show negligible correlations
\citep[see][]{bakos:2009}.  The relation between $\zrstar$ and the
quantity \arstar, used in \refsecl{stelparam}, is given by
\begin{equation}
    \arstar = P/2\pi (\zrstar) \sqrt{1-b^2} \sqrt{1-e^2}/(1+e \sin\omega)
\end{equation}
\citep[see, e.g.,][]{tingley:2005}. Note the subtle dependency of
\arstar\ on the $k \equiv e \cos\omega$ and $h \equiv e \sin\omega$
Lagrangian orbital parameters that are typically derived from the RV
data ($\omega$ is the longitude of periastron).  This dependency is
often ignored in the literature, and \arstar\ is quoted as a ``pure''
\lc\ parameter.  Of course, if high quality secondary eclipse
observations are available that determine both the location and
duration of the occultation, then $k$ and $h$ can be determined without
RV data.  Our model for the HATNet data was the simplified ``P1P3''
version of the \citet{mandel:2002} analytic functions (an expansion in
terms of Legendre polynomials), for the reasons described in
\citet{bakos:2009}.
Following the formalism presented by \citet{pal:2009}, the RVs were
fitted with an eccentric Keplerian model parametrized by the
semi-amplitude $K$ and Lagrangian elements $k$ and $h$.  Note that we
allowed for an eccentric orbit for all planets, even if the results
were consistent with a circular orbit.  There are several reasons for
this: i) many of the close-in hot Jupiters show eccentric orbits, thus
the assumption of fixing $e=0$ has no physical justification (while
this has been customary in early discoveries relying on very few
data-points) ii) the error-bars on various other derived quantities
(including $\arstar$) are more realistic with the inclusion of
eccentricity, and iii) non-zero eccentricities can be very important in
proper interpretation of these systems.

We assumed that there is a strict periodicity in the individual
transit times.  For each system we assigned the transit number $N_{tr} =
0$ to a complete follow-up \lc.  For
\setcounter{planetcounter}{1}
\loopand\hatcurb{20} this was the \lc\ gathered on 2009 Oct
21\loopcomma \setcounter{planetcounter}{2}
\loopand for \hatcurb{21}: 2010 Feb
19\loopcomma \setcounter{planetcounter}{3}
\loopand\hatcurb{22}: 2009 Feb
28\loopcomma \setcounter{planetcounter}{4}
\loopand\hatcurb{23}: 2008 Sep
13\loopcomma
The adjustable parameters in the fit that determine the ephemeris were
chosen to be the time of the first transit center observed with HATNet
($T_{c,-252}$, $T_{c,-286}$, $T_{c,-135}$, and $T_{c,-312}$ for
\hatcurb{20} through \hatcurb{23} respectively) and that of the last
transit center observed with the \flwof\ telescope ($T_{c,0}$,
$T_{c,1}$, $T_{c,19}$, and $T_{c,250}$ for \hatcurb{20} through
\hatcurb{23} respectively).  We used these as opposed to period and
reference epoch in order to minimize correlations between parameters
\citep[see][]{pal:2008}.  Times of mid-transit for intermediate events
were interpolated using these two epochs and the corresponding transit
number of each event, $N_{tr}$.  The eight main parameters describing
the physical model for each system were thus the first and last transit
center times, $\rpl/\rstar$, $b^2$, $\zrstar$, $K$, $k \equiv
e\cos\omega$, and $h \equiv e\sin\omega$.  For \hatcurb{20},
\hatcurb{22}, and \hatcurb{23} three additional parameters were
included (for each system) that have to do with the instrumental
configuration (blend factor, out-of-transit magnitudes, gamma
velocities; see later).  For \hatcurb{21} seven additional parameters
were included, because it was observed in 3 different HATNet fields. 
These are the HATNet blend factor $B_{\rm inst}$ (one for each HATNet
field), which accounts for possible dilution of the transit in the
HATNet \lc\ from background stars due to the broad PSF (20\arcsec\
FWHM), the HATNet out-of-transit magnitude $M_{\rm 0,HATNet}$ (one for
each HATNet field), and the relative zero-point $\gamma_{\rm rel}$ of
the Keck RVs.

We extended our physical model with an instrumental model that
describes brightness variations caused by systematic errors in the
measurements.  This was done in a similar fashion to the analysis
presented by \citet{bakos:2009}.  The HATNet photometry has already
been EPD- and TFA-corrected before the global modeling, so we only
considered corrections for systematics in the follow-up \lcs.  We chose
the ``ELTG'' method, i.e., EPD was performed in ``local'' mode with EPD
coefficients defined for each night, and TFA was performed in
``global'' mode using the same set of stars and TFA coefficients for
all nights.  The five EPD parameters were the hour angle (representing
a monotonic trend that changes linearly over time), the square of the
hour angle (reflecting elevation), and the stellar profile parameters
(equivalent to FWHM, elongation, and position angle of the image).
The functional forms of the above parameters contained six
coefficients, including the auxiliary out-of-transit magnitude of the
individual events.  For each system the EPD parameters were
independent for all nights, implying 12, 18, 12, and 36 additional
coefficients in the global fit for \hatcurb{20} through \hatcurb{23}
respectively.  For the global TFA analysis we chose 20, 3, 10, and 20
template stars for \hatcurb{20} through \hatcurb{23} that had good
quality measurements for all nights and on all frames, implying an
additional 20, 3, 10, and 20 parameters in the fit for each system. In
all cases the total number of fitted parameters (43, 36, 33 and 67 for
\hatcurb{20} through \hatcurb{23}) was much smaller than the number of
data points (755, 1172, 892 and 953, counting only RV measurements and
follow-up photometry measurements).

The joint fit was performed as described in \citet{bakos:2009}.  We
minimized \chisq\ in the space of parameters by using a hybrid
algorithm, combining the downhill simplex method \citep[AMOEBA;
  see][]{press:1992} with a classical linear least squares algorithm.
Uncertainties for the parameters were derived by applying the Markov
Chain Monte-Carlo method \citep[MCMC, see][]{ford:2006}.
This provided the full {\em a posteriori} probability distributions of
all adjusted variables.  The {\em a priori} distributions of the
parameters for these chains were chosen to be Gaussian, with
eigenvalues and eigenvectors derived from the Fisher covariance matrix
for the best-fit solution.  The Fisher covariance matrix was calculated
analytically using the partial derivatives given by \citet{pal:2009}.

Following this procedure we obtained the {\em a posteriori}
distributions for all fitted variables, and other quantities of
interest such as \arstar. As described in \refsecl{stelparam},
\arstar\ was used together with stellar evolution models to infer a
value for \loggstar\ that is significantly more accurate
than the spectroscopic value. The improved estimate was in turn
applied to a second iteration of the SME analysis, as explained
previously, in order to obtain better estimates of \teffstar\ and
\feh.  The global modeling was then repeated with updated
limb-darkening coefficients based on those new spectroscopic
determinations. The resulting geometric parameters pertaining to the
light curves and velocity curves for each system are listed in
\reftabl{planetparam}.

Included in each table is the RV ``jitter'', which is a noise term that
we added in quadrature to the internal errors for the RVs in order to
achieve $\chi^{2}/{\rm dof} = 1$ from the RV data for the global fit. 
The jitter is a combination of assumed astrophysical noise intrinsic to
the star, plus instrumental noise rising from uncorrected intstrumental
effects (such as a template spectrum taken under sub-optimal conditions).

The planetary parameters and their uncertainties were derived by
combining the {\em a posteriori} distributions for the stellar, \lc,
and RV parameters.  In this way we find masses and radii for each
planet.  These and other planetary parameters are listed at the bottom
of \reftabl{planetparam}, and further discussed in
\refsec{discussion}.

%
%
\ifthenelse{\boolean{emulateapj}}{
    \begin{deluxetable*}{lcccc}
}{
    \begin{deluxetable}{lcccc}
}
\tabletypesize{\tiny}
\tablecaption{Orbital and planetary parameters for 
\hatcurb{20}--\hatcurb{23}\label{tab:planetparam}}
\tablehead{
    \colhead{~~~~~~~~~~~~~~~Parameter~~~~~~~~~~~~~~~} &
    \colhead{\hatcurb{20}}                            &
    \colhead{\hatcurb{21}}                            &
    \colhead{\hatcurb{22}}                            &
    \colhead{\hatcurb{23}}                            
}
\startdata
\noalign{\vskip -3pt}
\sidehead{\Lc{} parameters}
~~~$P$ (days)             \dotfill    & $\hatcurLCP{20}$ & $\hatcurLCP{21}$ & $\hatcurLCP{22}$ & $\hatcurLCP{23}$              \\
~~~$T_c$ ($\mathrm{BJD_{UTC}}$)    
      \tablenotemark{a}   \dotfill    & $\hatcurLCT{20}$ & $\hatcurLCT{21}$ & $\hatcurLCT{22}$ & $\hatcurLCT{23}$              \\
~~~$T_{14}$ (days)
      \tablenotemark{a}   \dotfill    & $\hatcurLCdur{20}$ & $\hatcurLCdur{21}$ & $\hatcurLCdur{22}$ & $\hatcurLCdur{23}$            \\
~~~$T_{12} = T_{34}$ (days)
      \tablenotemark{a}   \dotfill    & $\hatcurLCingdur{20}$ & $\hatcurLCingdur{21}$ & $\hatcurLCingdur{22}$ & $\hatcurLCingdur{23}$         \\
~~~$\arstar$              \dotfill    & $\hatcurPPar{20}$ & $\hatcurPPar{21}$ & $\hatcurPPar{22}$ & $\hatcurPPar{23}$             \\
~~~$\zrstar$              \dotfill    & $\hatcurLCzeta{20}$ & $\hatcurLCzeta{21}$ & $\hatcurLCzeta{22}$ & $\hatcurLCzeta{23}$          \\
~~~$\rpl/\rstar$          \dotfill    & $\hatcurLCrprstar{20}$ & $\hatcurLCrprstar{21}$ & $\hatcurLCrprstar{22}$ & $\hatcurLCrprstar{23}$        \\
~~~$b^2$                  \dotfill    & $\hatcurLCbsq{20}$ & $\hatcurLCbsq{21}$ & $\hatcurLCbsq{22}$ & $\hatcurLCbsq{23}$            \\
~~~$b \equiv a \cos i/\rstar$
                          \dotfill    & $\hatcurLCimp{20}$ & $\hatcurLCimp{21}$ & $\hatcurLCimp{22}$ & $\hatcurLCimp{23}$            \\
~~~$i$ (deg)              \dotfill    & $\hatcurPPi{20}$ & $\hatcurPPi{21}$ & $\hatcurPPi{22}$ & $\hatcurPPi{23}$             \\

\sidehead{Limb-darkening coefficients \tablenotemark{b}}
~~~$a_i$ (linear term)  \dotfill    & $\hatcurLBii{20}$ & $\hatcurLBii{21}$ & $\hatcurLBii{22}$ & $\hatcurLBii{23}$             \\
~~~$b_i$ (quadratic term) \dotfill  & $\hatcurLBiii{20}$ & $\hatcurLBiii{21}$ & $\hatcurLBiii{22}$ & $\hatcurLBiii{23}$            \\
~~~$a_g$               \dotfill    & $\cdots$ & $\cdots$ & $\hatcurLBig{22}$ & $\hatcurLBig{23}$             \\
~~~$b_g$               \dotfill    & $\cdots$ & $\cdots$ & $\hatcurLBiig{22}$ & $\hatcurLBiig{23}$            \\

\sidehead{RV parameters}
~~~$K$ (\ms)              \dotfill    & $\hatcurRVK{20}$ & $\hatcurRVK{21}$ & $\hatcurRVK{22}$ & $\hatcurRVK{23}$              \\
~~~$k_{\rm RV}$\tablenotemark{c} 
                          \dotfill    & $\hatcurRVk{20}$ & $\hatcurRVk{21}$ & $\hatcurRVk{22}$ & $\hatcurRVk{23}$              \\
~~~$h_{\rm RV}$\tablenotemark{c}
                          \dotfill    & $\hatcurRVh{20}$ & $\hatcurRVh{21}$ & $\hatcurRVh{22}$ & $\hatcurRVh{23}$              \\
~~~$e$                    \dotfill    & $\hatcurRVeccen{20}$ & $\hatcurRVeccen{21}$ & $\hatcurRVeccen{22}$ & $\hatcurRVeccen{23}$          \\
%
~~~$\omega$ (deg)         \dotfill    & $\hatcurRVomega{20}$ & $\hatcurRVomega{21}$ & $\hatcurRVomega{22}$ & $\hatcurRVomega{23}$          \\
~~~RV jitter (\ms)        \dotfill    & \hatcurRVjitter{20}  & \hatcurRVjitter{21}  & \hatcurRVjitter{22}  & \hatcurRVjitter{23}           \\

\sidehead{Secondary eclipse parameters (derived)}
~~~$T_s$ ($\mathrm{BJD_{UTC}}$)            \dotfill    & $\hatcurXsecondary{20}$ & $\hatcurXsecondary{21}$ & $\hatcurXsecondary{22}$ & $\hatcurXsecondary{23}$       \\
~~~$T_{s,14}$              \dotfill   & $\hatcurXsecdur{20}$ & $\hatcurXsecdur{21}$ & $\hatcurXsecdur{22}$ & $\hatcurXsecdur{23}$          \\
~~~$T_{s,12}$              \dotfill   & $\hatcurXsecingdur{20}$ & $\hatcurXsecingdur{21}$ & $\hatcurXsecingdur{22}$ & $\hatcurXsecingdur{23}$       \\

\sidehead{Planetary parameters}
~~~$\mpl$ ($\mjup$)       \dotfill    & $\hatcurPPmlong{20}$ & $\hatcurPPmlong{21}$ & $\hatcurPPmlong{22}$ & $\hatcurPPmlong{23}$          \\
~~~$\rpl$ ($\rjup$)       \dotfill    & $\hatcurPPrlong{20}$ & $\hatcurPPrlong{21}$ & $\hatcurPPrlong{22}$ & $\hatcurPPrlong{23}$          \\
~~~$C(\mpl,\rpl)$
    \tablenotemark{d}     \dotfill    & $\hatcurPPmrcorr{20}$ & $\hatcurPPmrcorr{21}$ & $\hatcurPPmrcorr{22}$ & $\hatcurPPmrcorr{23}$         \\
~~~$\rhopl$ (\gcmc)       \dotfill    & $\hatcurPPrho{20}$ & $\hatcurPPrho{21}$ & $\hatcurPPrho{22}$ & $\hatcurPPrho{23}$            \\
~~~$\log g_p$ (cgs)       \dotfill    & $\hatcurPPlogg{20}$ & $\hatcurPPlogg{21}$ & $\hatcurPPlogg{22}$ & $\hatcurPPlogg{23}$           \\
~~~$a$ (AU)               \dotfill    & $\hatcurPParel{20}$ & $\hatcurPParel{21}$ & $\hatcurPParel{22}$ & $\hatcurPParel{23}$           \\
~~~$T_{\rm eq}$ (K)        \dotfill   & $\hatcurPPteff{20}$ & $\hatcurPPteff{21}$ & $\hatcurPPteff{22}$ & $\hatcurPPteff{23}$           \\
~~~$\Theta$\tablenotemark{e} \dotfill & $\hatcurPPtheta{20}$ & $\hatcurPPtheta{21}$ & $\hatcurPPtheta{22}$ & $\hatcurPPtheta{23}$          \\
%
~~~$F_{per}$ ($10^{\hatcurPPfluxperidim{20}}$\ergscmsq) \tablenotemark{f}
                          \dotfill    & $\hatcurPPfluxperi{20}$ & $\hatcurPPfluxperi{21}$ & $\hatcurPPfluxperi{22}$ & $\hatcurPPfluxperi{23}$      \\
~~~$F_{ap}$  ($10^{\hatcurPPfluxapdim{20}}$\ergscmsq) \tablenotemark{f} 
                          \dotfill    & $\hatcurPPfluxap{20}$ & $\hatcurPPfluxap{21}$ & $\hatcurPPfluxap{22}$ & $\hatcurPPfluxap{23}$        \\
~~~$\langle F \rangle$ ($10^{\hatcurPPfluxavgdim{20}}$\ergscmsq) \tablenotemark{f}
                          \dotfill    & $\hatcurPPfluxavg{20}$ & $\hatcurPPfluxavg{21}$ & $\hatcurPPfluxavg{22}$ & $\hatcurPPfluxavg{23}$        \\
[-1.0ex]
\enddata
\tablenotetext{a}{
    \ensuremath{T_c}: Reference epoch of mid transit that
    minimizes the correlation with the orbital period. It
    corresponds to $N_{tr} = -16$. BJD is calculated from UTC.
    \ensuremath{T_{14}}: total transit duration, time
    between first to last contact;
    \ensuremath{T_{12}=T_{34}}: ingress/egress time, time between first
    and second, or third and fourth contact.
}
\tablenotetext{b}{
    Values for a quadratic law, adopted from the tabulations by
    \cite{claret:2004} according to the spectroscopic (SME) parameters
    listed in \reftabl{stellar}.
}
\tablenotetext{c}{
    Lagrangian orbital parameters derived from the global modeling, 
    and primarily determined by the RV data. 
}
\tablenotetext{d}{
    Correlation coefficient between the planetary mass \mpl\ and radius
    \rpl.
}
\tablenotetext{e}{
    The Safronov number is given by $\Theta = \frac{1}{2}(V_{\rm
    esc}/V_{\rm orb})^2 = (a/\rpl)(\mpl / \mstar )$
    \citep[see][]{hansen:2007}.
}
\tablenotetext{f}{
    Incoming flux per unit surface area. $\langle F \rangle$ is
	averaged over the orbit.
}
\ifthenelse{\boolean{emulateapj}}{
    \end{deluxetable*}
}{
    \end{deluxetable}
}


\section{Discussion}
\label{sec:discussion}

\subsection{\hatcurb{20}}
\label{sec:disc20}

\hatcurb{20} is a very massive ($\mpl =
\hatcurPPmlong{20}\,\mjup=\hatcurPPme{20}\,\mearth$) and very compact
($\rpl=\hatcurPPrlong{20}\,\rjup$) hot Jupiter orbiting a
\hatcurISOspec{20} \citep{skiff:2009} star.  \hatcurb{20} is the sixth
most massive, and second most dense transiting planet with
$\rho_p=$\hatcurPPrho{20}\,\gcmc (see \reffigl{exomr}).  The only
planet (or brown dwarf) denser than \hatcurb{20} is CoRoT-3b
\citep{deleuil:2008} with $\rhopl \approx 27\gcmc$.  Modeling
\hatcurb{20} may be a challenge, as the oldest (4\,Gyr, i.e.~yielding
the most compact planets) \citet{fortney:2007} models with
$\mpl=2154\,\mearth$ total mass and 100\,\mearth\ core-mass predict a much
bigger radius (1.04\,\rjup).  The observed radius of
\hatcurPPrshort{20}\,\mjup\ would require a very high metal content. 
We note that the host star is one of the most metal rich stars that
have a transiting planet ($\feh=\hatcurSMEzfeh{20}$).  Curiously,
\hatcurb{20} orbits a fairly late type star (\hatcurISOspec{20}), as
compared to most of the massive hot Jupiters that orbit $\sim$F5
dwarfs.  It is also different from the rest of the population in that
the orbit is consistent with circular at the 3$\sigma$ level.  The
irradiation \hatcurb{20} receives is one of the smallest, clearly
making it a pL class exoplanet \citep{fortney:2008}: $\langle F \rangle
= (\hatcurPPfluxavg{20})\cdot 10^{\hatcurPPfluxavgdim{20}}$\ergscmsq,
comparable to the mean flux per orbit for another ``heavy'' planet
\hd{17156b} on a 21\,d period orbit.  The only other massive planet
that receives less average flux (integrated over an orbit) is
\hd{80606b}.  \hatcur{20} is an extreme outlier in the \mpl--\mstar\
plane; it is a relatively small mass star harboring a very massive
planet.  Another outlier (albeit to a much lesser extent) with similar
planetary radius and stellar mass is WASP-10b
\citep{johnson:2008b,christian:2009}, but this planet has less than
half of the mass of \hatcurb{20} (3.09\,\mjup).  We also calculated the
maximum mass of a stable moon for both the prograde and retrograde
cases, and derived $0.128\,\mearth$ and $8.31\,\mearth$, respectively,
i.e.~\hatcurb{20} can harbor a fairly massive moon.  An $8.31\,\mearth$
retrograde moon would cause $\sim 10\,s$ variations in the transit
times, which is marginally detectable from the ground.

\hatcur{20} has a close-by faint and red companion star at $\sim
6.86\arcsec$ separation.  Based on the Palomar sky survey archival
plates, we confirm that they form a close common-proper motion pair,
thus it is very likely that the two stars are physically associated. 
The binary has appeared in the Washington Double Star compilation (WDS)
as POU2795, and was discovered by \citet{pourteau:1933}.  Furthermore,
based on the summary of observations in the WDS, there is already a
hint of orbital motion of the companion to \hatcur{20} over the last
century.  The position angle of the companion changed from
PA=323\arcdeg\ to PA=320\arcdeg\ over the course of 89 years (between
1909 and 1998), and it seems to be retrograde on the sky (clockwise). 
Thus, \hatcur{20} is yet another example of a massive planet in a
binary system \citep{udry:2002}.  The binary companion makes this
system ideal for high precision ground or space-based studies, as it
provides a natural comparison source, even though it has a later
spectral type.

\subsection{\hatcurb{21}}
\label{sec:disc21}
With a mass of $\mpl=\hatcurPPmlong{21}\,\mjup$, \hatcurb{21} is the 11th
most massive transiting planet.  \hatcurb{21} has a radius of
$\rpl=\hatcurPPrlong{21}\,\rjup$, mean density
$\rhopl=\hatcurPPrho{21}\,\gcmc$, and orbits on a moderately eccentric
orbit with $e=\hatcurRVeccen{21}$, $\omega=\hatcurRVomega{21}\arcdeg$. 
The transits occur near apastron.  As noted by
\citet{buchhave:2010}, 4\,\mjup\ mass planets are very rare in the
sample of currently known transiting exoplanets, and the only siblings of
\hatcurb{21} are \hd{80606b} \citep[4.08\,\mjup;][]{naef:2001} and
HAT-P-16b \citep[4.19\,\mjup;][]{buchhave:2010}.  Among these,
HAT-P-16b has a shorter period, also an eccentric orbit, and a much larger
radius (1.29\,\rjup).  \hd{80606b}, on the other hand, has a similar
radius, and orbits on an extremely eccentric ($e=0.93$) orbit at
111\,day period.  It appears that \hatcurb{21} is thus an unusual, short
period, eccentric, massive and compact planet.

The only models from \citet{fortney:2007} consistent with the observed
radius are 4\,Gyr models with 100\,\mearth\ core mass, yielding
1.05\,\rjup\ radius.  Probably \hatcurb{21} has an even higher metal
content.  \hatcurb{21} has a very high mean density; it is 8th among
all TEPs, and very similar to \hd{80606b} and WASP-14b.

The flux received by the planet varies between
$(\hatcurPPfluxperi{21})\cdot 10^{\hatcurPPfluxperidim{21}}\,\ergscmsq$ and 
$(\hatcurPPfluxap{21})\cdot 10^{\hatcurPPfluxapdim{21}}\,\ergscmsq$.
Interestingly, this puts \hatcurb{21} on the bordlerline between pL
(low irradiation) and pM (high irradiation) planets. At the time of
occultation, \hatcurb{21} is just approaching its periastron, thus
entering the irradiation level quoted for pM type planets.

\subsection{\hatcurb{22}}
\label{sec:disc22}
\hatcurb{22} has a mass of $\mpl=\hatcurPPmlong{22}\,\mjup$, radius of
$\rpl=\hatcurPPrlong{22}\,\rjup$, and mean density of
$\rho_p=\hatcurPPrho{22}\,\gcmc$.  \hatcurb{22} orbits a fairly metal
rich ($\feh=\hatcurSMEzfeh{22}$), bright (V=\hatcurCCtassmv{22}), and close-by
(\hatcurXdist{22}\,pc) star.  Similarly to \hatcur{20}, the host star has a
faint and red neighbor at 9\arcsec\ separation that is co-moving with
\hatcur{22} (based on the POSS plates and recent Keck/HIRES snapshots),
thus they are likely to form a physical pair.

\begin{figure*}[!ht]
\plotone{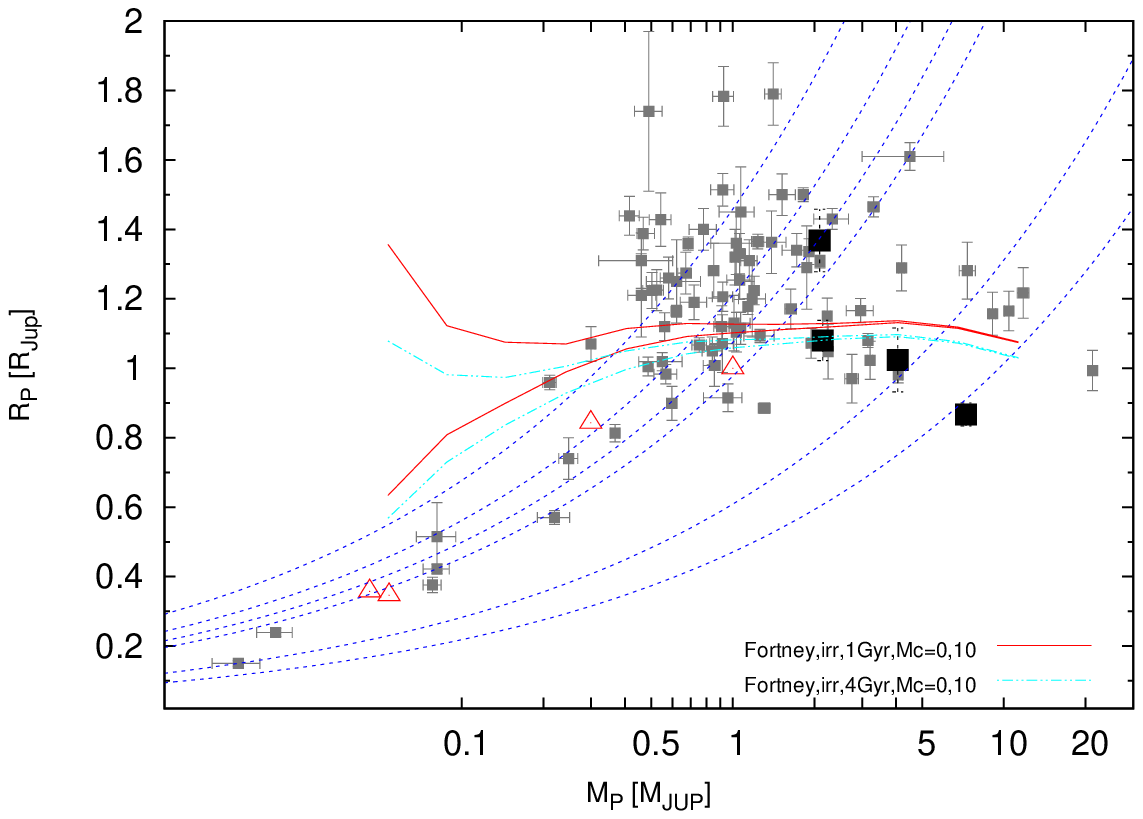}
\caption{ Mass--radius diagram of known TEPs (small filled
  squares). \hatcurb{20}--\hatcurb{23}\ are shown as a large filled
  squares.  Overlaid are \citet{fortney:2007} planetary isochrones
  interpolated to the solar equivalent semi-major axis of \hatcurb{20}
  for ages of 1.0\,Gyr (upper, solid lines) and 4\,Gyr (lower
  dashed-dotted lines) and core masses of 0 and 10\,\mearth (upper and
  lower lines respectively), as well as isodensity lines for 0.4, 0.7,
  1.0, 1.33, 5.5 and 11.9\,\gcmc (dashed lines). Solar system planets
  are shown with open triangles.
\label{fig:exomr}}
\end{figure*}

\hatcurb{22} belongs to the moderately massive ($\sim 2\,\mjup$) and
compact ($\rpl \approx 1\,\rjup$) hot Jupiters, such as 
HAT-P-15b \citep[\mpl=1.95\mjup, \rpl=1.07\,\rjup; ][]{kovacs:2010}, 
HAT-P-14b \citep[\mpl=2.23\mjup, \rpl=1.15\,\rjup; ][]{torres:2010},
and WASP-8b \citep[\mpl=2.25\mjup, \rpl=1.05\,\rjup; ][]{queloz:2010}.
The radius distribution is almost bi-modal for these planets (see
\reffigl{exomr}), with members of the inflated
($\rpl\approx1.3\,\rjup$) group being:
HAT-P-23b (\mpl=2.09\mjup, \rpl=1.37\,\rjup; this work), 
Kepler-5b \citep[\mpl=2.10\mjup, \rpl=1.31\,\rjup;][]{kipping:2010,koch:2010}, 
CoRoT-11b  \citep[\mpl=2.33\mjup, \rpl=1.43\,\rjup; ][]{gandolfi:2010}.

\hatcurb{22} is broadly consistent with the models of
\citet{fortney:2008}.  For 300\,Myr, 1\,Gyr and 4\,Gyr models it
requires a 100\,\mearth, 50\,\mearth\ and 25\,\mearth\ core,
respectively, to have a radius of $\sim\hatcurPPrshort{22}\,\rjup$. 
The low incoming flux (see \reftabl{planetparam}) means that
\hatcurb{22} is a pL class planet.  \hatcurb{22} can harbor a
$0.96\,\mearth$ mass retrograde moon, which would cause transit timing
variations (TTVs) of $\sim 2$\,seconds.

\subsection{\hatcurb{23}}
\label{sec:disc23}
\hatcurb{23} belongs to the inflated group of $2\,\mjup$ planets (see
discussion above for \hatcurb{22}).  This planet has a mass of
$\mpl=\hatcurPPmlong{23}\,\mjup$, radius
$\rpl=\hatcurPPrlong{23}\,\rjup$, and mean density
$\rhopl=\hatcurPPrho{23}\,\gcmc$. 
The orbit is nearly circular, with the eccentricity being marginally
significant.  The reason for the somewhat higher than usual errors in
the RV parameters is the high jitter of the star
(\hatcurRVjitter{23}\,\ms), which may be related to the moderately high
$\vsini=\hatcurSMEvsin{23}\,\kms$ and the very close-in orbit of
\hatcurb{23}.  The \citet{fortney:2008} models can not reproduce the
observed radius of \hatcurb{23}; even for the youngest, (300\,Myr)
core-less models, the theoretical radius for its mass is 1.25\,\rjup. 
\hatcurb{23} orbits its host star on a very close-in orbit.  The
orbital period is only \hatcurLCPshort{23}\,days; almost identical to
that of OGLE-TR-56b (1.21192\,days).  The nominal planetary radius of
the two objects is also the same within 1\%, but OGLE-TR-56b is much
less massive (1.39\,\mjup).  The flux falling on \hatcurb{23} from its
host star is one of the highest (i.e.~belongs to the pM class objects),
and is similar to that of HAT-P-7b and OGLE-TR-56b.  We also calculated
the spiral in-fall timescale for each new discovery based on
\citet{levrard:2009} and \citet{dobbs-dixon:2004}.  By assuming that
the stellar dissipation factor is $Q_\star=10^6$, the infall time for
\hatcurb{23} is $\tau_{infall} = \hatcurPPtinfall{23}$\,Myr, one of the
shortest among exoplanets. 

The Rossiter-McLaughlin effect for \hatcurb{23} should be quite
significant, given the moderately high
$\vsini=\hatcurSMEvsin{23}\,\kms$ of the host star, and the $\Delta i =
17$\,mmag deep transit.  The impact parameter is also ``ideal''
($b=\hatcurLCimp{23}$), i.e.~it is not equatorial ($b=0$), where there
would be a strong degeneracy between the stellar rotational velocity
\vsini\ and the sky-projected angle of the stellar spin axis and the
orbital normal, $\lambda$, and is also far from grazing, where the
transit is short, and other system parameters have lower accuracy.  The
effective temperature of the star ($\teffstar = \hatcurSMEteff{23}\,K$)
is close to the critical temperature of 6250\,K noted recently by
\citet{winn:2010}, which may be a border-line between systems where the
stellar spin axes and planetary orbital normals are preferentially
aligned ($\teffstar < 6250\,K$) and those that are misaligned
($\teffstar > 6250\,K$).  An alternative hypothesis has been brought up
by \citet{schlaufman:2010}, where misaligned stellar spin axes and
orbital normals are related to the mass of the host star.  The mass of
\hatcur{23} (\hatcurISOm{23}) is sufficiently close to the suggested
dividing line of $\mstar = 1.2\,\msun$, thus it will provide an
excellent additional test for these ideas.

\subsection{Summary}
We presented the discovery of four new massive transiting planets, and
provided accurate characterization of the host star and planetary
parameters.  These 4 new systems are very diverse, and significantly
expand the sample of $\sim13$ other massive ($\mpl\gtrsim2\,\mjup$)
planets.  Two of the new discoveries orbit stars that have fainter,
most probably physically associated companions.  The new discoveries do
not tend to enhance the mass--eccentricity correlation, since only one
(\hatcurb{21}) is significantly eccentric.  Also, the tentative
mass--\vsini\ correlation noted in the Introduction is weakened by the
new discoveries.  The heavier mass planets (\hatcurb{20} and
\hatcurb{21}) seem to be inconsistent with current theoretical models
in that they are too dense, and would require a huge core (or metal
content) to have such small radii.  One planet (\hatcurb{23}) is also
inconsistent with the models (unless we assume that the planet is very
young), but in the opposite sense; it has an inflated radius.  It has
been noted by \citet{winn:2010} and \citet{schlaufman:2010} that
systems exhibiting stellar spin axis--planetary orbital normal
misalignment are preferentially eccentric and heavy mass planets (in
addition to the key parameter being the effective temperature or mass
of the host star, respectively).  The four new planets presented in
this work will provide additional important tests for checking these
conjectures.  The host stars are all bright ($9.7<V<12.4$), and thus
enable in-depth future characterization of these systems.

\acknowledgements 

HATNet operations have been funded by NASA grants NNG04GN74G,
NNX08AF23G and SAO IR\&D grants.  Work of G.\'A.B.~and J.~Johnson were
supported by the Postdoctoral Fellowship of the NSF Astronomy and
Astrophysics Program (AST-0702843 and AST-0702821, respectively).  GT
acknowledges partial support from NASA grant NNX09AF59G.  We
acknowledge partial support also from the Kepler Mission under NASA
Cooperative Agreement NCC2-1390 (D.W.L., PI).  G.K.~thanks the
Hungarian Scientific Research Foundation (OTKA) for support through
grant K-81373.  This research has made use of Keck telescope time
granted through NOAO and NASA, and uses observations
obtained with facilities of the Las Cumbres Observatory Global
Telescope. 


\input{biblio.tex}

\end{document}

%% file: newcommand_gen.tex
\newcommand{\forloop}[5][1]%
{%
\setcounter{#2}{#3}%
\ifthenelse{#4}%
	{%
	#5%
	\addtocounter{#2}{#1}%
	\forloop[#1]{#2}{\value{#2}}{#4}{#5}%
	}%
	{%
	}%
}%

\newcommand{\ctbd}[1]{}

\newcommand{\lc}{light curve}
\newcommand{\lcs}{light curves}
\newcommand{\Lc}{Light curve}

\newcommand{\band}[1]{\ensuremath{#1}~band}

\newcommand{\chisq}{\ensuremath{\chi^2}}

\newcommand{\kms}{\ensuremath{\rm km\,s^{-1}}}
\newcommand{\ms}{\ensuremath{\rm m\,s^{-1}}}

\newcommand{\gcmc}{\ensuremath{\rm g\,cm^{-3}}}
\newcommand{\ergscmsq}{\ensuremath{\rm erg\,s^{-1}\,cm^{-2}}}

\newcommand{\pxs}{\ensuremath{\rm \arcsec pixel^{-1}}}

\newcommand{\vsini}{\ensuremath{v \sin{i}}}
\newcommand{\feh}{\ensuremath{\rm [Fe/H]}}

\newcommand{\vmac}{\ensuremath{v_{\rm mac}}}
\newcommand{\vmic}{\ensuremath{v_{\rm mic}}}

\newcommand{\vic}{\ensuremath{V\!-\!I_C}}

\newcommand{\rsun}{\ensuremath{R_\sun}}
\newcommand{\msun}{\ensuremath{M_\sun}}
\newcommand{\lsun}{\ensuremath{L_\sun}}

\newcommand{\rstar}{\ensuremath{R_\star}}
\newcommand{\mstar}{\ensuremath{M_\star}}
\newcommand{\lstar}{\ensuremath{L_\star}}

\newcommand{\teffstar}{\ensuremath{T_{\rm eff\star}}}
\newcommand{\rhostar}{\ensuremath{\rho_\star}}
\newcommand{\loggstar}{\ensuremath{\log{g_{\star}}}}

\newcommand{\mearth}{\ensuremath{M_\earth}}

\newcommand{\rpl}{\ensuremath{R_{p}}}
\newcommand{\mpl}{\ensuremath{M_{p}}}

\newcommand{\rhopl}{\ensuremath{\rho_{p}}}

\newcommand{\arstar}{\ensuremath{a/\rstar}}
\newcommand{\zrstar}{\ensuremath{\zeta/\rstar}}

\newcommand{\rjup}{\ensuremath{R_{\rm J}}}
\newcommand{\mjup}{\ensuremath{M_{\rm J}}}

\newcommand{\refsec}[1]{\mbox{\S\ \ref{sec:#1}}}

\newcommand{\reffigl}[1]{Figure~\ref{fig:#1}}
\newcommand{\refsecl}[1]{\mbox{Section \ref{sec:#1}}}

\newcommand{\reftabl}[1]{Table~\ref{tab:#1}}

\newcommand{\flwof}{\mbox{FLWO 1.2\,m}}

\newcommand{\flwos}{\mbox{FLWO 1.5\,m}}

\newcommand{\hd}[1]{\mbox{HD #1}}



%% file: newcommand_spec_multi.tex
\input{planetinfo_multi.tex}
\input{newcommand_spec1.tex}
\input{newcommand_spec2.tex}
\input{newcommand_spec3.tex}
\input{newcommand_spec4.tex}
\newcommand{\hatcur}[1]{\ifnum#1=20 %
\hatcurxxxA
\else
\ifnum#1=21 %
\hatcurxxxB
\else
\ifnum#1=22 %
\hatcurxxxC
\else
\ifnum#1=23 %
\hatcurxxxD
\else
??????\fi
\fi
\fi
\fi
}
\newcommand{\hatcurb}[1]{\ifnum#1=20 %
\hatcurbxxxA
\else
\ifnum#1=21 %
\hatcurbxxxB
\else
\ifnum#1=22 %
\hatcurbxxxC
\else
\ifnum#1=23 %
\hatcurbxxxD
\else
??????\fi
\fi
\fi
\fi
}
\newcommand{\hatcurc}[1]{\ifnum#1=20 %
\hatcurcxxxA
\else
\ifnum#1=21 %
\hatcurcxxxB
\else
\ifnum#1=22 %
\hatcurcxxxC
\else
\ifnum#1=23 %
\hatcurcxxxD
\else
??????\fi
\fi
\fi
\fi
}
\newcommand{\hatcurCChd}[1]{\ifnum#1=22 %
\hatcurCChdxxxC
\else
??????\fi
}
\newcommand{\hatcurCCtassvi}[1]{\ifnum#1=20 %
\hatcurCCtassvixxxA
\else
\ifnum#1=21 %
\hatcurCCtassvixxxB
\else
\ifnum#1=22 %
\hatcurCCtassvixxxC
\else
\ifnum#1=23 %
\hatcurCCtassvixxxD
\else
??????\fi
\fi
\fi
\fi
}
\newcommand{\hatcurisocite}[1]{\ifnum#1=20 %
\hatcurisocitexxxA
\else
\ifnum#1=21 %
\hatcurisocitexxxB
\else
\ifnum#1=22 %
\hatcurisocitexxxC
\else
\ifnum#1=23 %
\hatcurisocitexxxD
\else
??????\fi
\fi
\fi
\fi
}
\newcommand{\hatcurisofull}[1]{\ifnum#1=20 %
\hatcurisofullxxxA
\else
\ifnum#1=21 %
\hatcurisofullxxxB
\else
\ifnum#1=22 %
\hatcurisofullxxxC
\else
\ifnum#1=23 %
\hatcurisofullxxxD
\else
??????\fi
\fi
\fi
\fi
}
\newcommand{\hatcurisoshort}[1]{\ifnum#1=20 %
\hatcurisoshortxxxA
\else
\ifnum#1=21 %
\hatcurisoshortxxxB
\else
\ifnum#1=22 %
\hatcurisoshortxxxC
\else
\ifnum#1=23 %
\hatcurisoshortxxxD
\else
??????\fi
\fi
\fi
\fi
}
\newcommand{\hatcurjhkfilset}[1]{\ifnum#1=20 %
\hatcurjhkfilsetxxxA
\else
\ifnum#1=21 %
\hatcurjhkfilsetxxxB
\else
\ifnum#1=22 %
\hatcurjhkfilsetxxxC
\else
\ifnum#1=23 %
\hatcurjhkfilsetxxxD
\else
??????\fi
\fi
\fi
\fi
}
\newcommand{\hatcurlumind}[1]{\ifnum#1=20 %
\hatcurlumindxxxA
\else
\ifnum#1=21 %
\hatcurlumindxxxB
\else
\ifnum#1=22 %
\hatcurlumindxxxC
\else
\ifnum#1=23 %
\hatcurlumindxxxD
\else
??????\fi
\fi
\fi
\fi
}
\newcommand{\hatcurplanetnum}[1]{\ifnum#1=20 %
\hatcurplanetnumxxxA
\else
\ifnum#1=21 %
\hatcurplanetnumxxxB
\else
\ifnum#1=22 %
\hatcurplanetnumxxxC
\else
\ifnum#1=23 %
\hatcurplanetnumxxxD
\else
??????\fi
\fi
\fi
\fi
}
\newcommand{\hatcurRVgammaabs}[1]{\ifnum#1=20 %
\hatcurRVgammaabsxxxA
\else
\ifnum#1=21 %
\hatcurRVgammaabsxxxB
\else
\ifnum#1=22 %
\hatcurRVgammaabsxxxC
\else
\ifnum#1=23 %
\hatcurRVgammaabsxxxD
\else
??????\fi
\fi
\fi
\fi
}
\newcommand{\hatcurRVgammarel}[1]{\ifnum#1=20 %
\hatcurRVgammarelxxxA
\else
\ifnum#1=21 %
\hatcurRVgammarelxxxB
\else
\ifnum#1=22 %
\hatcurRVgammarelxxxC
\else
\ifnum#1=23 %
\hatcurRVgammarelxxxD
\else
??????\fi
\fi
\fi
\fi
}
\newcommand{\hatcurSMElogg}[1]{\ifnum#1=20 %
\hatcurSMEloggxxxA
\else
\ifnum#1=21 %
\hatcurSMEloggxxxB
\else
\ifnum#1=22 %
\hatcurSMEloggxxxC
\else
\ifnum#1=23 %
\hatcurSMEloggxxxD
\else
??????\fi
\fi
\fi
\fi
}
\newcommand{\hatcurSMEteff}[1]{\ifnum#1=20 %
\hatcurSMEteffxxxA
\else
\ifnum#1=21 %
\hatcurSMEteffxxxB
\else
\ifnum#1=22 %
\hatcurSMEteffxxxC
\else
\ifnum#1=23 %
\hatcurSMEteffxxxD
\else
??????\fi
\fi
\fi
\fi
}
\newcommand{\hatcurSMEversion}[1]{\ifnum#1=20 %
\hatcurSMEversionxxxA
\else
\ifnum#1=21 %
\hatcurSMEversionxxxB
\else
\ifnum#1=22 %
\hatcurSMEversionxxxC
\else
\ifnum#1=23 %
\hatcurSMEversionxxxD
\else
??????\fi
\fi
\fi
\fi
}
\newcommand{\hatcurSMEvmac}[1]{\ifnum#1=20 %
\hatcurSMEvmacxxxA
\else
\ifnum#1=21 %
\hatcurSMEvmacxxxB
\else
\ifnum#1=22 %
\hatcurSMEvmacxxxC
\else
\ifnum#1=23 %
\hatcurSMEvmacxxxD
\else
??????\fi
\fi
\fi
\fi
}
\newcommand{\hatcurSMEvmic}[1]{\ifnum#1=20 %
\hatcurSMEvmicxxxA
\else
\ifnum#1=21 %
\hatcurSMEvmicxxxB
\else
\ifnum#1=22 %
\hatcurSMEvmicxxxC
\else
\ifnum#1=23 %
\hatcurSMEvmicxxxD
\else
??????\fi
\fi
\fi
\fi
}
\newcommand{\hatcurSMEvsin}[1]{\ifnum#1=20 %
\hatcurSMEvsinxxxA
\else
\ifnum#1=21 %
\hatcurSMEvsinxxxB
\else
\ifnum#1=22 %
\hatcurSMEvsinxxxC
\else
\ifnum#1=23 %
\hatcurSMEvsinxxxD
\else
??????\fi
\fi
\fi
\fi
}
\newcommand{\hatcurSMEzfeh}[1]{\ifnum#1=20 %
\hatcurSMEzfehxxxA
\else
\ifnum#1=21 %
\hatcurSMEzfehxxxB
\else
\ifnum#1=22 %
\hatcurSMEzfehxxxC
\else
\ifnum#1=23 %
\hatcurSMEzfehxxxD
\else
??????\fi
\fi
\fi
\fi
}
\newcommand{\hatcurSMEzfehshort}[1]{\ifnum#1=20 %
\hatcurSMEzfehshortxxxA
\else
\ifnum#1=21 %
\hatcurSMEzfehshortxxxB
\else
\ifnum#1=22 %
\hatcurSMEzfehshortxxxC
\else
\ifnum#1=23 %
\hatcurSMEzfehshortxxxD
\else
??????\fi
\fi
\fi
\fi
}

%% file: planetinfo_multi.tex
\input{HTR267-010.tex}
\input{HTR183-008.tex}
\input{HTR102-001.tex}
\input{HTR341-004.tex}
\newcommand{\hatcurCCbbHmag}[1]{\ifnum#1=20 %
\hatcurCCbbHmagxxxA
\else
\ifnum#1=21 %
\hatcurCCbbHmagxxxB
\else
\ifnum#1=22 %
\hatcurCCbbHmagxxxC
\else
\ifnum#1=23 %
\hatcurCCbbHmagxxxD
\else
??????\fi
\fi
\fi
\fi
}
\newcommand{\hatcurCCbbJmag}[1]{\ifnum#1=20 %
\hatcurCCbbJmagxxxA
\else
\ifnum#1=21 %
\hatcurCCbbJmagxxxB
\else
\ifnum#1=22 %
\hatcurCCbbJmagxxxC
\else
\ifnum#1=23 %
\hatcurCCbbJmagxxxD
\else
??????\fi
\fi
\fi
\fi
}
\newcommand{\hatcurCCbbKmag}[1]{\ifnum#1=20 %
\hatcurCCbbKmagxxxA
\else
\ifnum#1=21 %
\hatcurCCbbKmagxxxB
\else
\ifnum#1=22 %
\hatcurCCbbKmagxxxC
\else
\ifnum#1=23 %
\hatcurCCbbKmagxxxD
\else
??????\fi
\fi
\fi
\fi
}
\newcommand{\hatcurCCcitHmag}[1]{\ifnum#1=20 %
\hatcurCCcitHmagxxxA
\else
\ifnum#1=21 %
\hatcurCCcitHmagxxxB
\else
\ifnum#1=22 %
\hatcurCCcitHmagxxxC
\else
\ifnum#1=23 %
\hatcurCCcitHmagxxxD
\else
??????\fi
\fi
\fi
\fi
}
\newcommand{\hatcurCCcitJmag}[1]{\ifnum#1=20 %
\hatcurCCcitJmagxxxA
\else
\ifnum#1=21 %
\hatcurCCcitJmagxxxB
\else
\ifnum#1=22 %
\hatcurCCcitJmagxxxC
\else
\ifnum#1=23 %
\hatcurCCcitJmagxxxD
\else
??????\fi
\fi
\fi
\fi
}
\newcommand{\hatcurCCcitKmag}[1]{\ifnum#1=20 %
\hatcurCCcitKmagxxxA
\else
\ifnum#1=21 %
\hatcurCCcitKmagxxxB
\else
\ifnum#1=22 %
\hatcurCCcitKmagxxxC
\else
\ifnum#1=23 %
\hatcurCCcitKmagxxxD
\else
??????\fi
\fi
\fi
\fi
}
\newcommand{\hatcurCCdec}[1]{\ifnum#1=20 %
\hatcurCCdecxxxA
\else
\ifnum#1=21 %
\hatcurCCdecxxxB
\else
\ifnum#1=22 %
\hatcurCCdecxxxC
\else
\ifnum#1=23 %
\hatcurCCdecxxxD
\else
??????\fi
\fi
\fi
\fi
}
\newcommand{\hatcurCCesoHKmag}[1]{\ifnum#1=20 %
\hatcurCCesoHKmagxxxA
\else
\ifnum#1=21 %
\hatcurCCesoHKmagxxxB
\else
\ifnum#1=22 %
\hatcurCCesoHKmagxxxC
\else
\ifnum#1=23 %
\hatcurCCesoHKmagxxxD
\else
??????\fi
\fi
\fi
\fi
}
\newcommand{\hatcurCCesoHmag}[1]{\ifnum#1=20 %
\hatcurCCesoHmagxxxA
\else
\ifnum#1=21 %
\hatcurCCesoHmagxxxB
\else
\ifnum#1=22 %
\hatcurCCesoHmagxxxC
\else
\ifnum#1=23 %
\hatcurCCesoHmagxxxD
\else
??????\fi
\fi
\fi
\fi
}
\newcommand{\hatcurCCesoJHmag}[1]{\ifnum#1=20 %
\hatcurCCesoJHmagxxxA
\else
\ifnum#1=21 %
\hatcurCCesoJHmagxxxB
\else
\ifnum#1=22 %
\hatcurCCesoJHmagxxxC
\else
\ifnum#1=23 %
\hatcurCCesoJHmagxxxD
\else
??????\fi
\fi
\fi
\fi
}
\newcommand{\hatcurCCesoJKmag}[1]{\ifnum#1=20 %
\hatcurCCesoJKmagxxxA
\else
\ifnum#1=21 %
\hatcurCCesoJKmagxxxB
\else
\ifnum#1=22 %
\hatcurCCesoJKmagxxxC
\else
\ifnum#1=23 %
\hatcurCCesoJKmagxxxD
\else
??????\fi
\fi
\fi
\fi
}
\newcommand{\hatcurCCesoJmag}[1]{\ifnum#1=20 %
\hatcurCCesoJmagxxxA
\else
\ifnum#1=21 %
\hatcurCCesoJmagxxxB
\else
\ifnum#1=22 %
\hatcurCCesoJmagxxxC
\else
\ifnum#1=23 %
\hatcurCCesoJmagxxxD
\else
??????\fi
\fi
\fi
\fi
}
\newcommand{\hatcurCCesoKmag}[1]{\ifnum#1=20 %
\hatcurCCesoKmagxxxA
\else
\ifnum#1=21 %
\hatcurCCesoKmagxxxB
\else
\ifnum#1=22 %
\hatcurCCesoKmagxxxC
\else
\ifnum#1=23 %
\hatcurCCesoKmagxxxD
\else
??????\fi
\fi
\fi
\fi
}
\newcommand{\hatcurCCgsc}[1]{\ifnum#1=20 %
\hatcurCCgscxxxA
\else
\ifnum#1=21 %
\hatcurCCgscxxxB
\else
\ifnum#1=22 %
\hatcurCCgscxxxC
\else
\ifnum#1=23 %
\hatcurCCgscxxxD
\else
??????\fi
\fi
\fi
\fi
}
\newcommand{\hatcurCCmag}[1]{\ifnum#1=20 %
\hatcurCCmagxxxA
\else
\ifnum#1=21 %
\hatcurCCmagxxxB
\else
\ifnum#1=22 %
\hatcurCCmagxxxC
\else
\ifnum#1=23 %
\hatcurCCmagxxxD
\else
??????\fi
\fi
\fi
\fi
}
\newcommand{\hatcurCCpm}[1]{\ifnum#1=20 %
\hatcurCCpmxxxA
\else
\ifnum#1=21 %
\hatcurCCpmxxxB
\else
\ifnum#1=22 %
\hatcurCCpmxxxC
\else
\ifnum#1=23 %
\hatcurCCpmxxxD
\else
??????\fi
\fi
\fi
\fi
}
\newcommand{\hatcurCCpmdec}[1]{\ifnum#1=20 %
\hatcurCCpmdecxxxA
\else
\ifnum#1=21 %
\hatcurCCpmdecxxxB
\else
\ifnum#1=22 %
\hatcurCCpmdecxxxC
\else
\ifnum#1=23 %
\hatcurCCpmdecxxxD
\else
??????\fi
\fi
\fi
\fi
}
\newcommand{\hatcurCCpmra}[1]{\ifnum#1=20 %
\hatcurCCpmraxxxA
\else
\ifnum#1=21 %
\hatcurCCpmraxxxB
\else
\ifnum#1=22 %
\hatcurCCpmraxxxC
\else
\ifnum#1=23 %
\hatcurCCpmraxxxD
\else
??????\fi
\fi
\fi
\fi
}
\newcommand{\hatcurCCra}[1]{\ifnum#1=20 %
\hatcurCCraxxxA
\else
\ifnum#1=21 %
\hatcurCCraxxxB
\else
\ifnum#1=22 %
\hatcurCCraxxxC
\else
\ifnum#1=23 %
\hatcurCCraxxxD
\else
??????\fi
\fi
\fi
\fi
}
\newcommand{\hatcurCCtassmv}[1]{\ifnum#1=20 %
\hatcurCCtassmvxxxA
\else
\ifnum#1=21 %
\hatcurCCtassmvxxxB
\else
\ifnum#1=22 %
\hatcurCCtassmvxxxC
\else
\ifnum#1=23 %
\hatcurCCtassmvxxxD
\else
??????\fi
\fi
\fi
\fi
}
\newcommand{\hatcurCCtwomass}[1]{\ifnum#1=20 %
\hatcurCCtwomassxxxA
\else
\ifnum#1=21 %
\hatcurCCtwomassxxxB
\else
\ifnum#1=22 %
\hatcurCCtwomassxxxC
\else
\ifnum#1=23 %
\hatcurCCtwomassxxxD
\else
??????\fi
\fi
\fi
\fi
}
\newcommand{\hatcurCCtwomassHmag}[1]{\ifnum#1=20 %
\hatcurCCtwomassHmagxxxA
\else
\ifnum#1=21 %
\hatcurCCtwomassHmagxxxB
\else
\ifnum#1=22 %
\hatcurCCtwomassHmagxxxC
\else
\ifnum#1=23 %
\hatcurCCtwomassHmagxxxD
\else
??????\fi
\fi
\fi
\fi
}
\newcommand{\hatcurCCtwomassJmag}[1]{\ifnum#1=20 %
\hatcurCCtwomassJmagxxxA
\else
\ifnum#1=21 %
\hatcurCCtwomassJmagxxxB
\else
\ifnum#1=22 %
\hatcurCCtwomassJmagxxxC
\else
\ifnum#1=23 %
\hatcurCCtwomassJmagxxxD
\else
??????\fi
\fi
\fi
\fi
}
\newcommand{\hatcurCCtwomassKmag}[1]{\ifnum#1=20 %
\hatcurCCtwomassKmagxxxA
\else
\ifnum#1=21 %
\hatcurCCtwomassKmagxxxB
\else
\ifnum#1=22 %
\hatcurCCtwomassKmagxxxC
\else
\ifnum#1=23 %
\hatcurCCtwomassKmagxxxD
\else
??????\fi
\fi
\fi
\fi
}
\newcommand{\hatcurDSgamma}[1]{\ifnum#1=20 %
\hatcurDSgammaxxxA
\else
\ifnum#1=21 %
\hatcurDSgammaxxxB
\else
\ifnum#1=22 %
\hatcurDSgammaxxxC
\else
\ifnum#1=23 %
\hatcurDSgammaxxxD
\else
??????\fi
\fi
\fi
\fi
}
\newcommand{\hatcurDSlogg}[1]{\ifnum#1=20 %
\hatcurDSloggxxxA
\else
\ifnum#1=21 %
\hatcurDSloggxxxB
\else
\ifnum#1=22 %
\hatcurDSloggxxxC
\else
\ifnum#1=23 %
\hatcurDSloggxxxD
\else
??????\fi
\fi
\fi
\fi
}
\newcommand{\hatcurDSnumspec}[1]{\ifnum#1=20 %
\hatcurDSnumspecxxxA
\else
\ifnum#1=21 %
\hatcurDSnumspecxxxB
\else
\ifnum#1=22 %
\hatcurDSnumspecxxxC
\else
\ifnum#1=23 %
\hatcurDSnumspecxxxD
\else
??????\fi
\fi
\fi
\fi
}
\newcommand{\hatcurDSrvrms}[1]{\ifnum#1=20 %
\hatcurDSrvrmsxxxA
\else
\ifnum#1=21 %
\hatcurDSrvrmsxxxB
\else
\ifnum#1=22 %
\hatcurDSrvrmsxxxC
\else
\ifnum#1=23 %
\hatcurDSrvrmsxxxD
\else
??????\fi
\fi
\fi
\fi
}
\newcommand{\hatcurDSspan}[1]{\ifnum#1=20 %
\hatcurDSspanxxxA
\else
\ifnum#1=21 %
\hatcurDSspanxxxB
\else
\ifnum#1=22 %
\hatcurDSspanxxxC
\else
\ifnum#1=23 %
\hatcurDSspanxxxD
\else
??????\fi
\fi
\fi
\fi
}
\newcommand{\hatcurDSteff}[1]{\ifnum#1=20 %
\hatcurDSteffxxxA
\else
\ifnum#1=21 %
\hatcurDSteffxxxB
\else
\ifnum#1=22 %
\hatcurDSteffxxxC
\else
\ifnum#1=23 %
\hatcurDSteffxxxD
\else
??????\fi
\fi
\fi
\fi
}
\newcommand{\hatcurDSvsini}[1]{\ifnum#1=20 %
\hatcurDSvsinixxxA
\else
\ifnum#1=21 %
\hatcurDSvsinixxxB
\else
\ifnum#1=22 %
\hatcurDSvsinixxxC
\else
\ifnum#1=23 %
\hatcurDSvsinixxxD
\else
??????\fi
\fi
\fi
\fi
}
\newcommand{\hatcurDSzfeh}[1]{\ifnum#1=20 %
\hatcurDSzfehxxxA
\else
\ifnum#1=21 %
\hatcurDSzfehxxxB
\else
\ifnum#1=22 %
\hatcurDSzfehxxxC
\else
\ifnum#1=23 %
\hatcurDSzfehxxxD
\else
??????\fi
\fi
\fi
\fi
}
\newcommand{\hatcurfield}[1]{\ifnum#1=20 %
\hatcurfieldxxxA
\else
\ifnum#1=21 %
\hatcurfieldxxxB
\else
\ifnum#1=22 %
\hatcurfieldxxxC
\else
\ifnum#1=23 %
\hatcurfieldxxxD
\else
??????\fi
\fi
\fi
\fi
}
\newcommand{\hatcurFIESgamma}[1]{\ifnum#1=20 %
\hatcurFIESgammaxxxA
\else
\ifnum#1=21 %
\hatcurFIESgammaxxxB
\else
\ifnum#1=22 %
\hatcurFIESgammaxxxC
\else
\ifnum#1=23 %
\hatcurFIESgammaxxxD
\else
??????\fi
\fi
\fi
\fi
}
\newcommand{\hatcurFIESlogg}[1]{\ifnum#1=20 %
\hatcurFIESloggxxxA
\else
\ifnum#1=21 %
\hatcurFIESloggxxxB
\else
\ifnum#1=22 %
\hatcurFIESloggxxxC
\else
\ifnum#1=23 %
\hatcurFIESloggxxxD
\else
??????\fi
\fi
\fi
\fi
}
\newcommand{\hatcurFIESnumspec}[1]{\ifnum#1=20 %
\hatcurFIESnumspecxxxA
\else
\ifnum#1=21 %
\hatcurFIESnumspecxxxB
\else
\ifnum#1=22 %
\hatcurFIESnumspecxxxC
\else
\ifnum#1=23 %
\hatcurFIESnumspecxxxD
\else
??????\fi
\fi
\fi
\fi
}
\newcommand{\hatcurFIESrvrms}[1]{\ifnum#1=20 %
\hatcurFIESrvrmsxxxA
\else
\ifnum#1=21 %
\hatcurFIESrvrmsxxxB
\else
\ifnum#1=22 %
\hatcurFIESrvrmsxxxC
\else
\ifnum#1=23 %
\hatcurFIESrvrmsxxxD
\else
??????\fi
\fi
\fi
\fi
}
\newcommand{\hatcurFIESspan}[1]{\ifnum#1=20 %
\hatcurFIESspanxxxA
\else
\ifnum#1=21 %
\hatcurFIESspanxxxB
\else
\ifnum#1=22 %
\hatcurFIESspanxxxC
\else
\ifnum#1=23 %
\hatcurFIESspanxxxD
\else
??????\fi
\fi
\fi
\fi
}
\newcommand{\hatcurFIESteff}[1]{\ifnum#1=20 %
\hatcurFIESteffxxxA
\else
\ifnum#1=21 %
\hatcurFIESteffxxxB
\else
\ifnum#1=22 %
\hatcurFIESteffxxxC
\else
\ifnum#1=23 %
\hatcurFIESteffxxxD
\else
??????\fi
\fi
\fi
\fi
}
\newcommand{\hatcurFIESvsini}[1]{\ifnum#1=20 %
\hatcurFIESvsinixxxA
\else
\ifnum#1=21 %
\hatcurFIESvsinixxxB
\else
\ifnum#1=22 %
\hatcurFIESvsinixxxC
\else
\ifnum#1=23 %
\hatcurFIESvsinixxxD
\else
??????\fi
\fi
\fi
\fi
}
\newcommand{\hatcurFIESzfeh}[1]{\ifnum#1=20 %
\hatcurFIESzfehxxxA
\else
\ifnum#1=21 %
\hatcurFIESzfehxxxB
\else
\ifnum#1=22 %
\hatcurFIESzfehxxxC
\else
\ifnum#1=23 %
\hatcurFIESzfehxxxD
\else
??????\fi
\fi
\fi
\fi
}
\newcommand{\hatcurhtr}[1]{\ifnum#1=20 %
\hatcurhtrxxxA
\else
\ifnum#1=21 %
\hatcurhtrxxxB
\else
\ifnum#1=22 %
\hatcurhtrxxxC
\else
\ifnum#1=23 %
\hatcurhtrxxxD
\else
??????\fi
\fi
\fi
\fi
}
\newcommand{\hatcurISOage}[1]{\ifnum#1=20 %
\hatcurISOagexxxA
\else
\ifnum#1=21 %
\hatcurISOagexxxB
\else
\ifnum#1=22 %
\hatcurISOagexxxC
\else
\ifnum#1=23 %
\hatcurISOagexxxD
\else
??????\fi
\fi
\fi
\fi
}
\newcommand{\hatcurISOJK}[1]{\ifnum#1=20 %
\hatcurISOJKxxxA
\else
\ifnum#1=21 %
\hatcurISOJKxxxB
\else
\ifnum#1=22 %
\hatcurISOJKxxxC
\else
\ifnum#1=23 %
\hatcurISOJKxxxD
\else
??????\fi
\fi
\fi
\fi
}
\newcommand{\hatcurISOlogg}[1]{\ifnum#1=20 %
\hatcurISOloggxxxA
\else
\ifnum#1=21 %
\hatcurISOloggxxxB
\else
\ifnum#1=22 %
\hatcurISOloggxxxC
\else
\ifnum#1=23 %
\hatcurISOloggxxxD
\else
??????\fi
\fi
\fi
\fi
}
\newcommand{\hatcurISOlum}[1]{\ifnum#1=20 %
\hatcurISOlumxxxA
\else
\ifnum#1=21 %
\hatcurISOlumxxxB
\else
\ifnum#1=22 %
\hatcurISOlumxxxC
\else
\ifnum#1=23 %
\hatcurISOlumxxxD
\else
??????\fi
\fi
\fi
\fi
}
\newcommand{\hatcurISOlumshort}[1]{\ifnum#1=20 %
\hatcurISOlumshortxxxA
\else
\ifnum#1=21 %
\hatcurISOlumshortxxxB
\else
\ifnum#1=22 %
\hatcurISOlumshortxxxC
\else
\ifnum#1=23 %
\hatcurISOlumshortxxxD
\else
??????\fi
\fi
\fi
\fi
}
\newcommand{\hatcurISOm}[1]{\ifnum#1=20 %
\hatcurISOmxxxA
\else
\ifnum#1=21 %
\hatcurISOmxxxB
\else
\ifnum#1=22 %
\hatcurISOmxxxC
\else
\ifnum#1=23 %
\hatcurISOmxxxD
\else
??????\fi
\fi
\fi
\fi
}
\newcommand{\hatcurISOMH}[1]{\ifnum#1=20 %
\hatcurISOMHxxxA
\else
\ifnum#1=21 %
\hatcurISOMHxxxB
\else
\ifnum#1=22 %
\hatcurISOMHxxxC
\else
\ifnum#1=23 %
\hatcurISOMHxxxD
\else
??????\fi
\fi
\fi
\fi
}
\newcommand{\hatcurISOMJ}[1]{\ifnum#1=20 %
\hatcurISOMJxxxA
\else
\ifnum#1=21 %
\hatcurISOMJxxxB
\else
\ifnum#1=22 %
\hatcurISOMJxxxC
\else
\ifnum#1=23 %
\hatcurISOMJxxxD
\else
??????\fi
\fi
\fi
\fi
}
\newcommand{\hatcurISOMK}[1]{\ifnum#1=20 %
\hatcurISOMKxxxA
\else
\ifnum#1=21 %
\hatcurISOMKxxxB
\else
\ifnum#1=22 %
\hatcurISOMKxxxC
\else
\ifnum#1=23 %
\hatcurISOMKxxxD
\else
??????\fi
\fi
\fi
\fi
}
\newcommand{\hatcurISOmlong}[1]{\ifnum#1=20 %
\hatcurISOmlongxxxA
\else
\ifnum#1=21 %
\hatcurISOmlongxxxB
\else
\ifnum#1=22 %
\hatcurISOmlongxxxC
\else
\ifnum#1=23 %
\hatcurISOmlongxxxD
\else
??????\fi
\fi
\fi
\fi
}
\newcommand{\hatcurISOmshort}[1]{\ifnum#1=20 %
\hatcurISOmshortxxxA
\else
\ifnum#1=21 %
\hatcurISOmshortxxxB
\else
\ifnum#1=22 %
\hatcurISOmshortxxxC
\else
\ifnum#1=23 %
\hatcurISOmshortxxxD
\else
??????\fi
\fi
\fi
\fi
}
\newcommand{\hatcurISOmv}[1]{\ifnum#1=20 %
\hatcurISOmvxxxA
\else
\ifnum#1=21 %
\hatcurISOmvxxxB
\else
\ifnum#1=22 %
\hatcurISOmvxxxC
\else
\ifnum#1=23 %
\hatcurISOmvxxxD
\else
??????\fi
\fi
\fi
\fi
}
\newcommand{\hatcurISOr}[1]{\ifnum#1=20 %
\hatcurISOrxxxA
\else
\ifnum#1=21 %
\hatcurISOrxxxB
\else
\ifnum#1=22 %
\hatcurISOrxxxC
\else
\ifnum#1=23 %
\hatcurISOrxxxD
\else
??????\fi
\fi
\fi
\fi
}
\newcommand{\hatcurISOrho}[1]{\ifnum#1=20 %
\hatcurISOrhoxxxA
\else
\ifnum#1=21 %
\hatcurISOrhoxxxB
\else
\ifnum#1=22 %
\hatcurISOrhoxxxC
\else
\ifnum#1=23 %
\hatcurISOrhoxxxD
\else
??????\fi
\fi
\fi
\fi
}
\newcommand{\hatcurISOrlong}[1]{\ifnum#1=20 %
\hatcurISOrlongxxxA
\else
\ifnum#1=21 %
\hatcurISOrlongxxxB
\else
\ifnum#1=22 %
\hatcurISOrlongxxxC
\else
\ifnum#1=23 %
\hatcurISOrlongxxxD
\else
??????\fi
\fi
\fi
\fi
}
\newcommand{\hatcurISOrshort}[1]{\ifnum#1=20 %
\hatcurISOrshortxxxA
\else
\ifnum#1=21 %
\hatcurISOrshortxxxB
\else
\ifnum#1=22 %
\hatcurISOrshortxxxC
\else
\ifnum#1=23 %
\hatcurISOrshortxxxD
\else
??????\fi
\fi
\fi
\fi
}
\newcommand{\hatcurISOsigma}[1]{\ifnum#1=20 %
\hatcurISOsigmaxxxA
\else
\ifnum#1=21 %
\hatcurISOsigmaxxxB
\else
\ifnum#1=22 %
\hatcurISOsigmaxxxC
\else
\ifnum#1=23 %
\hatcurISOsigmaxxxD
\else
??????\fi
\fi
\fi
\fi
}
\newcommand{\hatcurISOspec}[1]{\ifnum#1=20 %
\hatcurISOspecxxxA
\else
\ifnum#1=21 %
\hatcurISOspecxxxB
\else
\ifnum#1=22 %
\hatcurISOspecxxxC
\else
\ifnum#1=23 %
\hatcurISOspecxxxD
\else
??????\fi
\fi
\fi
\fi
}
\newcommand{\hatcurISOvi}[1]{\ifnum#1=20 %
\hatcurISOvixxxA
\else
\ifnum#1=21 %
\hatcurISOvixxxB
\else
\ifnum#1=22 %
\hatcurISOvixxxC
\else
\ifnum#1=23 %
\hatcurISOvixxxD
\else
??????\fi
\fi
\fi
\fi
}
\newcommand{\hatcurLBig}[1]{\ifnum#1=20 %
\hatcurLBigxxxA
\else
\ifnum#1=21 %
\hatcurLBigxxxB
\else
\ifnum#1=22 %
\hatcurLBigxxxC
\else
\ifnum#1=23 %
\hatcurLBigxxxD
\else
??????\fi
\fi
\fi
\fi
}
\newcommand{\hatcurLBii}[1]{\ifnum#1=20 %
\hatcurLBiixxxA
\else
\ifnum#1=21 %
\hatcurLBiixxxB
\else
\ifnum#1=22 %
\hatcurLBiixxxC
\else
\ifnum#1=23 %
\hatcurLBiixxxD
\else
??????\fi
\fi
\fi
\fi
}
\newcommand{\hatcurLBiI}[1]{\ifnum#1=20 %
\hatcurLBiIxxxA
\else
\ifnum#1=21 %
\hatcurLBiIxxxB
\else
\ifnum#1=22 %
\hatcurLBiIxxxC
\else
\ifnum#1=23 %
\hatcurLBiIxxxD
\else
??????\fi
\fi
\fi
\fi
}
\newcommand{\hatcurLBiig}[1]{\ifnum#1=20 %
\hatcurLBiigxxxA
\else
\ifnum#1=21 %
\hatcurLBiigxxxB
\else
\ifnum#1=22 %
\hatcurLBiigxxxC
\else
\ifnum#1=23 %
\hatcurLBiigxxxD
\else
??????\fi
\fi
\fi
\fi
}
\newcommand{\hatcurLBiii}[1]{\ifnum#1=20 %
\hatcurLBiiixxxA
\else
\ifnum#1=21 %
\hatcurLBiiixxxB
\else
\ifnum#1=22 %
\hatcurLBiiixxxC
\else
\ifnum#1=23 %
\hatcurLBiiixxxD
\else
??????\fi
\fi
\fi
\fi
}
\newcommand{\hatcurLBiiI}[1]{\ifnum#1=20 %
\hatcurLBiiIxxxA
\else
\ifnum#1=21 %
\hatcurLBiiIxxxB
\else
\ifnum#1=22 %
\hatcurLBiiIxxxC
\else
\ifnum#1=23 %
\hatcurLBiiIxxxD
\else
??????\fi
\fi
\fi
\fi
}
\newcommand{\hatcurLBiikep}[1]{\ifnum#1=20 %
\hatcurLBiikepxxxA
\else
\ifnum#1=21 %
\hatcurLBiikepxxxB
\else
\ifnum#1=22 %
\hatcurLBiikepxxxC
\else
\ifnum#1=23 %
\hatcurLBiikepxxxD
\else
??????\fi
\fi
\fi
\fi
}
\newcommand{\hatcurLBiiz}[1]{\ifnum#1=20 %
\hatcurLBiizxxxA
\else
\ifnum#1=21 %
\hatcurLBiizxxxB
\else
\ifnum#1=22 %
\hatcurLBiizxxxC
\else
\ifnum#1=23 %
\hatcurLBiizxxxD
\else
??????\fi
\fi
\fi
\fi
}
\newcommand{\hatcurLBikep}[1]{\ifnum#1=20 %
\hatcurLBikepxxxA
\else
\ifnum#1=21 %
\hatcurLBikepxxxB
\else
\ifnum#1=22 %
\hatcurLBikepxxxC
\else
\ifnum#1=23 %
\hatcurLBikepxxxD
\else
??????\fi
\fi
\fi
\fi
}
\newcommand{\hatcurLBiz}[1]{\ifnum#1=20 %
\hatcurLBizxxxA
\else
\ifnum#1=21 %
\hatcurLBizxxxB
\else
\ifnum#1=22 %
\hatcurLBizxxxC
\else
\ifnum#1=23 %
\hatcurLBizxxxD
\else
??????\fi
\fi
\fi
\fi
}
\newcommand{\hatcurLCbsq}[1]{\ifnum#1=20 %
\hatcurLCbsqxxxA
\else
\ifnum#1=21 %
\hatcurLCbsqxxxB
\else
\ifnum#1=22 %
\hatcurLCbsqxxxC
\else
\ifnum#1=23 %
\hatcurLCbsqxxxD
\else
??????\fi
\fi
\fi
\fi
}
\newcommand{\hatcurLCdip}[1]{\ifnum#1=20 %
\hatcurLCdipxxxA
\else
\ifnum#1=21 %
\hatcurLCdipxxxB
\else
\ifnum#1=22 %
\hatcurLCdipxxxC
\else
\ifnum#1=23 %
\hatcurLCdipxxxD
\else
??????\fi
\fi
\fi
\fi
}
\newcommand{\hatcurLCdur}[1]{\ifnum#1=20 %
\hatcurLCdurxxxA
\else
\ifnum#1=21 %
\hatcurLCdurxxxB
\else
\ifnum#1=22 %
\hatcurLCdurxxxC
\else
\ifnum#1=23 %
\hatcurLCdurxxxD
\else
??????\fi
\fi
\fi
\fi
}
\newcommand{\hatcurLCdurhr}[1]{\ifnum#1=20 %
\hatcurLCdurhrxxxA
\else
\ifnum#1=21 %
\hatcurLCdurhrxxxB
\else
\ifnum#1=22 %
\hatcurLCdurhrxxxC
\else
\ifnum#1=23 %
\hatcurLCdurhrxxxD
\else
??????\fi
\fi
\fi
\fi
}
\newcommand{\hatcurLCdurhrshort}[1]{\ifnum#1=20 %
\hatcurLCdurhrshortxxxA
\else
\ifnum#1=21 %
\hatcurLCdurhrshortxxxB
\else
\ifnum#1=22 %
\hatcurLCdurhrshortxxxC
\else
\ifnum#1=23 %
\hatcurLCdurhrshortxxxD
\else
??????\fi
\fi
\fi
\fi
}
\newcommand{\hatcurLCdurshort}[1]{\ifnum#1=20 %
\hatcurLCdurshortxxxA
\else
\ifnum#1=21 %
\hatcurLCdurshortxxxB
\else
\ifnum#1=22 %
\hatcurLCdurshortxxxC
\else
\ifnum#1=23 %
\hatcurLCdurshortxxxD
\else
??????\fi
\fi
\fi
\fi
}
\newcommand{\hatcurLChatnetm}[1]{\ifnum#1=20 %
\hatcurLChatnetmxxxA
\else
\ifnum#1=22 %
\hatcurLChatnetmxxxC
\else
\ifnum#1=23 %
\hatcurLChatnetmxxxD
\else
??????\fi
\fi
\fi
}
\newcommand{\hatcurLChatnetmA}[1]{\ifnum#1=21 %
\hatcurLChatnetmAxxxB
\else
??????\fi
}
\newcommand{\hatcurLChatnetmB}[1]{\ifnum#1=21 %
\hatcurLChatnetmBxxxB
\else
??????\fi
}
\newcommand{\hatcurLChatnetmC}[1]{\ifnum#1=21 %
\hatcurLChatnetmCxxxB
\else
??????\fi
}
\newcommand{\hatcurLCiblend}[1]{\ifnum#1=20 %
\hatcurLCiblendxxxA
\else
\ifnum#1=22 %
\hatcurLCiblendxxxC
\else
\ifnum#1=23 %
\hatcurLCiblendxxxD
\else
??????\fi
\fi
\fi
}
\newcommand{\hatcurLCiblendA}[1]{\ifnum#1=21 %
\hatcurLCiblendAxxxB
\else
??????\fi
}
\newcommand{\hatcurLCiblendB}[1]{\ifnum#1=21 %
\hatcurLCiblendBxxxB
\else
??????\fi
}
\newcommand{\hatcurLCiblendC}[1]{\ifnum#1=21 %
\hatcurLCiblendCxxxB
\else
??????\fi
}
\newcommand{\hatcurLCimp}[1]{\ifnum#1=20 %
\hatcurLCimpxxxA
\else
\ifnum#1=21 %
\hatcurLCimpxxxB
\else
\ifnum#1=22 %
\hatcurLCimpxxxC
\else
\ifnum#1=23 %
\hatcurLCimpxxxD
\else
??????\fi
\fi
\fi
\fi
}
\newcommand{\hatcurLCingdur}[1]{\ifnum#1=20 %
\hatcurLCingdurxxxA
\else
\ifnum#1=21 %
\hatcurLCingdurxxxB
\else
\ifnum#1=22 %
\hatcurLCingdurxxxC
\else
\ifnum#1=23 %
\hatcurLCingdurxxxD
\else
??????\fi
\fi
\fi
\fi
}
\newcommand{\hatcurLCP}[1]{\ifnum#1=20 %
\hatcurLCPxxxA
\else
\ifnum#1=21 %
\hatcurLCPxxxB
\else
\ifnum#1=22 %
\hatcurLCPxxxC
\else
\ifnum#1=23 %
\hatcurLCPxxxD
\else
??????\fi
\fi
\fi
\fi
}
\newcommand{\hatcurLCPprec}[1]{\ifnum#1=20 %
\hatcurLCPprecxxxA
\else
\ifnum#1=21 %
\hatcurLCPprecxxxB
\else
\ifnum#1=22 %
\hatcurLCPprecxxxC
\else
\ifnum#1=23 %
\hatcurLCPprecxxxD
\else
??????\fi
\fi
\fi
\fi
}
\newcommand{\hatcurLCPshort}[1]{\ifnum#1=20 %
\hatcurLCPshortxxxA
\else
\ifnum#1=21 %
\hatcurLCPshortxxxB
\else
\ifnum#1=22 %
\hatcurLCPshortxxxC
\else
\ifnum#1=23 %
\hatcurLCPshortxxxD
\else
??????\fi
\fi
\fi
\fi
}
\newcommand{\hatcurLCq}[1]{\ifnum#1=20 %
\hatcurLCqxxxA
\else
\ifnum#1=21 %
\hatcurLCqxxxB
\else
\ifnum#1=22 %
\hatcurLCqxxxC
\else
\ifnum#1=23 %
\hatcurLCqxxxD
\else
??????\fi
\fi
\fi
\fi
}
\newcommand{\hatcurLCqshort}[1]{\ifnum#1=20 %
\hatcurLCqshortxxxA
\else
\ifnum#1=21 %
\hatcurLCqshortxxxB
\else
\ifnum#1=22 %
\hatcurLCqshortxxxC
\else
\ifnum#1=23 %
\hatcurLCqshortxxxD
\else
??????\fi
\fi
\fi
\fi
}
\newcommand{\hatcurLCrprstar}[1]{\ifnum#1=20 %
\hatcurLCrprstarxxxA
\else
\ifnum#1=21 %
\hatcurLCrprstarxxxB
\else
\ifnum#1=22 %
\hatcurLCrprstarxxxC
\else
\ifnum#1=23 %
\hatcurLCrprstarxxxD
\else
??????\fi
\fi
\fi
\fi
}
\newcommand{\hatcurLCT}[1]{\ifnum#1=20 %
\hatcurLCTxxxA
\else
\ifnum#1=21 %
\hatcurLCTxxxB
\else
\ifnum#1=22 %
\hatcurLCTxxxC
\else
\ifnum#1=23 %
\hatcurLCTxxxD
\else
??????\fi
\fi
\fi
\fi
}
\newcommand{\hatcurLCTA}[1]{\ifnum#1=20 %
\hatcurLCTAxxxA
\else
\ifnum#1=21 %
\hatcurLCTAxxxB
\else
\ifnum#1=22 %
\hatcurLCTAxxxC
\else
\ifnum#1=23 %
\hatcurLCTAxxxD
\else
??????\fi
\fi
\fi
\fi
}
\newcommand{\hatcurLCTB}[1]{\ifnum#1=20 %
\hatcurLCTBxxxA
\else
\ifnum#1=21 %
\hatcurLCTBxxxB
\else
\ifnum#1=22 %
\hatcurLCTBxxxC
\else
\ifnum#1=23 %
\hatcurLCTBxxxD
\else
??????\fi
\fi
\fi
\fi
}
\newcommand{\hatcurLCzeta}[1]{\ifnum#1=20 %
\hatcurLCzetaxxxA
\else
\ifnum#1=21 %
\hatcurLCzetaxxxB
\else
\ifnum#1=22 %
\hatcurLCzetaxxxC
\else
\ifnum#1=23 %
\hatcurLCzetaxxxD
\else
??????\fi
\fi
\fi
\fi
}
\newcommand{\hatcurPPaequiv}[1]{\ifnum#1=20 %
\hatcurPPaequivxxxA
\else
\ifnum#1=21 %
\hatcurPPaequivxxxB
\else
\ifnum#1=22 %
\hatcurPPaequivxxxC
\else
\ifnum#1=23 %
\hatcurPPaequivxxxD
\else
??????\fi
\fi
\fi
\fi
}
\newcommand{\hatcurPPar}[1]{\ifnum#1=20 %
\hatcurPParxxxA
\else
\ifnum#1=21 %
\hatcurPParxxxB
\else
\ifnum#1=22 %
\hatcurPParxxxC
\else
\ifnum#1=23 %
\hatcurPParxxxD
\else
??????\fi
\fi
\fi
\fi
}
\newcommand{\hatcurPParel}[1]{\ifnum#1=20 %
\hatcurPParelxxxA
\else
\ifnum#1=21 %
\hatcurPParelxxxB
\else
\ifnum#1=22 %
\hatcurPParelxxxC
\else
\ifnum#1=23 %
\hatcurPParelxxxD
\else
??????\fi
\fi
\fi
\fi
}
\newcommand{\hatcurPPfluxap}[1]{\ifnum#1=20 %
\hatcurPPfluxapxxxA
\else
\ifnum#1=21 %
\hatcurPPfluxapxxxB
\else
\ifnum#1=22 %
\hatcurPPfluxapxxxC
\else
\ifnum#1=23 %
\hatcurPPfluxapxxxD
\else
??????\fi
\fi
\fi
\fi
}
\newcommand{\hatcurPPfluxapdim}[1]{\ifnum#1=20 %
\hatcurPPfluxapdimxxxA
\else
\ifnum#1=21 %
\hatcurPPfluxapdimxxxB
\else
\ifnum#1=22 %
\hatcurPPfluxapdimxxxC
\else
\ifnum#1=23 %
\hatcurPPfluxapdimxxxD
\else
??????\fi
\fi
\fi
\fi
}
\newcommand{\hatcurPPfluxavg}[1]{\ifnum#1=20 %
\hatcurPPfluxavgxxxA
\else
\ifnum#1=21 %
\hatcurPPfluxavgxxxB
\else
\ifnum#1=22 %
\hatcurPPfluxavgxxxC
\else
\ifnum#1=23 %
\hatcurPPfluxavgxxxD
\else
??????\fi
\fi
\fi
\fi
}
\newcommand{\hatcurPPfluxavgdim}[1]{\ifnum#1=20 %
\hatcurPPfluxavgdimxxxA
\else
\ifnum#1=21 %
\hatcurPPfluxavgdimxxxB
\else
\ifnum#1=22 %
\hatcurPPfluxavgdimxxxC
\else
\ifnum#1=23 %
\hatcurPPfluxavgdimxxxD
\else
??????\fi
\fi
\fi
\fi
}
\newcommand{\hatcurPPfluxperi}[1]{\ifnum#1=20 %
\hatcurPPfluxperixxxA
\else
\ifnum#1=21 %
\hatcurPPfluxperixxxB
\else
\ifnum#1=22 %
\hatcurPPfluxperixxxC
\else
\ifnum#1=23 %
\hatcurPPfluxperixxxD
\else
??????\fi
\fi
\fi
\fi
}
\newcommand{\hatcurPPfluxperidim}[1]{\ifnum#1=20 %
\hatcurPPfluxperidimxxxA
\else
\ifnum#1=21 %
\hatcurPPfluxperidimxxxB
\else
\ifnum#1=22 %
\hatcurPPfluxperidimxxxC
\else
\ifnum#1=23 %
\hatcurPPfluxperidimxxxD
\else
??????\fi
\fi
\fi
\fi
}
\newcommand{\hatcurPPg}[1]{\ifnum#1=20 %
\hatcurPPgxxxA
\else
\ifnum#1=21 %
\hatcurPPgxxxB
\else
\ifnum#1=22 %
\hatcurPPgxxxC
\else
\ifnum#1=23 %
\hatcurPPgxxxD
\else
??????\fi
\fi
\fi
\fi
}
\newcommand{\hatcurPPi}[1]{\ifnum#1=20 %
\hatcurPPixxxA
\else
\ifnum#1=21 %
\hatcurPPixxxB
\else
\ifnum#1=22 %
\hatcurPPixxxC
\else
\ifnum#1=23 %
\hatcurPPixxxD
\else
??????\fi
\fi
\fi
\fi
}
\newcommand{\hatcurPPlogg}[1]{\ifnum#1=20 %
\hatcurPPloggxxxA
\else
\ifnum#1=21 %
\hatcurPPloggxxxB
\else
\ifnum#1=22 %
\hatcurPPloggxxxC
\else
\ifnum#1=23 %
\hatcurPPloggxxxD
\else
??????\fi
\fi
\fi
\fi
}
\newcommand{\hatcurPPm}[1]{\ifnum#1=20 %
\hatcurPPmxxxA
\else
\ifnum#1=21 %
\hatcurPPmxxxB
\else
\ifnum#1=22 %
\hatcurPPmxxxC
\else
\ifnum#1=23 %
\hatcurPPmxxxD
\else
??????\fi
\fi
\fi
\fi
}
\newcommand{\hatcurPPme}[1]{\ifnum#1=20 %
\hatcurPPmexxxA
\else
\ifnum#1=21 %
\hatcurPPmexxxB
\else
\ifnum#1=22 %
\hatcurPPmexxxC
\else
\ifnum#1=23 %
\hatcurPPmexxxD
\else
??????\fi
\fi
\fi
\fi
}
\newcommand{\hatcurPPmelong}[1]{\ifnum#1=20 %
\hatcurPPmelongxxxA
\else
\ifnum#1=21 %
\hatcurPPmelongxxxB
\else
\ifnum#1=22 %
\hatcurPPmelongxxxC
\else
\ifnum#1=23 %
\hatcurPPmelongxxxD
\else
??????\fi
\fi
\fi
\fi
}
\newcommand{\hatcurPPmeshort}[1]{\ifnum#1=20 %
\hatcurPPmeshortxxxA
\else
\ifnum#1=21 %
\hatcurPPmeshortxxxB
\else
\ifnum#1=22 %
\hatcurPPmeshortxxxC
\else
\ifnum#1=23 %
\hatcurPPmeshortxxxD
\else
??????\fi
\fi
\fi
\fi
}
\newcommand{\hatcurPPmlong}[1]{\ifnum#1=20 %
\hatcurPPmlongxxxA
\else
\ifnum#1=21 %
\hatcurPPmlongxxxB
\else
\ifnum#1=22 %
\hatcurPPmlongxxxC
\else
\ifnum#1=23 %
\hatcurPPmlongxxxD
\else
??????\fi
\fi
\fi
\fi
}
\newcommand{\hatcurPPmrcorr}[1]{\ifnum#1=20 %
\hatcurPPmrcorrxxxA
\else
\ifnum#1=21 %
\hatcurPPmrcorrxxxB
\else
\ifnum#1=22 %
\hatcurPPmrcorrxxxC
\else
\ifnum#1=23 %
\hatcurPPmrcorrxxxD
\else
??????\fi
\fi
\fi
\fi
}
\newcommand{\hatcurPPmshort}[1]{\ifnum#1=20 %
\hatcurPPmshortxxxA
\else
\ifnum#1=21 %
\hatcurPPmshortxxxB
\else
\ifnum#1=22 %
\hatcurPPmshortxxxC
\else
\ifnum#1=23 %
\hatcurPPmshortxxxD
\else
??????\fi
\fi
\fi
\fi
}
\newcommand{\hatcurPPperi}[1]{\ifnum#1=20 %
\hatcurPPperixxxA
\else
\ifnum#1=21 %
\hatcurPPperixxxB
\else
\ifnum#1=22 %
\hatcurPPperixxxC
\else
\ifnum#1=23 %
\hatcurPPperixxxD
\else
??????\fi
\fi
\fi
\fi
}
\newcommand{\hatcurPPphiconj}[1]{\ifnum#1=20 %
\hatcurPPphiconjxxxA
\else
\ifnum#1=21 %
\hatcurPPphiconjxxxB
\else
\ifnum#1=22 %
\hatcurPPphiconjxxxC
\else
\ifnum#1=23 %
\hatcurPPphiconjxxxD
\else
??????\fi
\fi
\fi
\fi
}
\newcommand{\hatcurPPr}[1]{\ifnum#1=20 %
\hatcurPPrxxxA
\else
\ifnum#1=21 %
\hatcurPPrxxxB
\else
\ifnum#1=22 %
\hatcurPPrxxxC
\else
\ifnum#1=23 %
\hatcurPPrxxxD
\else
??????\fi
\fi
\fi
\fi
}
\newcommand{\hatcurPPre}[1]{\ifnum#1=20 %
\hatcurPPrexxxA
\else
\ifnum#1=21 %
\hatcurPPrexxxB
\else
\ifnum#1=22 %
\hatcurPPrexxxC
\else
\ifnum#1=23 %
\hatcurPPrexxxD
\else
??????\fi
\fi
\fi
\fi
}
\newcommand{\hatcurPPrelong}[1]{\ifnum#1=20 %
\hatcurPPrelongxxxA
\else
\ifnum#1=21 %
\hatcurPPrelongxxxB
\else
\ifnum#1=22 %
\hatcurPPrelongxxxC
\else
\ifnum#1=23 %
\hatcurPPrelongxxxD
\else
??????\fi
\fi
\fi
\fi
}
\newcommand{\hatcurPPreshort}[1]{\ifnum#1=20 %
\hatcurPPreshortxxxA
\else
\ifnum#1=21 %
\hatcurPPreshortxxxB
\else
\ifnum#1=22 %
\hatcurPPreshortxxxC
\else
\ifnum#1=23 %
\hatcurPPreshortxxxD
\else
??????\fi
\fi
\fi
\fi
}
\newcommand{\hatcurPPrho}[1]{\ifnum#1=20 %
\hatcurPPrhoxxxA
\else
\ifnum#1=21 %
\hatcurPPrhoxxxB
\else
\ifnum#1=22 %
\hatcurPPrhoxxxC
\else
\ifnum#1=23 %
\hatcurPPrhoxxxD
\else
??????\fi
\fi
\fi
\fi
}
\newcommand{\hatcurPPrlong}[1]{\ifnum#1=20 %
\hatcurPPrlongxxxA
\else
\ifnum#1=21 %
\hatcurPPrlongxxxB
\else
\ifnum#1=22 %
\hatcurPPrlongxxxC
\else
\ifnum#1=23 %
\hatcurPPrlongxxxD
\else
??????\fi
\fi
\fi
\fi
}
\newcommand{\hatcurPPrshort}[1]{\ifnum#1=20 %
\hatcurPPrshortxxxA
\else
\ifnum#1=21 %
\hatcurPPrshortxxxB
\else
\ifnum#1=22 %
\hatcurPPrshortxxxC
\else
\ifnum#1=23 %
\hatcurPPrshortxxxD
\else
??????\fi
\fi
\fi
\fi
}
\newcommand{\hatcurPPtcirc}[1]{\ifnum#1=20 %
\hatcurPPtcircxxxA
\else
\ifnum#1=21 %
\hatcurPPtcircxxxB
\else
\ifnum#1=22 %
\hatcurPPtcircxxxC
\else
\ifnum#1=23 %
\hatcurPPtcircxxxD
\else
??????\fi
\fi
\fi
\fi
}
\newcommand{\hatcurPPteff}[1]{\ifnum#1=20 %
\hatcurPPteffxxxA
\else
\ifnum#1=21 %
\hatcurPPteffxxxB
\else
\ifnum#1=22 %
\hatcurPPteffxxxC
\else
\ifnum#1=23 %
\hatcurPPteffxxxD
\else
??????\fi
\fi
\fi
\fi
}
\newcommand{\hatcurPPtheta}[1]{\ifnum#1=20 %
\hatcurPPthetaxxxA
\else
\ifnum#1=21 %
\hatcurPPthetaxxxB
\else
\ifnum#1=22 %
\hatcurPPthetaxxxC
\else
\ifnum#1=23 %
\hatcurPPthetaxxxD
\else
??????\fi
\fi
\fi
\fi
}
\newcommand{\hatcurPPtinfall}[1]{\ifnum#1=20 %
\hatcurPPtinfallxxxA
\else
\ifnum#1=21 %
\hatcurPPtinfallxxxB
\else
\ifnum#1=22 %
\hatcurPPtinfallxxxC
\else
\ifnum#1=23 %
\hatcurPPtinfallxxxD
\else
??????\fi
\fi
\fi
\fi
}
\newcommand{\hatcurRVeccen}[1]{\ifnum#1=20 %
\hatcurRVeccenxxxA
\else
\ifnum#1=21 %
\hatcurRVeccenxxxB
\else
\ifnum#1=22 %
\hatcurRVeccenxxxC
\else
\ifnum#1=23 %
\hatcurRVeccenxxxD
\else
??????\fi
\fi
\fi
\fi
}
\newcommand{\hatcurRVfitrms}[1]{\ifnum#1=20 %
\hatcurRVfitrmsxxxA
\else
\ifnum#1=21 %
\hatcurRVfitrmsxxxB
\else
\ifnum#1=22 %
\hatcurRVfitrmsxxxC
\else
\ifnum#1=23 %
\hatcurRVfitrmsxxxD
\else
??????\fi
\fi
\fi
\fi
}
\newcommand{\hatcurRVgamma}[1]{\ifnum#1=20 %
\hatcurRVgammaxxxA
\else
\ifnum#1=21 %
\hatcurRVgammaxxxB
\else
\ifnum#1=22 %
\hatcurRVgammaxxxC
\else
\ifnum#1=23 %
\hatcurRVgammaxxxD
\else
??????\fi
\fi
\fi
\fi
}
\newcommand{\hatcurRVh}[1]{\ifnum#1=20 %
\hatcurRVhxxxA
\else
\ifnum#1=21 %
\hatcurRVhxxxB
\else
\ifnum#1=22 %
\hatcurRVhxxxC
\else
\ifnum#1=23 %
\hatcurRVhxxxD
\else
??????\fi
\fi
\fi
\fi
}
\newcommand{\hatcurRVjitter}[1]{\ifnum#1=20 %
\hatcurRVjitterxxxA
\else
\ifnum#1=21 %
\hatcurRVjitterxxxB
\else
\ifnum#1=22 %
\hatcurRVjitterxxxC
\else
\ifnum#1=23 %
\hatcurRVjitterxxxD
\else
??????\fi
\fi
\fi
\fi
}
\newcommand{\hatcurRVk}[1]{\ifnum#1=20 %
\hatcurRVkxxxA
\else
\ifnum#1=21 %
\hatcurRVkxxxB
\else
\ifnum#1=22 %
\hatcurRVkxxxC
\else
\ifnum#1=23 %
\hatcurRVkxxxD
\else
??????\fi
\fi
\fi
\fi
}
\newcommand{\hatcurRVK}[1]{\ifnum#1=20 %
\hatcurRVKxxxA
\else
\ifnum#1=21 %
\hatcurRVKxxxB
\else
\ifnum#1=22 %
\hatcurRVKxxxC
\else
\ifnum#1=23 %
\hatcurRVKxxxD
\else
??????\fi
\fi
\fi
\fi
}
\newcommand{\hatcurRVomega}[1]{\ifnum#1=20 %
\hatcurRVomegaxxxA
\else
\ifnum#1=21 %
\hatcurRVomegaxxxB
\else
\ifnum#1=22 %
\hatcurRVomegaxxxC
\else
\ifnum#1=23 %
\hatcurRVomegaxxxD
\else
??????\fi
\fi
\fi
\fi
}
\newcommand{\hatcurSMEiilogg}[1]{\ifnum#1=20 %
\hatcurSMEiiloggxxxA
\else
\ifnum#1=21 %
\hatcurSMEiiloggxxxB
\else
\ifnum#1=22 %
\hatcurSMEiiloggxxxC
\else
\ifnum#1=23 %
\hatcurSMEiiloggxxxD
\else
??????\fi
\fi
\fi
\fi
}
\newcommand{\hatcurSMEiiteff}[1]{\ifnum#1=20 %
\hatcurSMEiiteffxxxA
\else
\ifnum#1=21 %
\hatcurSMEiiteffxxxB
\else
\ifnum#1=22 %
\hatcurSMEiiteffxxxC
\else
\ifnum#1=23 %
\hatcurSMEiiteffxxxD
\else
??????\fi
\fi
\fi
\fi
}
\newcommand{\hatcurSMEiivmac}[1]{\ifnum#1=20 %
\hatcurSMEiivmacxxxA
\else
\ifnum#1=21 %
\hatcurSMEiivmacxxxB
\else
\ifnum#1=22 %
\hatcurSMEiivmacxxxC
\else
\ifnum#1=23 %
\hatcurSMEiivmacxxxD
\else
??????\fi
\fi
\fi
\fi
}
\newcommand{\hatcurSMEiivmic}[1]{\ifnum#1=20 %
\hatcurSMEiivmicxxxA
\else
\ifnum#1=21 %
\hatcurSMEiivmicxxxB
\else
\ifnum#1=22 %
\hatcurSMEiivmicxxxC
\else
\ifnum#1=23 %
\hatcurSMEiivmicxxxD
\else
??????\fi
\fi
\fi
\fi
}
\newcommand{\hatcurSMEiivsin}[1]{\ifnum#1=20 %
\hatcurSMEiivsinxxxA
\else
\ifnum#1=21 %
\hatcurSMEiivsinxxxB
\else
\ifnum#1=22 %
\hatcurSMEiivsinxxxC
\else
\ifnum#1=23 %
\hatcurSMEiivsinxxxD
\else
??????\fi
\fi
\fi
\fi
}
\newcommand{\hatcurSMEiizfeh}[1]{\ifnum#1=20 %
\hatcurSMEiizfehxxxA
\else
\ifnum#1=21 %
\hatcurSMEiizfehxxxB
\else
\ifnum#1=22 %
\hatcurSMEiizfehxxxC
\else
\ifnum#1=23 %
\hatcurSMEiizfehxxxD
\else
??????\fi
\fi
\fi
\fi
}
\newcommand{\hatcurSMEiizfehshort}[1]{\ifnum#1=20 %
\hatcurSMEiizfehshortxxxA
\else
\ifnum#1=21 %
\hatcurSMEiizfehshortxxxB
\else
\ifnum#1=22 %
\hatcurSMEiizfehshortxxxC
\else
\ifnum#1=23 %
\hatcurSMEiizfehshortxxxD
\else
??????\fi
\fi
\fi
\fi
}
\newcommand{\hatcurSMEilogg}[1]{\ifnum#1=20 %
\hatcurSMEiloggxxxA
\else
\ifnum#1=21 %
\hatcurSMEiloggxxxB
\else
\ifnum#1=22 %
\hatcurSMEiloggxxxC
\else
\ifnum#1=23 %
\hatcurSMEiloggxxxD
\else
??????\fi
\fi
\fi
\fi
}
\newcommand{\hatcurSMEiteff}[1]{\ifnum#1=20 %
\hatcurSMEiteffxxxA
\else
\ifnum#1=21 %
\hatcurSMEiteffxxxB
\else
\ifnum#1=22 %
\hatcurSMEiteffxxxC
\else
\ifnum#1=23 %
\hatcurSMEiteffxxxD
\else
??????\fi
\fi
\fi
\fi
}
\newcommand{\hatcurSMEivmac}[1]{\ifnum#1=20 %
\hatcurSMEivmacxxxA
\else
\ifnum#1=21 %
\hatcurSMEivmacxxxB
\else
\ifnum#1=22 %
\hatcurSMEivmacxxxC
\else
\ifnum#1=23 %
\hatcurSMEivmacxxxD
\else
??????\fi
\fi
\fi
\fi
}
\newcommand{\hatcurSMEivmic}[1]{\ifnum#1=20 %
\hatcurSMEivmicxxxA
\else
\ifnum#1=21 %
\hatcurSMEivmicxxxB
\else
\ifnum#1=22 %
\hatcurSMEivmicxxxC
\else
\ifnum#1=23 %
\hatcurSMEivmicxxxD
\else
??????\fi
\fi
\fi
\fi
}
\newcommand{\hatcurSMEivsin}[1]{\ifnum#1=20 %
\hatcurSMEivsinxxxA
\else
\ifnum#1=21 %
\hatcurSMEivsinxxxB
\else
\ifnum#1=22 %
\hatcurSMEivsinxxxC
\else
\ifnum#1=23 %
\hatcurSMEivsinxxxD
\else
??????\fi
\fi
\fi
\fi
}
\newcommand{\hatcurSMEizfeh}[1]{\ifnum#1=20 %
\hatcurSMEizfehxxxA
\else
\ifnum#1=21 %
\hatcurSMEizfehxxxB
\else
\ifnum#1=22 %
\hatcurSMEizfehxxxC
\else
\ifnum#1=23 %
\hatcurSMEizfehxxxD
\else
??????\fi
\fi
\fi
\fi
}
\newcommand{\hatcurSMEizfehshort}[1]{\ifnum#1=20 %
\hatcurSMEizfehshortxxxA
\else
\ifnum#1=21 %
\hatcurSMEizfehshortxxxB
\else
\ifnum#1=22 %
\hatcurSMEizfehshortxxxC
\else
\ifnum#1=23 %
\hatcurSMEizfehshortxxxD
\else
??????\fi
\fi
\fi
\fi
}
\newcommand{\hatcurTRESgamma}[1]{\ifnum#1=20 %
\hatcurTRESgammaxxxA
\else
\ifnum#1=21 %
\hatcurTRESgammaxxxB
\else
\ifnum#1=22 %
\hatcurTRESgammaxxxC
\else
\ifnum#1=23 %
\hatcurTRESgammaxxxD
\else
??????\fi
\fi
\fi
\fi
}
\newcommand{\hatcurTRESlogg}[1]{\ifnum#1=20 %
\hatcurTRESloggxxxA
\else
\ifnum#1=21 %
\hatcurTRESloggxxxB
\else
\ifnum#1=22 %
\hatcurTRESloggxxxC
\else
\ifnum#1=23 %
\hatcurTRESloggxxxD
\else
??????\fi
\fi
\fi
\fi
}
\newcommand{\hatcurTRESnumspec}[1]{\ifnum#1=20 %
\hatcurTRESnumspecxxxA
\else
\ifnum#1=21 %
\hatcurTRESnumspecxxxB
\else
\ifnum#1=22 %
\hatcurTRESnumspecxxxC
\else
\ifnum#1=23 %
\hatcurTRESnumspecxxxD
\else
??????\fi
\fi
\fi
\fi
}
\newcommand{\hatcurTRESrvrms}[1]{\ifnum#1=20 %
\hatcurTRESrvrmsxxxA
\else
\ifnum#1=21 %
\hatcurTRESrvrmsxxxB
\else
\ifnum#1=22 %
\hatcurTRESrvrmsxxxC
\else
\ifnum#1=23 %
\hatcurTRESrvrmsxxxD
\else
??????\fi
\fi
\fi
\fi
}
\newcommand{\hatcurTRESspan}[1]{\ifnum#1=20 %
\hatcurTRESspanxxxA
\else
\ifnum#1=21 %
\hatcurTRESspanxxxB
\else
\ifnum#1=22 %
\hatcurTRESspanxxxC
\else
\ifnum#1=23 %
\hatcurTRESspanxxxD
\else
??????\fi
\fi
\fi
\fi
}
\newcommand{\hatcurTRESteff}[1]{\ifnum#1=20 %
\hatcurTRESteffxxxA
\else
\ifnum#1=21 %
\hatcurTRESteffxxxB
\else
\ifnum#1=22 %
\hatcurTRESteffxxxC
\else
\ifnum#1=23 %
\hatcurTRESteffxxxD
\else
??????\fi
\fi
\fi
\fi
}
\newcommand{\hatcurTRESvsini}[1]{\ifnum#1=20 %
\hatcurTRESvsinixxxA
\else
\ifnum#1=21 %
\hatcurTRESvsinixxxB
\else
\ifnum#1=22 %
\hatcurTRESvsinixxxC
\else
\ifnum#1=23 %
\hatcurTRESvsinixxxD
\else
??????\fi
\fi
\fi
\fi
}
\newcommand{\hatcurTRESzfeh}[1]{\ifnum#1=20 %
\hatcurTRESzfehxxxA
\else
\ifnum#1=21 %
\hatcurTRESzfehxxxB
\else
\ifnum#1=22 %
\hatcurTRESzfehxxxC
\else
\ifnum#1=23 %
\hatcurTRESzfehxxxD
\else
??????\fi
\fi
\fi
\fi
}
\newcommand{\hatcurXdist}[1]{\ifnum#1=20 %
\hatcurXdistxxxA
\else
\ifnum#1=21 %
\hatcurXdistxxxB
\else
\ifnum#1=22 %
\hatcurXdistxxxC
\else
\ifnum#1=23 %
\hatcurXdistxxxD
\else
??????\fi
\fi
\fi
\fi
}
\newcommand{\hatcurXsecdur}[1]{\ifnum#1=20 %
\hatcurXsecdurxxxA
\else
\ifnum#1=21 %
\hatcurXsecdurxxxB
\else
\ifnum#1=22 %
\hatcurXsecdurxxxC
\else
\ifnum#1=23 %
\hatcurXsecdurxxxD
\else
??????\fi
\fi
\fi
\fi
}
\newcommand{\hatcurXsecingdur}[1]{\ifnum#1=20 %
\hatcurXsecingdurxxxA
\else
\ifnum#1=21 %
\hatcurXsecingdurxxxB
\else
\ifnum#1=22 %
\hatcurXsecingdurxxxC
\else
\ifnum#1=23 %
\hatcurXsecingdurxxxD
\else
??????\fi
\fi
\fi
\fi
}
\newcommand{\hatcurXsecondary}[1]{\ifnum#1=20 %
\hatcurXsecondaryxxxA
\else
\ifnum#1=21 %
\hatcurXsecondaryxxxB
\else
\ifnum#1=22 %
\hatcurXsecondaryxxxC
\else
\ifnum#1=23 %
\hatcurXsecondaryxxxD
\else
??????\fi
\fi
\fi
\fi
}
\newcommand{\hatcurXsecphase}[1]{\ifnum#1=20 %
\hatcurXsecphasexxxA
\else
\ifnum#1=21 %
\hatcurXsecphasexxxB
\else
\ifnum#1=22 %
\hatcurXsecphasexxxC
\else
\ifnum#1=23 %
\hatcurXsecphasexxxD
\else
??????\fi
\fi
\fi
\fi
}

%% file: HTR267-010.tex
\newcommand{\hatcurhtrxxxA}{HTR267-010}                                
\newcommand{\hatcurfieldxxxA}{267}                                     
\newcommand{\hatcurCCraxxxA}{\ensuremath{07^{\mathrm h}27^{\mathrm m}39.96^{\mathrm s}}}                              
\newcommand{\hatcurCCdecxxxA}{\ensuremath{+24{\arcdeg}20{\arcmin}11.9{\arcsec}}}                             
\newcommand{\hatcurCCmagxxxA}{11.339}                                  
\newcommand{\hatcurCCtwomassxxxA}{2MASS~07273995+2420118}              
\newcommand{\hatcurCCgscxxxA}{GSC~1910-00239}                          
\newcommand{\hatcurCCtassmvxxxA}{11.339}                               
\newcommand{\hatcurCCtwomassJmagxxxA}{\ensuremath{9.276\pm0.022}}      
\newcommand{\hatcurCCtwomassHmagxxxA}{\ensuremath{8.743\pm0.021}}      
\newcommand{\hatcurCCtwomassKmagxxxA}{\ensuremath{8.601\pm0.019}}      
\newcommand{\hatcurCCcitJmagxxxA}{\ensuremath{9.277\pm0.023}}          
\newcommand{\hatcurCCcitHmagxxxA}{\ensuremath{8.736\pm0.022}}          
\newcommand{\hatcurCCcitKmagxxxA}{\ensuremath{8.625\pm0.020}}          
\newcommand{\hatcurCCbbJmagxxxA}{\ensuremath{9.351\pm0.025}}           
\newcommand{\hatcurCCbbHmagxxxA}{\ensuremath{8.760\pm0.023}}           
\newcommand{\hatcurCCbbKmagxxxA}{\ensuremath{8.645\pm0.020}}           
\newcommand{\hatcurCCesoJmagxxxA}{\ensuremath{9.357\pm0.028}}          
\newcommand{\hatcurCCesoHmagxxxA}{\ensuremath{8.756\pm0.030}}          
\newcommand{\hatcurCCesoKmagxxxA}{\ensuremath{8.642\pm0.021}}          
\newcommand{\hatcurCCesoJHmagxxxA}{\ensuremath{0.600\pm0.038}}         
\newcommand{\hatcurCCesoJKmagxxxA}{\ensuremath{0.715\pm0.033}}         
\newcommand{\hatcurCCesoHKmagxxxA}{\ensuremath{0.114\pm0.035}}         
\newcommand{\hatcurLCdipxxxA}{\ensuremath{10.8}}                       
\newcommand{\hatcurLCrprstarxxxA}{\ensuremath{0.1284\pm0.0016}}        
\newcommand{\hatcurLCbsqxxxA}{\ensuremath{0.398_{-0.034}^{+0.032}}}    
\newcommand{\hatcurLCimpxxxA}{\ensuremath{0.631_{-0.028}^{+0.025}}}    
\newcommand{\hatcurLCzetaxxxA}{\ensuremath{31.32\pm0.22}}              
\newcommand{\hatcurLCdurxxxA}{\ensuremath{0.0770\pm0.0008}}            
\newcommand{\hatcurLCdurshortxxxA}{\ensuremath{0.0770}}                
\newcommand{\hatcurLCdurhrxxxA}{\ensuremath{1.848\pm0.019}}            
\newcommand{\hatcurLCdurhrshortxxxA}{\ensuremath{1.848}}               
\newcommand{\hatcurLCqxxxA}{\ensuremath{0.0268\pm0.0003}}              
\newcommand{\hatcurLCqshortxxxA}{\ensuremath{0.027}}                   
\newcommand{\hatcurLCingdurxxxA}{\ensuremath{0.0137\pm0.0009}}         
\newcommand{\hatcurLCPxxxA}{\ensuremath{2.875317\pm0.000004}}          
\newcommand{\hatcurLCPprecxxxA}{\ensuremath{2.8753172}}                
\newcommand{\hatcurLCPshortxxxA}{\ensuremath{2.8753}}                  
\newcommand{\hatcurLCTxxxA}{\ensuremath{2455080.92661\pm0.00021}}      
\newcommand{\hatcurLCTAxxxA}{\ensuremath{2454402.35175\pm0.00098}}     
\newcommand{\hatcurLCTBxxxA}{\ensuremath{2455126.93168\pm0.00025}}     
\newcommand{\hatcurLChatnetmxxxA}{\ensuremath{10.6939\pm0.0001}}       
\newcommand{\hatcurLCiblendxxxA}{\ensuremath{0.66\pm0.05}}             
\newcommand{\hatcurSMEiteffxxxA}{\ensuremath{4626\pm104}}              
\newcommand{\hatcurSMEizfehxxxA}{\ensuremath{+0.31\pm0.08}}            
\newcommand{\hatcurSMEizfehshortxxxA}{\ensuremath{+0.31}}              
\newcommand{\hatcurSMEiloggxxxA}{\ensuremath{4.80\pm0.06}}             
\newcommand{\hatcurSMEivsinxxxA}{\ensuremath{3.1\pm0.5}}               
\newcommand{\hatcurSMEivmacxxxA}{\ensuremath{2.26}}                    
\newcommand{\hatcurSMEivmicxxxA}{\ensuremath{0.85}}                    
\newcommand{\hatcurSMEiiteffxxxA}{\ensuremath{4595\pm80}}              
\newcommand{\hatcurSMEiizfehxxxA}{\ensuremath{+0.35\pm0.08}}           
\newcommand{\hatcurSMEiizfehshortxxxA}{\ensuremath{+0.35}}             
\newcommand{\hatcurSMEiiloggxxxA}{\ensuremath{4.64\pm0.06}}            
\newcommand{\hatcurSMEiivsinxxxA}{\ensuremath{2.1\pm0.5}}              
\newcommand{\hatcurSMEiivmacxxxA}{\ensuremath{2.21}}                   
\newcommand{\hatcurSMEiivmicxxxA}{\ensuremath{0.85}}                   
\newcommand{\hatcurDSteffxxxA}{\ensuremath{4500\pm125}}                
\newcommand{\hatcurDSzfehxxxA}{\ensuremath{0.0\pm0.0}}                 
\newcommand{\hatcurDSloggxxxA}{\ensuremath{4.0\pm0.25}}                
\newcommand{\hatcurDSvsinixxxA}{\ensuremath{0\pm^{4}_{0}}}             
\newcommand{\hatcurDSgammaxxxA}{\ensuremath{-18.81\pm0.68}}            
\newcommand{\hatcurDSnumspecxxxA}{\ensuremath{3}}                      
\newcommand{\hatcurDSspanxxxA}{\ensuremath{4}}                         
\newcommand{\hatcurDSrvrmsxxxA}{\ensuremath{1.18}}                     
\newcommand{\hatcurTRESteffxxxA}{\ensuremath{}}                        
\newcommand{\hatcurTRESzfehxxxA}{\ensuremath{}}                        
\newcommand{\hatcurTRESloggxxxA}{\ensuremath{}}                        
\newcommand{\hatcurTRESvsinixxxA}{\ensuremath{}}                       
\newcommand{\hatcurTRESgammaxxxA}{\ensuremath{}}                       
\newcommand{\hatcurTRESnumspecxxxA}{\ensuremath{}}             
\newcommand{\hatcurTRESspanxxxA}{\ensuremath{}}                
\newcommand{\hatcurTRESrvrmsxxxA}{\ensuremath{}}               
\newcommand{\hatcurFIESteffxxxA}{\ensuremath{4500}}                            
\newcommand{\hatcurFIESzfehxxxA}{\ensuremath{}}                            
\newcommand{\hatcurFIESloggxxxA}{\ensuremath{4.0}}                            
\newcommand{\hatcurFIESvsinixxxA}{\ensuremath{4}}                           
\newcommand{\hatcurFIESgammaxxxA}{\ensuremath{-16.76\pm0.1}}                           
\newcommand{\hatcurFIESnumspecxxxA}{\ensuremath{1}}                 
\newcommand{\hatcurFIESspanxxxA}{\ensuremath{}}                    
\newcommand{\hatcurFIESrvrmsxxxA}{\ensuremath{}}                   
\newcommand{\hatcurLBizxxxA}{\ensuremath{0.3688}}                      
\newcommand{\hatcurLBiizxxxA}{\ensuremath{0.2494}}                     
\newcommand{\hatcurLBiixxxA}{\ensuremath{0.4719}}                      
\newcommand{\hatcurLBiiixxxA}{\ensuremath{0.2174}}                     
\newcommand{\hatcurLBiIxxxA}{\ensuremath{0.4348}}                      
\newcommand{\hatcurLBiiIxxxA}{\ensuremath{0.2297}}                     
\newcommand{\hatcurLBigxxxA}{\ensuremath{0.9096}}                      
\newcommand{\hatcurLBiigxxxA}{\ensuremath{-0.0641}}                    
\newcommand{\hatcurLBikepxxxA}{\ensuremath{0.1000}}                    
\newcommand{\hatcurLBiikepxxxA}{\ensuremath{0.1000}}                   
\newcommand{\hatcurISOmxxxA}{\ensuremath{0.76\pm0.03}}                 
\newcommand{\hatcurISOmshortxxxA}{\ensuremath{0.76}}                   
\newcommand{\hatcurISOmlongxxxA}{\ensuremath{0.756\pm0.028}}           
\newcommand{\hatcurISOrxxxA}{\ensuremath{0.69\pm0.02}}                 
\newcommand{\hatcurISOrshortxxxA}{\ensuremath{0.69}}                   
\newcommand{\hatcurISOrlongxxxA}{\ensuremath{0.694\pm0.021}}           
\newcommand{\hatcurISOrhoxxxA}{\ensuremath{3.19\pm0.25}}               
\newcommand{\hatcurISOloggxxxA}{\ensuremath{4.63\pm0.02}}              
\newcommand{\hatcurISOlumxxxA}{\ensuremath{0.19\pm0.02}}               
\newcommand{\hatcurISOlumshortxxxA}{\ensuremath{0.19}}                 
\newcommand{\hatcurISOmvxxxA}{\ensuremath{7.07\pm0.17}}                
\newcommand{\hatcurISOvixxxA}{\ensuremath{1.155\pm0.054}}              
\newcommand{\hatcurISOagexxxA}{\ensuremath{6.7_{-3.8}^{+5.7}}}         
\newcommand{\hatcurISOsigmaxxxA}{\ensuremath{0.02070\pm0.00139}}       
\newcommand{\hatcurISOMJxxxA}{\ensuremath{5.08\pm0.11}}                
\newcommand{\hatcurISOMHxxxA}{\ensuremath{4.51\pm0.09}}                
\newcommand{\hatcurISOMKxxxA}{\ensuremath{4.42\pm0.09}}                
\newcommand{\hatcurISOJKxxxA}{\ensuremath{0.66\pm0.02}}                
\newcommand{\hatcurISOspecxxxA}{K3}                                    
\newcommand{\hatcurRVKxxxA}{\ensuremath{1246.0\pm8.1}}                 
\newcommand{\hatcurRVkxxxA}{\ensuremath{0.012\pm0.004}}                
\newcommand{\hatcurRVhxxxA}{\ensuremath{-0.007\pm0.008}}               
\newcommand{\hatcurRVgammaxxxA}{\ensuremath{269.2\pm7.0}}              
\newcommand{\hatcurRVjitterxxxA}{\ensuremath{15.7}}                    
\newcommand{\hatcurRVeccenxxxA}{\ensuremath{0.015\pm0.005}}            
\newcommand{\hatcurRVomegaxxxA}{\ensuremath{317\pm130}}                
\newcommand{\hatcurRVfitrmsxxxA}{\ensuremath{16.2}}                    
\newcommand{\hatcurPPixxxA}{\ensuremath{86.8\pm0.2}}                   
\newcommand{\hatcurPPgxxxA}{\ensuremath{238.8\pm17.0}}                 
\newcommand{\hatcurPPloggxxxA}{\ensuremath{4.38\pm0.03}}               
\newcommand{\hatcurPParxxxA}{\ensuremath{11.17\pm0.29}}                
\newcommand{\hatcurPParelxxxA}{\ensuremath{0.0361\pm0.0005}}           
\newcommand{\hatcurPPrhoxxxA}{\ensuremath{13.78\pm1.50}}               
\newcommand{\hatcurPPmxxxA}{\ensuremath{7.25\pm0.19}}                  
\newcommand{\hatcurPPmshortxxxA}{\ensuremath{7.25}}                    
\newcommand{\hatcurPPmlongxxxA}{\ensuremath{7.246\pm0.187}}            
\newcommand{\hatcurPPmexxxA}{\ensuremath{2302.9\pm59.5}}               
\newcommand{\hatcurPPmeshortxxxA}{\ensuremath{2302.9}}                 
\newcommand{\hatcurPPmelongxxxA}{\ensuremath{2302.91\pm59.49}}         
\newcommand{\hatcurPPrxxxA}{\ensuremath{0.87\pm0.03}}                  
\newcommand{\hatcurPPrshortxxxA}{\ensuremath{0.87}}                    
\newcommand{\hatcurPPrlongxxxA}{\ensuremath{0.867\pm0.033}}            
\newcommand{\hatcurPPrexxxA}{\ensuremath{9.7\pm0.4}}                   
\newcommand{\hatcurPPreshortxxxA}{\ensuremath{9.7}}                    
\newcommand{\hatcurPPrelongxxxA}{\ensuremath{9.72\pm0.37}}             
\newcommand{\hatcurPPmrcorrxxxA}{\ensuremath{0.50}}                    
\newcommand{\hatcurPPteffxxxA}{\ensuremath{970\pm23}}                  
\newcommand{\hatcurPPthetaxxxA}{\ensuremath{0.794\pm0.031}}            
\newcommand{\hatcurPPfluxperixxxA}{\ensuremath{2.06\pm0.20}}           
\newcommand{\hatcurPPfluxperidimxxxA}{\ensuremath{8}}                  
\newcommand{\hatcurPPfluxapxxxA}{\ensuremath{1.94\pm0.19}}             
\newcommand{\hatcurPPfluxapdimxxxA}{\ensuremath{8}}                    
\newcommand{\hatcurPPfluxavgxxxA}{\ensuremath{2.00\pm0.19}}            
\newcommand{\hatcurPPfluxavgdimxxxA}{\ensuremath{8}}                   
\newcommand{\hatcurXsecphasexxxA}{\ensuremath{0.5073\pm0.0024}}        
\newcommand{\hatcurXsecondaryxxxA}{\ensuremath{2455082.385\pm0.007}}   
\newcommand{\hatcurXsecdurxxxA}{\ensuremath{0.0764\pm0.0010}}          
\newcommand{\hatcurXsecingdurxxxA}{\ensuremath{0.0134\pm0.0009}}       
\newcommand{\hatcurPPphiconjxxxA}{\ensuremath{0.3314_{-0.1029}^{+0.0566}}} 
\newcommand{\hatcurPPperixxxA}{\ensuremath{2455079.97\pm0.24}}         
\newcommand{\hatcurPPaequivxxxA}{\ensuremath{0.0826\pm0.0040}}         
\newcommand{\hatcurPPtcircxxxA}{\ensuremath{3618.5\pm657.4}}           
\newcommand{\hatcurPPtinfallxxxA}{\ensuremath{495.9\pm65.0}}           
\newcommand{\hatcurXdistxxxA}{\ensuremath{70\pm3}}                     
\newcommand{\hatcurCCpmraxxxA}{\ensuremath{-8.00\pm2.0}}               
\newcommand{\hatcurCCpmdecxxxA}{\ensuremath{-96.0\pm8.0}}              
\newcommand{\hatcurCCpmxxxA}{\ensuremath{96.3328\pm8.24621}}           

%% file: HTR183-008.tex
\newcommand{\hatcurhtrxxxB}{HTR183-008}                                
\newcommand{\hatcurfieldxxxB}{183}                                     
\newcommand{\hatcurCCraxxxB}{\ensuremath{11^{\mathrm h}25^{\mathrm m}05.88^{\mathrm s}}}                              
\newcommand{\hatcurCCdecxxxB}{\ensuremath{+41{\arcdeg}01{\arcmin}40.6{\arcsec}}}                             
\newcommand{\hatcurCCmagxxxB}{11.685}                                  
\newcommand{\hatcurCCtwomassxxxB}{2MASS~11250598+4101406}              
\newcommand{\hatcurCCgscxxxB}{GSC~3013-01229}                          
\newcommand{\hatcurCCtassmvxxxB}{11.685}                               
\newcommand{\hatcurCCtwomassJmagxxxB}{\ensuremath{10.503\pm0.022}}     
\newcommand{\hatcurCCtwomassHmagxxxB}{\ensuremath{10.154\pm0.019}}     
\newcommand{\hatcurCCtwomassKmagxxxB}{\ensuremath{10.111\pm0.018}}     
\newcommand{\hatcurCCcitJmagxxxB}{\ensuremath{10.519\pm0.022}}         
\newcommand{\hatcurCCcitHmagxxxB}{\ensuremath{10.149\pm0.020}}         
\newcommand{\hatcurCCcitKmagxxxB}{\ensuremath{10.135\pm0.018}}         
\newcommand{\hatcurCCbbJmagxxxB}{\ensuremath{10.570\pm0.024}}          
\newcommand{\hatcurCCbbHmagxxxB}{\ensuremath{10.170\pm0.020}}          
\newcommand{\hatcurCCbbKmagxxxB}{\ensuremath{10.155\pm0.018}}          
\newcommand{\hatcurCCesoJmagxxxB}{\ensuremath{10.572\pm0.026}}         
\newcommand{\hatcurCCesoHmagxxxB}{\ensuremath{10.164\pm0.023}}         
\newcommand{\hatcurCCesoKmagxxxB}{\ensuremath{10.154\pm0.019}}         
\newcommand{\hatcurCCesoJHmagxxxB}{\ensuremath{0.408\pm0.032}}         
\newcommand{\hatcurCCesoJKmagxxxB}{\ensuremath{0.419\pm0.031}}         
\newcommand{\hatcurCCesoHKmagxxxB}{\ensuremath{0.010\pm0.029}}         
\newcommand{\hatcurLCdipxxxB}{\ensuremath{8.0}}                        
\newcommand{\hatcurLCrprstarxxxB}{\ensuremath{0.0950\pm0.0022}}        
\newcommand{\hatcurLCbsqxxxB}{\ensuremath{0.298_{-0.118}^{+0.089}}}    
\newcommand{\hatcurLCimpxxxB}{\ensuremath{0.546_{-0.139}^{+0.074}}}    
\newcommand{\hatcurLCzetaxxxB}{\ensuremath{14.81\pm0.15}}              
\newcommand{\hatcurLCdurxxxB}{\ensuremath{0.1530\pm0.0027}}            
\newcommand{\hatcurLCdurshortxxxB}{\ensuremath{0.1530}}                
\newcommand{\hatcurLCdurhrxxxB}{\ensuremath{3.672\pm0.065}}            
\newcommand{\hatcurLCdurhrshortxxxB}{\ensuremath{3.672}}               
\newcommand{\hatcurLCqxxxB}{\ensuremath{0.0371\pm0.0007}}              
\newcommand{\hatcurLCqshortxxxB}{\ensuremath{0.037}}                   
\newcommand{\hatcurLCingdurxxxB}{\ensuremath{0.0184\pm0.0029}}         
\newcommand{\hatcurLCPxxxB}{\ensuremath{4.124481\pm0.000007}}          
\newcommand{\hatcurLCPprecxxxB}{\ensuremath{4.1244808}}                
\newcommand{\hatcurLCPshortxxxB}{\ensuremath{4.1245}}                  
\newcommand{\hatcurLCTxxxB}{\ensuremath{2454996.41312\pm0.00069}}      
\newcommand{\hatcurLCTAxxxB}{\ensuremath{2454064.28044\pm0.00167}}     
\newcommand{\hatcurLCTBxxxB}{\ensuremath{2455248.00646\pm0.00087}}     
\newcommand{\hatcurLChatnetmAxxxB}{\ensuremath{11.3326\pm0.0001}}      
\newcommand{\hatcurLCiblendAxxxB}{\ensuremath{0.81\pm0.06}}            
\newcommand{\hatcurLChatnetmBxxxB}{\ensuremath{10.9964\pm0.0001}}      
\newcommand{\hatcurLCiblendBxxxB}{\ensuremath{0.95\pm0.04}}            
\newcommand{\hatcurLChatnetmCxxxB}{\ensuremath{10.9972\pm0.0001}}      
\newcommand{\hatcurLCiblendCxxxB}{\ensuremath{0.87\pm0.08}}            
\newcommand{\hatcurSMEiteffxxxB}{\ensuremath{5701\pm144}}              
\newcommand{\hatcurSMEizfehxxxB}{\ensuremath{+0.04\pm0.1}}             
\newcommand{\hatcurSMEizfehshortxxxB}{\ensuremath{+0.04}}              
\newcommand{\hatcurSMEiloggxxxB}{\ensuremath{4.48\pm0.14}}             
\newcommand{\hatcurSMEivsinxxxB}{\ensuremath{4.1\pm0.5}}               
\newcommand{\hatcurSMEivmacxxxB}{\ensuremath{3.91}}                    
\newcommand{\hatcurSMEivmicxxxB}{\ensuremath{0.85}}                    
\newcommand{\hatcurSMEiiteffxxxB}{\ensuremath{5588\pm80}}              
\newcommand{\hatcurSMEiizfehxxxB}{\ensuremath{+0.01\pm0.08}}           
\newcommand{\hatcurSMEiizfehshortxxxB}{\ensuremath{+0.01}}             
\newcommand{\hatcurSMEiiloggxxxB}{\ensuremath{4.31\pm0.06}}            
\newcommand{\hatcurSMEiivsinxxxB}{\ensuremath{3.5\pm0.5}}              
\newcommand{\hatcurSMEiivmacxxxB}{\ensuremath{3.74}}                   
\newcommand{\hatcurSMEiivmicxxxB}{\ensuremath{0.85}}                   
\newcommand{\hatcurDSteffxxxB}{\ensuremath{5750\pm125}}                
\newcommand{\hatcurDSzfehxxxB}{\ensuremath{0.0\pm0.0}}                 
\newcommand{\hatcurDSloggxxxB}{\ensuremath{4.5\pm0.25}}                
\newcommand{\hatcurDSvsinixxxB}{\ensuremath{3.0\pm3.0}}                
\newcommand{\hatcurDSgammaxxxB}{\ensuremath{-53.19\pm0.09}}            
\newcommand{\hatcurDSnumspecxxxB}{\ensuremath{3}}                      
\newcommand{\hatcurDSspanxxxB}{\ensuremath{28}}                        
\newcommand{\hatcurDSrvrmsxxxB}{\ensuremath{0.16}}                     
\newcommand{\hatcurTRESteffxxxB}{\ensuremath{}}                        
\newcommand{\hatcurTRESzfehxxxB}{\ensuremath{}}                        
\newcommand{\hatcurTRESloggxxxB}{\ensuremath{}}                        
\newcommand{\hatcurTRESvsinixxxB}{\ensuremath{}}                       
\newcommand{\hatcurTRESgammaxxxB}{\ensuremath{}}                       
\newcommand{\hatcurTRESnumspecxxxB}{\ensuremath{}}             
\newcommand{\hatcurTRESspanxxxB}{\ensuremath{}}                
\newcommand{\hatcurTRESrvrmsxxxB}{\ensuremath{}}               
\newcommand{\hatcurFIESteffxxxB}{\ensuremath{}}                        
\newcommand{\hatcurFIESzfehxxxB}{\ensuremath{}}                        
\newcommand{\hatcurFIESloggxxxB}{\ensuremath{}}                        
\newcommand{\hatcurFIESvsinixxxB}{\ensuremath{}}                       
\newcommand{\hatcurFIESgammaxxxB}{\ensuremath{}}                       
\newcommand{\hatcurFIESnumspecxxxB}{\ensuremath{}}             
\newcommand{\hatcurFIESspanxxxB}{\ensuremath{}}                
\newcommand{\hatcurFIESrvrmsxxxB}{\ensuremath{}}               
\newcommand{\hatcurLBizxxxB}{\ensuremath{0.2331}}                      
\newcommand{\hatcurLBiizxxxB}{\ensuremath{0.3156}}                     
\newcommand{\hatcurLBiixxxB}{\ensuremath{0.2976}}                      
\newcommand{\hatcurLBiiixxxB}{\ensuremath{0.3131}}                     
\newcommand{\hatcurLBiIxxxB}{\ensuremath{0.2760}}                      
\newcommand{\hatcurLBiiIxxxB}{\ensuremath{0.3146}}                     
\newcommand{\hatcurLBigxxxB}{\ensuremath{0.5983}}                      
\newcommand{\hatcurLBiigxxxB}{\ensuremath{0.2016}}                     
\newcommand{\hatcurLBikepxxxB}{\ensuremath{0.1000}}                    
\newcommand{\hatcurLBiikepxxxB}{\ensuremath{0.1000}}                   
\newcommand{\hatcurISOmxxxB}{\ensuremath{0.95\pm0.04}}                 
\newcommand{\hatcurISOmshortxxxB}{\ensuremath{0.95}}                   
\newcommand{\hatcurISOmlongxxxB}{\ensuremath{0.947\pm0.042}}           
\newcommand{\hatcurISOrxxxB}{\ensuremath{1.10\pm0.08}}                 
\newcommand{\hatcurISOrshortxxxB}{\ensuremath{1.10}}                   
\newcommand{\hatcurISOrlongxxxB}{\ensuremath{1.105\pm0.083}}           
\newcommand{\hatcurISOrhoxxxB}{\ensuremath{0.98_{-0.18}^{+0.27}}}      
\newcommand{\hatcurISOloggxxxB}{\ensuremath{4.33\pm0.06}}              
\newcommand{\hatcurISOlumxxxB}{\ensuremath{1.06_{-0.16}^{+0.20}}}      
\newcommand{\hatcurISOlumshortxxxB}{\ensuremath{1.06}}                 
\newcommand{\hatcurISOmvxxxB}{\ensuremath{4.80\pm0.19}}                
\newcommand{\hatcurISOvixxxB}{\ensuremath{0.756\pm0.024}}              
\newcommand{\hatcurISOagexxxB}{\ensuremath{10.2\pm2.5}}                
\newcommand{\hatcurISOsigmaxxxB}{\ensuremath{0.00450\pm0.00066}}       
\newcommand{\hatcurISOMJxxxB}{\ensuremath{3.56\pm0.17}}                
\newcommand{\hatcurISOMHxxxB}{\ensuremath{3.19\pm0.17}}                
\newcommand{\hatcurISOMKxxxB}{\ensuremath{3.12\pm0.16}}                
\newcommand{\hatcurISOJKxxxB}{\ensuremath{0.44\pm0.02}}                
\newcommand{\hatcurISOspecxxxB}{G3}                                    
\newcommand{\hatcurRVKxxxB}{\ensuremath{548.3\pm14.2}}                 
\newcommand{\hatcurRVkxxxB}{\ensuremath{0.147\pm0.011}}                
\newcommand{\hatcurRVhxxxB}{\ensuremath{-0.175\pm0.019}}               
\newcommand{\hatcurRVgammaxxxB}{\ensuremath{51.5\pm8.1}}               
\newcommand{\hatcurRVjitterxxxB}{\ensuremath{26.4}}                    
\newcommand{\hatcurRVeccenxxxB}{\ensuremath{0.228\pm0.016}}            
\newcommand{\hatcurRVomegaxxxB}{\ensuremath{309\pm3}}                  
\newcommand{\hatcurRVfitrmsxxxB}{\ensuremath{26.6}}                    
\newcommand{\hatcurPPixxxB}{\ensuremath{87.2\pm0.7}}                   
\newcommand{\hatcurPPgxxxB}{\ensuremath{95.8_{-14.1}^{+20.2}}}         
\newcommand{\hatcurPPloggxxxB}{\ensuremath{3.98\pm0.08}}               
\newcommand{\hatcurPParxxxB}{\ensuremath{9.60\pm0.71}}                 
\newcommand{\hatcurPParelxxxB}{\ensuremath{0.0494\pm0.0007}}           
\newcommand{\hatcurPPrhoxxxB}{\ensuremath{4.68_{-0.99}^{+1.59}}}       
\newcommand{\hatcurPPmxxxB}{\ensuremath{4.06\pm0.16}}                  
\newcommand{\hatcurPPmshortxxxB}{\ensuremath{4.06}}                    
\newcommand{\hatcurPPmlongxxxB}{\ensuremath{4.063\pm0.161}}            
\newcommand{\hatcurPPmexxxB}{\ensuremath{1291.3\pm51.1}}               
\newcommand{\hatcurPPmeshortxxxB}{\ensuremath{1291.3}}                 
\newcommand{\hatcurPPmelongxxxB}{\ensuremath{1291.32\pm51.13}}         
\newcommand{\hatcurPPrxxxB}{\ensuremath{1.02\pm0.09}}                  
\newcommand{\hatcurPPrshortxxxB}{\ensuremath{1.02}}                    
\newcommand{\hatcurPPrlongxxxB}{\ensuremath{1.024\pm0.092}}            
\newcommand{\hatcurPPrexxxB}{\ensuremath{11.5\pm1.0}}                  
\newcommand{\hatcurPPreshortxxxB}{\ensuremath{11.5}}                   
\newcommand{\hatcurPPrelongxxxB}{\ensuremath{11.47\pm1.03}}            
\newcommand{\hatcurPPmrcorrxxxB}{\ensuremath{0.28}}                    
\newcommand{\hatcurPPteffxxxB}{\ensuremath{1283\pm50}}                 
\newcommand{\hatcurPPthetaxxxB}{\ensuremath{0.413\pm0.038}}            
\newcommand{\hatcurPPfluxperixxxB}{\ensuremath{10.0\pm1.5}}           
\newcommand{\hatcurPPfluxperidimxxxB}{\ensuremath{8}}                  
\newcommand{\hatcurPPfluxapxxxB}{\ensuremath{3.96\pm0.66}}             
\newcommand{\hatcurPPfluxapdimxxxB}{\ensuremath{8}}                    
\newcommand{\hatcurPPfluxavgxxxB}{\ensuremath{6.12\pm0.97}}            
\newcommand{\hatcurPPfluxavgdimxxxB}{\ensuremath{8}}                   
\newcommand{\hatcurXsecphasexxxB}{\ensuremath{0.5944\pm0.0071}}        
\newcommand{\hatcurXsecondaryxxxB}{\ensuremath{2454998.865\pm0.029}}   
\newcommand{\hatcurXsecdurxxxB}{\ensuremath{0.1163\pm0.0068}}          
\newcommand{\hatcurXsecingdurxxxB}{\ensuremath{0.0117\pm0.0015}}       
\newcommand{\hatcurPPphiconjxxxB}{\ensuremath{0.3353\pm0.0124}}        
\newcommand{\hatcurPPperixxxB}{\ensuremath{2454995.03\pm0.05}}         
\newcommand{\hatcurPPaequivxxxB}{\ensuremath{0.0478\pm0.0038}}         
\newcommand{\hatcurPPtcircxxxB}{\ensuremath{3448.7_{-1073.0}^{+2239.1}}} 
\newcommand{\hatcurPPtinfallxxxB}{\ensuremath{744.2_{-206.3}^{+388.2}}} 
\newcommand{\hatcurXdistxxxB}{\ensuremath{254\pm19}}                   
\newcommand{\hatcurCCpmraxxxB}{\ensuremath{-0.6\pm1.5}}                
\newcommand{\hatcurCCpmdecxxxB}{\ensuremath{11.9\pm1.0}}               
\newcommand{\hatcurCCpmxxxB}{\ensuremath{11.9151\pm1.80278}}           

%% file: HTR102-001.tex
\newcommand{\hatcurhtrxxxC}{HTR102-001}                                
\newcommand{\hatcurfieldxxxC}{139}                                     
\newcommand{\hatcurCCraxxxC}{\ensuremath{10^{\mathrm h}22^{\mathrm m}43.68^{\mathrm s}}}                              
\newcommand{\hatcurCCdecxxxC}{\ensuremath{+50{\arcdeg}07{\arcmin}42.0{\arcsec}}}                             
\newcommand{\hatcurCCmagxxxC}{9.732}                                   
\newcommand{\hatcurCCtwomassxxxC}{2MASS~10224361+5007420}              
\newcommand{\hatcurCCgscxxxC}{GSC~03441-00925}                         
\newcommand{\hatcurCCtassmvxxxC}{9.732}                                
\newcommand{\hatcurCCtwomassJmagxxxC}{\ensuremath{8.293\pm0.023}}      
\newcommand{\hatcurCCtwomassHmagxxxC}{\ensuremath{7.935\pm0.029}}      
\newcommand{\hatcurCCtwomassKmagxxxC}{\ensuremath{7.837\pm0.021}}      
\newcommand{\hatcurCCcitJmagxxxC}{\ensuremath{8.305\pm0.023}}          
\newcommand{\hatcurCCcitHmagxxxC}{\ensuremath{7.929\pm0.029}}          
\newcommand{\hatcurCCcitKmagxxxC}{\ensuremath{7.861\pm0.021}}          
\newcommand{\hatcurCCbbJmagxxxC}{\ensuremath{8.362\pm0.025}}           
\newcommand{\hatcurCCbbHmagxxxC}{\ensuremath{7.951\pm0.030}}           
\newcommand{\hatcurCCbbKmagxxxC}{\ensuremath{7.881\pm0.021}}           
\newcommand{\hatcurCCesoJmagxxxC}{\ensuremath{8.365\pm0.027}}          
\newcommand{\hatcurCCesoHmagxxxC}{\ensuremath{7.947\pm0.034}}          
\newcommand{\hatcurCCesoKmagxxxC}{\ensuremath{7.880\pm0.022}}          
\newcommand{\hatcurCCesoJHmagxxxC}{\ensuremath{0.417\pm0.041}}         
\newcommand{\hatcurCCesoJKmagxxxC}{\ensuremath{0.486\pm0.034}}         
\newcommand{\hatcurCCesoHKmagxxxC}{\ensuremath{0.068\pm0.041}}         
\newcommand{\hatcurLCdipxxxC}{\ensuremath{9.7}}                        
\newcommand{\hatcurLCrprstarxxxC}{\ensuremath{0.1065\pm0.0017}}        
\newcommand{\hatcurLCbsqxxxC}{\ensuremath{0.217_{-0.065}^{+0.052}}}    
\newcommand{\hatcurLCimpxxxC}{\ensuremath{0.466_{-0.083}^{+0.052}}}    
\newcommand{\hatcurLCzetaxxxC}{\ensuremath{18.97\pm0.09}}              
\newcommand{\hatcurLCdurxxxC}{\ensuremath{0.1196\pm0.0014}}            
\newcommand{\hatcurLCdurshortxxxC}{\ensuremath{0.1196}}                
\newcommand{\hatcurLCdurhrxxxC}{\ensuremath{2.869\pm0.033}}            
\newcommand{\hatcurLCdurhrshortxxxC}{\ensuremath{2.869}}               
\newcommand{\hatcurLCqxxxC}{\ensuremath{0.0372\pm0.0004}}              
\newcommand{\hatcurLCqshortxxxC}{\ensuremath{0.037}}                   
\newcommand{\hatcurLCingdurxxxC}{\ensuremath{0.0144\pm0.0013}}         
\newcommand{\hatcurLCPxxxC}{\ensuremath{3.212220\pm0.000009}}          
\newcommand{\hatcurLCPprecxxxC}{\ensuremath{3.2122198}}                
\newcommand{\hatcurLCPshortxxxC}{\ensuremath{3.2122}}                  
\newcommand{\hatcurLCTxxxC}{\ensuremath{2454930.22001\pm0.00025}}      
\newcommand{\hatcurLCTAxxxC}{\ensuremath{2454458.02370\pm0.00125}}     
\newcommand{\hatcurLCTBxxxC}{\ensuremath{2454952.70554\pm0.00028}}     
\newcommand{\hatcurLChatnetmxxxC}{\ensuremath{9.2791\pm0.0001}}        
\newcommand{\hatcurLCiblendxxxC}{\ensuremath{0.80\pm0.04}}             
\newcommand{\hatcurSMEiteffxxxC}{\ensuremath{5338\pm88}}               
\newcommand{\hatcurSMEizfehxxxC}{\ensuremath{+0.26\pm0.08}}            
\newcommand{\hatcurSMEizfehshortxxxC}{\ensuremath{+0.26}}              
\newcommand{\hatcurSMEiloggxxxC}{\ensuremath{4.52\pm0.14}}             
\newcommand{\hatcurSMEivsinxxxC}{\ensuremath{1.5\pm0.5}}               
\newcommand{\hatcurSMEivmacxxxC}{\ensuremath{3.35}}                    
\newcommand{\hatcurSMEivmicxxxC}{\ensuremath{0.85}}                    
\newcommand{\hatcurSMEiiteffxxxC}{\ensuremath{5302\pm80}}              
\newcommand{\hatcurSMEiizfehxxxC}{\ensuremath{+0.24\pm0.08}}           
\newcommand{\hatcurSMEiizfehshortxxxC}{\ensuremath{+0.24}}             
\newcommand{\hatcurSMEiiloggxxxC}{\ensuremath{4.37\pm0.06}}            
\newcommand{\hatcurSMEiivsinxxxC}{\ensuremath{0.5\pm0.5}}              
\newcommand{\hatcurSMEiivmacxxxC}{\ensuremath{3.30}}                   
\newcommand{\hatcurSMEiivmicxxxC}{\ensuremath{0.85}}                   
\newcommand{\hatcurDSteffxxxC}{\ensuremath{5250\pm125}}                
\newcommand{\hatcurDSzfehxxxC}{\ensuremath{0.0\pm0.0}}                 
\newcommand{\hatcurDSloggxxxC}{\ensuremath{4.5\pm0.25}}                
\newcommand{\hatcurDSvsinixxxC}{\ensuremath{2.0\pm2.0}}                
\newcommand{\hatcurDSgammaxxxC}{\ensuremath{+12.49\pm0.28}}             
\newcommand{\hatcurDSnumspecxxxC}{\ensuremath{4}}                      
\newcommand{\hatcurDSspanxxxC}{\ensuremath{5}}                         
\newcommand{\hatcurDSrvrmsxxxC}{\ensuremath{0.56}}                     
\newcommand{\hatcurTRESteffxxxC}{\ensuremath{}}                        
\newcommand{\hatcurTRESzfehxxxC}{\ensuremath{}}                        
\newcommand{\hatcurTRESloggxxxC}{\ensuremath{}}                        
\newcommand{\hatcurTRESvsinixxxC}{\ensuremath{}}                       
\newcommand{\hatcurTRESgammaxxxC}{\ensuremath{}}                       
\newcommand{\hatcurTRESnumspecxxxC}{\ensuremath{}}             
\newcommand{\hatcurTRESspanxxxC}{\ensuremath{}}                
\newcommand{\hatcurTRESrvrmsxxxC}{\ensuremath{}}               
\newcommand{\hatcurFIESteffxxxC}{\ensuremath{}}                        
\newcommand{\hatcurFIESzfehxxxC}{\ensuremath{}}                        
\newcommand{\hatcurFIESloggxxxC}{\ensuremath{}}                        
\newcommand{\hatcurFIESvsinixxxC}{\ensuremath{}}                       
\newcommand{\hatcurFIESgammaxxxC}{\ensuremath{}}                       
\newcommand{\hatcurFIESnumspecxxxC}{\ensuremath{}}             
\newcommand{\hatcurFIESspanxxxC}{\ensuremath{}}                
\newcommand{\hatcurFIESrvrmsxxxC}{\ensuremath{}}               
\newcommand{\hatcurLBizxxxC}{\ensuremath{0.2772}}                      
\newcommand{\hatcurLBiizxxxC}{\ensuremath{0.2989}}                     
\newcommand{\hatcurLBiixxxC}{\ensuremath{0.3587}}                      
\newcommand{\hatcurLBiiixxxC}{\ensuremath{0.2831}}                     
\newcommand{\hatcurLBiIxxxC}{\ensuremath{0.3320}}                      
\newcommand{\hatcurLBiiIxxxC}{\ensuremath{0.2883}}                     
\newcommand{\hatcurLBigxxxC}{\ensuremath{0.7080}}                      
\newcommand{\hatcurLBiigxxxC}{\ensuremath{0.1165}}                     
\newcommand{\hatcurLBikepxxxC}{\ensuremath{0.1000}}                    
\newcommand{\hatcurLBiikepxxxC}{\ensuremath{0.1000}}                   
\newcommand{\hatcurISOmxxxC}{\ensuremath{0.92\pm0.03}}                 
\newcommand{\hatcurISOmshortxxxC}{\ensuremath{0.92}}                   
\newcommand{\hatcurISOmlongxxxC}{\ensuremath{0.916\pm0.035}}           
\newcommand{\hatcurISOrxxxC}{\ensuremath{1.04\pm0.04}}                 
\newcommand{\hatcurISOrshortxxxC}{\ensuremath{1.04}}                   
\newcommand{\hatcurISOrlongxxxC}{\ensuremath{1.040\pm0.044}}           
\newcommand{\hatcurISOrhoxxxC}{\ensuremath{1.14\pm0.14}}               
\newcommand{\hatcurISOloggxxxC}{\ensuremath{4.36\pm0.04}}              
\newcommand{\hatcurISOlumxxxC}{\ensuremath{0.77\pm0.09}}               
\newcommand{\hatcurISOlumshortxxxC}{\ensuremath{0.77}}                 
\newcommand{\hatcurISOmvxxxC}{\ensuremath{5.22\pm0.14}}                
\newcommand{\hatcurISOvixxxC}{\ensuremath{0.840\pm0.019}}              
\newcommand{\hatcurISOagexxxC}{\ensuremath{12.4\pm2.6}}                
\newcommand{\hatcurISOsigmaxxxC}{\ensuremath{0.00270\pm0.00022}}       
\newcommand{\hatcurISOMJxxxC}{\ensuremath{3.81\pm0.11}}                
\newcommand{\hatcurISOMHxxxC}{\ensuremath{3.38\pm0.10}}                
\newcommand{\hatcurISOMKxxxC}{\ensuremath{3.30\pm0.10}}                
\newcommand{\hatcurISOJKxxxC}{\ensuremath{0.50\pm0.02}}                
\newcommand{\hatcurISOspecxxxC}{G5}                                    
\newcommand{\hatcurRVKxxxC}{\ensuremath{313.3\pm4.2}}                  
\newcommand{\hatcurRVkxxxC}{\ensuremath{-0.008\pm0.008}}               
\newcommand{\hatcurRVhxxxC}{\ensuremath{0.004\pm0.016}}                
\newcommand{\hatcurRVgammaxxxC}{\ensuremath{5.3\pm3.1}}                
\newcommand{\hatcurRVjitterxxxC}{\ensuremath{9.9}}                     
\newcommand{\hatcurRVeccenxxxC}{\ensuremath{0.016\pm0.009}}            
\newcommand{\hatcurRVomegaxxxC}{\ensuremath{156\pm66}}                 
\newcommand{\hatcurRVfitrmsxxxC}{\ensuremath{10.0}}                    
\newcommand{\hatcurPPixxxC}{\ensuremath{86.9_{-0.5}^{+0.6}}}           
\newcommand{\hatcurPPgxxxC}{\ensuremath{45.6\pm4.9}}                   
\newcommand{\hatcurPPloggxxxC}{\ensuremath{3.66\pm0.05}}               
\newcommand{\hatcurPParxxxC}{\ensuremath{8.55\pm0.35}}                 
\newcommand{\hatcurPParelxxxC}{\ensuremath{0.0414\pm0.0005}}           
\newcommand{\hatcurPPrhoxxxC}{\ensuremath{2.11_{-0.29}^{+0.40}}}       
\newcommand{\hatcurPPmxxxC}{\ensuremath{2.15\pm0.06}}                  
\newcommand{\hatcurPPmshortxxxC}{\ensuremath{2.15}}                    
\newcommand{\hatcurPPmlongxxxC}{\ensuremath{2.147\pm0.061}}            
\newcommand{\hatcurPPmexxxC}{\ensuremath{682.4\pm19.4}}                
\newcommand{\hatcurPPmeshortxxxC}{\ensuremath{682.4}}                  
\newcommand{\hatcurPPmelongxxxC}{\ensuremath{682.44\pm19.36}}          
\newcommand{\hatcurPPrxxxC}{\ensuremath{1.08\pm0.06}}                  
\newcommand{\hatcurPPrshortxxxC}{\ensuremath{1.08}}                    
\newcommand{\hatcurPPrlongxxxC}{\ensuremath{1.080\pm0.058}}            
\newcommand{\hatcurPPrexxxC}{\ensuremath{12.1\pm0.7}}                  
\newcommand{\hatcurPPreshortxxxC}{\ensuremath{12.1}}                   
\newcommand{\hatcurPPrelongxxxC}{\ensuremath{12.10\pm0.66}}            
\newcommand{\hatcurPPmrcorrxxxC}{\ensuremath{0.36}}                    
\newcommand{\hatcurPPteffxxxC}{\ensuremath{1283\pm32}}                 
\newcommand{\hatcurPPthetaxxxC}{\ensuremath{0.179\pm0.010}}            
\newcommand{\hatcurPPfluxperixxxC}{\ensuremath{6.33\pm0.67}}           
\newcommand{\hatcurPPfluxperidimxxxC}{\ensuremath{8}}                  
\newcommand{\hatcurPPfluxapxxxC}{\ensuremath{5.91\pm0.60}}             
\newcommand{\hatcurPPfluxapdimxxxC}{\ensuremath{8}}                    
\newcommand{\hatcurPPfluxavgxxxC}{\ensuremath{6.12\pm0.62}}            
\newcommand{\hatcurPPfluxavgdimxxxC}{\ensuremath{8}}                   
\newcommand{\hatcurXsecphasexxxC}{\ensuremath{0.4946\pm0.0049}}        
\newcommand{\hatcurXsecondaryxxxC}{\ensuremath{2454931.809\pm0.016}}   
\newcommand{\hatcurXsecdurxxxC}{\ensuremath{0.1204\pm0.0032}}          
\newcommand{\hatcurXsecingdurxxxC}{\ensuremath{0.0145\pm0.0014}}       
\newcommand{\hatcurPPphiconjxxxC}{\ensuremath{-0.1431\pm0.2218}}       
\newcommand{\hatcurPPperixxxC}{\ensuremath{2454930.68\pm0.71}}         
\newcommand{\hatcurPPaequivxxxC}{\ensuremath{0.0473\pm0.0024}}         
\newcommand{\hatcurPPtcircxxxC}{\ensuremath{656.9_{-139.0}^{+228.2}}}  
\newcommand{\hatcurPPtinfallxxxC}{\ensuremath{592.9_{-102.3}^{+153.4}}} 
\newcommand{\hatcurXdistxxxC}{\ensuremath{82\pm3}}                     
\newcommand{\hatcurCCpmraxxxC}{\ensuremath{-26.6\pm1.3}}               
\newcommand{\hatcurCCpmdecxxxC}{\ensuremath{81.9\pm1.3}}               
\newcommand{\hatcurCCpmxxxC}{\ensuremath{86.1114\pm1.83848}}           

%% file: HTR341-004.tex
\newcommand{\hatcurhtrxxxD}{HTR341-004}                                
\newcommand{\hatcurfieldxxxD}{341}                                     
\newcommand{\hatcurCCraxxxD}{\ensuremath{20^{\mathrm h}24^{\mathrm m}29.88^{\mathrm s}}}                              
\newcommand{\hatcurCCdecxxxD}{\ensuremath{+16{\arcdeg}45{\arcmin}43.7{\arcsec}}}                             
\newcommand{\hatcurCCmagxxxD}{12.432}                                  
\newcommand{\hatcurCCtwomassxxxD}{2MASS~20242972+1645437}              
\newcommand{\hatcurCCgscxxxD}{GSC~1632-01396}                          
\newcommand{\hatcurCCtassmvxxxD}{12.432}                               
\newcommand{\hatcurCCtwomassJmagxxxD}{\ensuremath{11.103\pm0.022}}     
\newcommand{\hatcurCCtwomassHmagxxxD}{\ensuremath{10.846\pm0.022}}     
\newcommand{\hatcurCCtwomassKmagxxxD}{\ensuremath{10.791\pm0.020}}     
\newcommand{\hatcurCCcitJmagxxxD}{\ensuremath{11.122\pm0.022}}         
\newcommand{\hatcurCCcitHmagxxxD}{\ensuremath{10.841\pm0.023}}         
\newcommand{\hatcurCCcitKmagxxxD}{\ensuremath{10.815\pm0.020}}         
\newcommand{\hatcurCCbbJmagxxxD}{\ensuremath{11.167\pm0.024}}          
\newcommand{\hatcurCCbbHmagxxxD}{\ensuremath{10.862\pm0.023}}          
\newcommand{\hatcurCCbbKmagxxxD}{\ensuremath{10.835\pm0.020}}          
\newcommand{\hatcurCCesoJmagxxxD}{\ensuremath{11.169\pm0.025}}         
\newcommand{\hatcurCCesoHmagxxxD}{\ensuremath{10.856\pm0.025}}         
\newcommand{\hatcurCCesoKmagxxxD}{\ensuremath{10.834\pm0.021}}         
\newcommand{\hatcurCCesoJHmagxxxD}{\ensuremath{0.313\pm0.008}}         
\newcommand{\hatcurCCesoJKmagxxxD}{\ensuremath{0.335\pm0.032}}         
\newcommand{\hatcurCCesoHKmagxxxD}{\ensuremath{0.022\pm0.033}}         
\newcommand{\hatcurLCdipxxxD}{\ensuremath{7.5}}                       
\newcommand{\hatcurLCrprstarxxxD}{\ensuremath{0.1169\pm0.0012}}        
\newcommand{\hatcurLCbsqxxxD}{\ensuremath{0.105_{-0.049}^{+0.053}}}    
\newcommand{\hatcurLCimpxxxD}{\ensuremath{0.324_{-0.101}^{+0.070}}}    
\newcommand{\hatcurLCzetaxxxD}{\ensuremath{24.90\pm0.12}}              
\newcommand{\hatcurLCdurxxxD}{\ensuremath{0.0908\pm0.0007}}            
\newcommand{\hatcurLCdurshortxxxD}{\ensuremath{0.0908}}                
\newcommand{\hatcurLCdurhrxxxD}{\ensuremath{2.178\pm0.017}}            
\newcommand{\hatcurLCdurhrshortxxxD}{\ensuremath{2.178}}               
\newcommand{\hatcurLCqxxxD}{\ensuremath{0.0748\pm0.0006}}              
\newcommand{\hatcurLCqshortxxxD}{\ensuremath{0.075}}                   
\newcommand{\hatcurLCingdurxxxD}{\ensuremath{0.0105\pm0.0007}}         
\newcommand{\hatcurLCPxxxD}{\ensuremath{1.212884\pm0.000002}}          
\newcommand{\hatcurLCPprecxxxD}{\ensuremath{1.2128841}}                
\newcommand{\hatcurLCPshortxxxD}{\ensuremath{1.2129}}                  
\newcommand{\hatcurLCTxxxD}{\ensuremath{2454852.26464\pm0.00018}}      
\newcommand{\hatcurLCTAxxxD}{\ensuremath{2454345.27907\pm0.00058}}     
\newcommand{\hatcurLCTBxxxD}{\ensuremath{2455026.91995\pm0.00034}}     
\newcommand{\hatcurLChatnetmxxxD}{\ensuremath{11.8026\pm0.0001}}       
\newcommand{\hatcurLCiblendxxxD}{\ensuremath{0.55\pm0.03}}             
\newcommand{\hatcurSMEiteffxxxD}{\ensuremath{5987\pm120}}              
\newcommand{\hatcurSMEizfehxxxD}{\ensuremath{+0.18\pm0.1}}             
\newcommand{\hatcurSMEizfehshortxxxD}{\ensuremath{+0.18}}              
\newcommand{\hatcurSMEiloggxxxD}{\ensuremath{4.48\pm0.12}}             
\newcommand{\hatcurSMEivsinxxxD}{\ensuremath{8.0\pm0.5}}               
\newcommand{\hatcurSMEivmacxxxD}{\ensuremath{0.0}}                     
\newcommand{\hatcurSMEivmicxxxD}{\ensuremath{0.0}}                     
\newcommand{\hatcurSMEiiteffxxxD}{\ensuremath{5905\pm80}}              
\newcommand{\hatcurSMEiizfehxxxD}{\ensuremath{+0.15\pm0.04}}           
\newcommand{\hatcurSMEiizfehshortxxxD}{\ensuremath{+0.15}}             
\newcommand{\hatcurSMEiiloggxxxD}{\ensuremath{4.33\pm0.06}}            
\newcommand{\hatcurSMEiivsinxxxD}{\ensuremath{8.1\pm0.5}}              
\newcommand{\hatcurSMEiivmacxxxD}{\ensuremath{4.22}}                   
\newcommand{\hatcurSMEiivmicxxxD}{\ensuremath{0.85}}                   
\newcommand{\hatcurDSteffxxxD}{\ensuremath{6000\pm125}}                
\newcommand{\hatcurDSzfehxxxD}{\ensuremath{0.0\pm0.0}}                 
\newcommand{\hatcurDSloggxxxD}{\ensuremath{4.5\pm0.25}}                
\newcommand{\hatcurDSvsinixxxD}{\ensuremath{9.0\pm1.0}}                
\newcommand{\hatcurDSgammaxxxD}{\ensuremath{-15.10\pm0.30}}            
\newcommand{\hatcurDSnumspecxxxD}{\ensuremath{5}}                      
\newcommand{\hatcurDSspanxxxD}{\ensuremath{118}}                       
\newcommand{\hatcurDSrvrmsxxxD}{\ensuremath{0.68}}                     
\newcommand{\hatcurTRESteffxxxD}{\ensuremath{}}                        
\newcommand{\hatcurTRESzfehxxxD}{\ensuremath{}}                        
\newcommand{\hatcurTRESloggxxxD}{\ensuremath{}}                        
\newcommand{\hatcurTRESvsinixxxD}{\ensuremath{}}                       
\newcommand{\hatcurTRESgammaxxxD}{\ensuremath{}}                       
\newcommand{\hatcurTRESnumspecxxxD}{\ensuremath{}}             
\newcommand{\hatcurTRESspanxxxD}{\ensuremath{}}                
\newcommand{\hatcurTRESrvrmsxxxD}{\ensuremath{}}               
\newcommand{\hatcurFIESteffxxxD}{\ensuremath{}}                        
\newcommand{\hatcurFIESzfehxxxD}{\ensuremath{}}                        
\newcommand{\hatcurFIESloggxxxD}{\ensuremath{}}                        
\newcommand{\hatcurFIESvsinixxxD}{\ensuremath{}}                       
\newcommand{\hatcurFIESgammaxxxD}{\ensuremath{}}                       
\newcommand{\hatcurFIESnumspecxxxD}{\ensuremath{}}             
\newcommand{\hatcurFIESspanxxxD}{\ensuremath{}}                
\newcommand{\hatcurFIESrvrmsxxxD}{\ensuremath{}}               
\newcommand{\hatcurLBizxxxD}{\ensuremath{0.1926}}                      
\newcommand{\hatcurLBiizxxxD}{\ensuremath{0.3391}}                     
\newcommand{\hatcurLBiixxxD}{\ensuremath{0.2524}}                      
\newcommand{\hatcurLBiiixxxD}{\ensuremath{0.3426}}                     
\newcommand{\hatcurLBiIxxxD}{\ensuremath{0.2319}}                      
\newcommand{\hatcurLBiiIxxxD}{\ensuremath{0.3424}}                     
\newcommand{\hatcurLBigxxxD}{\ensuremath{0.5311}}                      
\newcommand{\hatcurLBiigxxxD}{\ensuremath{0.2539}}                     
\newcommand{\hatcurLBikepxxxD}{\ensuremath{0.1000}}                    
\newcommand{\hatcurLBiikepxxxD}{\ensuremath{0.1000}}                   
\newcommand{\hatcurISOmxxxD}{\ensuremath{1.13\pm0.04}}                 
\newcommand{\hatcurISOmshortxxxD}{\ensuremath{1.13}}                   
\newcommand{\hatcurISOmlongxxxD}{\ensuremath{1.130\pm0.035}}           
\newcommand{\hatcurISOrxxxD}{\ensuremath{1.20\pm0.07}}                 
\newcommand{\hatcurISOrshortxxxD}{\ensuremath{1.20}}                   
\newcommand{\hatcurISOrlongxxxD}{\ensuremath{1.203\pm0.074}}           
\newcommand{\hatcurISOrhoxxxD}{\ensuremath{0.91\pm0.15}}               
\newcommand{\hatcurISOloggxxxD}{\ensuremath{4.33\pm0.05}}              
\newcommand{\hatcurISOlumxxxD}{\ensuremath{1.58\pm0.23}}               
\newcommand{\hatcurISOlumshortxxxD}{\ensuremath{1.58}}                 
\newcommand{\hatcurISOmvxxxD}{\ensuremath{4.31\pm0.16}}                
\newcommand{\hatcurISOvixxxD}{\ensuremath{0.660\pm0.024}}              
\newcommand{\hatcurISOagexxxD}{\ensuremath{4.0\pm1.0}}                 
\newcommand{\hatcurISOsigmaxxxD}{\ensuremath{0.00190\pm0.00021}}       
\newcommand{\hatcurISOMJxxxD}{\ensuremath{3.23\pm0.14}}                
\newcommand{\hatcurISOMHxxxD}{\ensuremath{2.91\pm0.14}}                
\newcommand{\hatcurISOMKxxxD}{\ensuremath{2.86\pm0.14}}                
\newcommand{\hatcurISOJKxxxD}{\ensuremath{0.37\pm0.01}}                
\newcommand{\hatcurISOspecxxxD}{G0}                                    
\newcommand{\hatcurRVKxxxD}{\ensuremath{368.5\pm17.6}}                 
\newcommand{\hatcurRVkxxxD}{\ensuremath{-0.048\pm0.023}}               
\newcommand{\hatcurRVhxxxD}{\ensuremath{0.090\pm0.052}}                
\newcommand{\hatcurRVgammaxxxD}{\ensuremath{83.3\pm12.3}}              
\newcommand{\hatcurRVjitterxxxD}{\ensuremath{34.7}}                    
\newcommand{\hatcurRVeccenxxxD}{\ensuremath{0.106\pm0.044}}            
\newcommand{\hatcurRVomegaxxxD}{\ensuremath{118\pm25}}                 
\newcommand{\hatcurRVfitrmsxxxD}{\ensuremath{35.1}}                    
\newcommand{\hatcurPPixxxD}{\ensuremath{85.1\pm1.5}}                   
\newcommand{\hatcurPPgxxxD}{\ensuremath{27.6\pm3.1}}                   
\newcommand{\hatcurPPloggxxxD}{\ensuremath{3.44\pm0.05}}               
\newcommand{\hatcurPParxxxD}{\ensuremath{4.14\pm0.23}}                 
\newcommand{\hatcurPParelxxxD}{\ensuremath{0.0232\pm0.0002}}           
\newcommand{\hatcurPPrhoxxxD}{\ensuremath{1.01\pm0.18}}                
\newcommand{\hatcurPPmxxxD}{\ensuremath{2.09\pm0.11}}                  
\newcommand{\hatcurPPmshortxxxD}{\ensuremath{2.09}}                    
\newcommand{\hatcurPPmlongxxxD}{\ensuremath{2.090\pm0.111}}            
\newcommand{\hatcurPPmexxxD}{\ensuremath{664.2\pm35.3}}                
\newcommand{\hatcurPPmeshortxxxD}{\ensuremath{664.2}}                  
\newcommand{\hatcurPPmelongxxxD}{\ensuremath{664.20\pm35.34}}          
\newcommand{\hatcurPPrxxxD}{\ensuremath{1.37\pm0.09}}                  
\newcommand{\hatcurPPrshortxxxD}{\ensuremath{1.37}}                    
\newcommand{\hatcurPPrlongxxxD}{\ensuremath{1.368\pm0.090}}            
\newcommand{\hatcurPPrexxxD}{\ensuremath{15.3\pm1.0}}                  
\newcommand{\hatcurPPreshortxxxD}{\ensuremath{15.3}}                   
\newcommand{\hatcurPPrelongxxxD}{\ensuremath{15.34\pm1.00}}            
\newcommand{\hatcurPPmrcorrxxxD}{\ensuremath{0.56}}                    
\newcommand{\hatcurPPteffxxxD}{\ensuremath{2056\pm66}}                 
\newcommand{\hatcurPPthetaxxxD}{\ensuremath{0.062\pm0.004}}            
\newcommand{\hatcurPPfluxperixxxD}{\ensuremath{50.0\pm11.4}}           
\newcommand{\hatcurPPfluxperidimxxxD}{\ensuremath{8}}                  
\newcommand{\hatcurPPfluxapxxxD}{\ensuremath{32.7\pm2.7}}             
\newcommand{\hatcurPPfluxapdimxxxD}{\ensuremath{8}}                    
\newcommand{\hatcurPPfluxavgxxxD}{\ensuremath{40.3\pm5.3}}            
\newcommand{\hatcurPPfluxavgdimxxxD}{\ensuremath{8}}                   
\newcommand{\hatcurXsecphasexxxD}{\ensuremath{0.4693\pm0.0145}}        
\newcommand{\hatcurXsecondaryxxxD}{\ensuremath{2454852.834\pm0.018}}   
\newcommand{\hatcurXsecdurxxxD}{\ensuremath{0.1064\pm0.0095}}          
\newcommand{\hatcurXsecingdurxxxD}{\ensuremath{0.0129\pm0.0020}}       
\newcommand{\hatcurPPphiconjxxxD}{\ensuremath{-0.0632_{-0.0887}^{+0.0357}}} 
\newcommand{\hatcurPPperixxxD}{\ensuremath{2454852.34\pm0.08}}         
\newcommand{\hatcurPPaequivxxxD}{\ensuremath{0.0185\pm0.0012}}         
\newcommand{\hatcurPPtcircxxxD}{\ensuremath{3.1_{-0.9}^{+1.3}}}        
\newcommand{\hatcurPPtinfallxxxD}{\ensuremath{7.5_{-1.8}^{+2.9}}}      
\newcommand{\hatcurXdistxxxD}{\ensuremath{393\pm25}}                   
\newcommand{\hatcurCCpmraxxxD}{\ensuremath{15.7\pm5.0}}                
\newcommand{\hatcurCCpmdecxxxD}{\ensuremath{0.5\pm5.1}}                
\newcommand{\hatcurCCpmxxxD}{\ensuremath{15.708\pm7.14213}}            

%% file: newcommand_spec1.tex
\newcommand{\hatcurxxxA}{HAT-P-20}
\newcommand{\hatcurbxxxA}{HAT-P-20b}
\newcommand{\hatcurcxxxA}{HAT-P-20c}

\newcommand{\hatcurplanetnumxxxA}{20}

\newcommand{\hatcurRVgammaabsxxxA}{\hatcurDSgamma{\hatcurplanetnumxxxA}}                           

\newcommand{\hatcurRVgammarelxxxA}{\hatcurRVgamma{\hatcurplanetnumxxxA}}                           

\newcommand{\hatcurCCtassvixxxA}{\ensuremath{1.50\pm0.12}}                  

\newcommand{\hatcurSMEversionxxxA}{ii}                                       

\newcommand{\hatcurisoshortxxxA}{YY}
\newcommand{\hatcurisofullxxxA}{Yonsei-Yale (YY)}
\newcommand{\hatcurisocitexxxA}{yi:2001}

\newcommand{\hatcurlumindxxxA}{\arstar}

\newcommand{\hatcurjhkfilsetxxxA}{ESO}

\newcommand{\hatcurSMEteffxxxA}{\ifthenelse{\equal{\hatcurSMEversionxxxA}{i}}{\hatcurSMEiteff{\hatcurplanetnumxxxA}}{\hatcurSMEiiteff{\hatcurplanetnumxxxA}}}
\newcommand{\hatcurSMEzfehxxxA}{\ifthenelse{\equal{\hatcurSMEversionxxxA}{i}}{\hatcurSMEizfeh{\hatcurplanetnumxxxA}}{\hatcurSMEiizfeh{\hatcurplanetnumxxxA}}}
\newcommand{\hatcurSMEzfehshortxxxA}{\ifthenelse{\equal{\hatcurSMEversionxxxA}{i}}{\hatcurSMEizfehshort{\hatcurplanetnumxxxA}}{\hatcurSMEiizfehshort{\hatcurplanetnumxxxA}}}
\newcommand{\hatcurSMEloggxxxA}{\ifthenelse{\equal{\hatcurSMEversionxxxA}{i}}{\hatcurSMEilogg{\hatcurplanetnumxxxA}}{\hatcurSMEiilogg{\hatcurplanetnumxxxA}}}
\newcommand{\hatcurSMEvsinxxxA}{\ifthenelse{\equal{\hatcurSMEversionxxxA}{i}}{\hatcurSMEivsin{\hatcurplanetnumxxxA}}{\hatcurSMEiivsin{\hatcurplanetnumxxxA}}}
\newcommand{\hatcurSMEvmacxxxA}{\ifthenelse{\equal{\hatcurSMEversionxxxA}{i}}{\hatcurSMEivmac{\hatcurplanetnumxxxA}}{\hatcurSMEiivmac{\hatcurplanetnumxxxA}}}
\newcommand{\hatcurSMEvmicxxxA}{\ifthenelse{\equal{\hatcurSMEversionxxxA}{i}}{\hatcurSMEivmic{\hatcurplanetnumxxxA}}{\hatcurSMEiivmic{\hatcurplanetnumxxxA}}}

%% file: newcommand_spec2.tex
\newcommand{\hatcurxxxB}{HAT-P-21}
\newcommand{\hatcurbxxxB}{HAT-P-21b}
\newcommand{\hatcurcxxxB}{HAT-P-21c}

\newcommand{\hatcurplanetnumxxxB}{21}

\newcommand{\hatcurRVgammaabsxxxB}{\hatcurDSgamma{\hatcurplanetnumxxxB}}                           

\newcommand{\hatcurRVgammarelxxxB}{\hatcurRVgamma{\hatcurplanetnumxxxB}}                           

\newcommand{\hatcurCCtassvixxxB}{\ensuremath{0.654\pm0.097}}                  

\newcommand{\hatcurSMEversionxxxB}{ii}                                       

\newcommand{\hatcurisoshortxxxB}{YY}
\newcommand{\hatcurisofullxxxB}{Yonsei-Yale (YY)}
\newcommand{\hatcurisocitexxxB}{yi:2001}

\newcommand{\hatcurlumindxxxB}{\arstar}

\newcommand{\hatcurjhkfilsetxxxB}{ESO}

\newcommand{\hatcurSMEteffxxxB}{\ifthenelse{\equal{\hatcurSMEversionxxxB}{i}}{\hatcurSMEiteff{\hatcurplanetnumxxxB}}{\hatcurSMEiiteff{\hatcurplanetnumxxxB}}}
\newcommand{\hatcurSMEzfehxxxB}{\ifthenelse{\equal{\hatcurSMEversionxxxB}{i}}{\hatcurSMEizfeh{\hatcurplanetnumxxxB}}{\hatcurSMEiizfeh{\hatcurplanetnumxxxB}}}
\newcommand{\hatcurSMEzfehshortxxxB}{\ifthenelse{\equal{\hatcurSMEversionxxxB}{i}}{\hatcurSMEizfehshort{\hatcurplanetnumxxxB}}{\hatcurSMEiizfehshort{\hatcurplanetnumxxxB}}}
\newcommand{\hatcurSMEloggxxxB}{\ifthenelse{\equal{\hatcurSMEversionxxxB}{i}}{\hatcurSMEilogg{\hatcurplanetnumxxxB}}{\hatcurSMEiilogg{\hatcurplanetnumxxxB}}}
\newcommand{\hatcurSMEvsinxxxB}{\ifthenelse{\equal{\hatcurSMEversionxxxB}{i}}{\hatcurSMEivsin{\hatcurplanetnumxxxB}}{\hatcurSMEiivsin{\hatcurplanetnumxxxB}}}
\newcommand{\hatcurSMEvmacxxxB}{\ifthenelse{\equal{\hatcurSMEversionxxxB}{i}}{\hatcurSMEivmac{\hatcurplanetnumxxxB}}{\hatcurSMEiivmac{\hatcurplanetnumxxxB}}}
\newcommand{\hatcurSMEvmicxxxB}{\ifthenelse{\equal{\hatcurSMEversionxxxB}{i}}{\hatcurSMEivmic{\hatcurplanetnumxxxB}}{\hatcurSMEiivmic{\hatcurplanetnumxxxB}}}

%% file: newcommand_spec3.tex
\newcommand{\hatcurxxxC}{HAT-P-22}
\newcommand{\hatcurbxxxC}{HAT-P-22b}
\newcommand{\hatcurcxxxC}{HAT-P-22c}

\newcommand{\hatcurplanetnumxxxC}{22}

\newcommand{\hatcurCChdxxxC}{HD~233731}

\newcommand{\hatcurRVgammaabsxxxC}{\hatcurDSgamma{\hatcurplanetnumxxxC}}                           

\newcommand{\hatcurRVgammarelxxxC}{\hatcurRVgamma{\hatcurplanetnumxxxC}}                           

\newcommand{\hatcurCCtassvixxxC}{\ensuremath{0.992\pm0.066}}                  

\newcommand{\hatcurSMEversionxxxC}{ii}                                       

\newcommand{\hatcurisoshortxxxC}{YY}
\newcommand{\hatcurisofullxxxC}{Yonsei-Yale (YY)}
\newcommand{\hatcurisocitexxxC}{yi:2001}

\newcommand{\hatcurlumindxxxC}{\arstar}

\newcommand{\hatcurjhkfilsetxxxC}{ESO}

\newcommand{\hatcurSMEteffxxxC}{\ifthenelse{\equal{\hatcurSMEversionxxxC}{i}}{\hatcurSMEiteff{\hatcurplanetnumxxxC}}{\hatcurSMEiiteff{\hatcurplanetnumxxxC}}}
\newcommand{\hatcurSMEzfehxxxC}{\ifthenelse{\equal{\hatcurSMEversionxxxC}{i}}{\hatcurSMEizfeh{\hatcurplanetnumxxxC}}{\hatcurSMEiizfeh{\hatcurplanetnumxxxC}}}
\newcommand{\hatcurSMEzfehshortxxxC}{\ifthenelse{\equal{\hatcurSMEversionxxxC}{i}}{\hatcurSMEizfehshort{\hatcurplanetnumxxxC}}{\hatcurSMEiizfehshort{\hatcurplanetnumxxxC}}}
\newcommand{\hatcurSMEloggxxxC}{\ifthenelse{\equal{\hatcurSMEversionxxxC}{i}}{\hatcurSMEilogg{\hatcurplanetnumxxxC}}{\hatcurSMEiilogg{\hatcurplanetnumxxxC}}}
\newcommand{\hatcurSMEvsinxxxC}{\ifthenelse{\equal{\hatcurSMEversionxxxC}{i}}{\hatcurSMEivsin{\hatcurplanetnumxxxC}}{\hatcurSMEiivsin{\hatcurplanetnumxxxC}}}
\newcommand{\hatcurSMEvmacxxxC}{\ifthenelse{\equal{\hatcurSMEversionxxxC}{i}}{\hatcurSMEivmac{\hatcurplanetnumxxxC}}{\hatcurSMEiivmac{\hatcurplanetnumxxxC}}}
\newcommand{\hatcurSMEvmicxxxC}{\ifthenelse{\equal{\hatcurSMEversionxxxC}{i}}{\hatcurSMEivmic{\hatcurplanetnumxxxC}}{\hatcurSMEiivmic{\hatcurplanetnumxxxC}}}

%% file: newcommand_spec4.tex
\newcommand{\hatcurxxxD}{HAT-P-23}
\newcommand{\hatcurbxxxD}{HAT-P-23b}
\newcommand{\hatcurcxxxD}{HAT-P-23c}

\newcommand{\hatcurplanetnumxxxD}{23}

\newcommand{\hatcurRVgammaabsxxxD}{\hatcurDSgamma{\hatcurplanetnumxxxD}}                           

\newcommand{\hatcurRVgammarelxxxD}{\hatcurRVgamma{\hatcurplanetnumxxxD}}                           

\newcommand{\hatcurCCtassvixxxD}{\ensuremath{0.68\pm0.12}}                  

\newcommand{\hatcurSMEversionxxxD}{ii}                                       

\newcommand{\hatcurisoshortxxxD}{YY}
\newcommand{\hatcurisofullxxxD}{Yonsei-Yale (YY)}
\newcommand{\hatcurisocitexxxD}{yi:2001}

\newcommand{\hatcurlumindxxxD}{\arstar}

\newcommand{\hatcurjhkfilsetxxxD}{ESO}

\newcommand{\hatcurSMEteffxxxD}{\ifthenelse{\equal{\hatcurSMEversionxxxD}{i}}{\hatcurSMEiteff{\hatcurplanetnumxxxD}}{\hatcurSMEiiteff{\hatcurplanetnumxxxD}}}
\newcommand{\hatcurSMEzfehxxxD}{\ifthenelse{\equal{\hatcurSMEversionxxxD}{i}}{\hatcurSMEizfeh{\hatcurplanetnumxxxD}}{\hatcurSMEiizfeh{\hatcurplanetnumxxxD}}}
\newcommand{\hatcurSMEzfehshortxxxD}{\ifthenelse{\equal{\hatcurSMEversionxxxD}{i}}{\hatcurSMEizfehshort{\hatcurplanetnumxxxD}}{\hatcurSMEiizfehshort{\hatcurplanetnumxxxD}}}
\newcommand{\hatcurSMEloggxxxD}{\ifthenelse{\equal{\hatcurSMEversionxxxD}{i}}{\hatcurSMEilogg{\hatcurplanetnumxxxD}}{\hatcurSMEiilogg{\hatcurplanetnumxxxD}}}
\newcommand{\hatcurSMEvsinxxxD}{\ifthenelse{\equal{\hatcurSMEversionxxxD}{i}}{\hatcurSMEivsin{\hatcurplanetnumxxxD}}{\hatcurSMEiivsin{\hatcurplanetnumxxxD}}}
\newcommand{\hatcurSMEvmacxxxD}{\ifthenelse{\equal{\hatcurSMEversionxxxD}{i}}{\hatcurSMEivmac{\hatcurplanetnumxxxD}}{\hatcurSMEiivmac{\hatcurplanetnumxxxD}}}
\newcommand{\hatcurSMEvmicxxxD}{\ifthenelse{\equal{\hatcurSMEversionxxxD}{i}}{\hatcurSMEivmic{\hatcurplanetnumxxxD}}{\hatcurSMEiivmic{\hatcurplanetnumxxxD}}}
